\documentclass[aps,prb,twocolumn,amsmath,amssymb,nofootinbib,eqsecnum,tighten,showpacs,floatfix]{revtex4}

\usepackage[dvips]{color} 
\usepackage{graphicx}
\usepackage{dcolumn} 
\usepackage{bm}      
\usepackage{epic}
\usepackage{longtable}
\usepackage{pslatex} 

\begin{document}

\title{
Density of states for the $\pi$-flux state 
with bipartite real random hopping only:
A weak disorder approach
}

\author{C.\ Mudry}
\affiliation{Paul Scherrer Institute,
             CH-5232 Villigen PSI, 
             Switzerland\\
             and Yukawa Institute for Theoretical Physics,
             Kyoto University,
             Kyoto 606-8502, 
             Japan}
\author{S.\ Ryu}
\affiliation{Department of Applied Physics,
             University of Tokyo,
             7-3-1 Hongo Bunkyo-ku,
             Tokyo 113-8656,
             Japan}
\author{A.\ Furusaki}
\affiliation{Yukawa Institute for Theoretical Physics,
             Kyoto University,
             Kyoto 606-8502, 
             Japan}

\date{\today}

\begin{abstract}

Gade [Nucl.\ Phys.\ B \textbf{398}, 499 (1993)]
has shown that the {}{density of states}
for a particle hopping on a two-dimensional 
bipartite lattice in the presence of weak disorder 
and in the absence of time-reversal symmetry
(chiral unitary universality class)
is anomalous in the vicinity of the band center $\varepsilon=0$
whenever the disorder preserves the sublattice symmetry.
More precisely, using a nonlinear $\sigma$ model that encodes the sublattice
(chiral) symmetry and the absence of time-reversal symmetry
she argues that the disorder average {}{density of states}
diverges as
$
 |\varepsilon|^{-1}\exp\left(-c|\ln\varepsilon|^\kappa\right)
$
with $c$ some nonuniversal positive constant and
$\kappa=1/2$ a universal exponent.
Her analysis has been extended to the 
case when time-reversal symmetry is present
(chiral orthogonal universality class)
for which the same exponent $\kappa=1/2$ was predicted.
{{}}{Motrunich \textit{et al.}}\ 
[Phys.\ Rev. B \textbf{65}, 064206 (2002)]
have argued that the exponent $\kappa=1/2$ does not apply
to the 
{}{density of states}
in the chiral orthogonal universality class.
They predict that
$\kappa=2/3$ instead.
We confirm the 
analysis of {{}}{Motrunich \textit{et al.}}\ within a field theory
for two flavors of Dirac fermions subjected to
two types of weak uncorrelated random potentials:
a purely imaginary vector potential and
a complex valued mass potential.
This model is the naive continuum limit of a model describing
a particle hopping on a square lattice in the background
of a $\pi$-flux phase and subjected to weak disorder 
that preserves the sublattice symmetry and time-reversal invariance.
By commonly held universality arguments, this model
is believed to belong to the chiral orthogonal universality class.
Our calculation relies in an
essential way on the existence of infinitely many local composite
operators with negative anomalous scaling dimensions.

\end{abstract}

\maketitle

\section{Introduction}

The effect of disorder on the density of states (DOS) 
in the problem of Anderson localization is known to be 
rather mild: In general, disorder smoothes the DOS.\cite{Edwards71,Wegner81}
For example, the logarithmic Van Hove singularity in the
DOS at the band center for a single-particle 
undergoing uniform nearest-neighbor
hopping on a square lattice is washed out by disorder 
in two dimensions (2D).\cite{Pryor92,Gruzberg01}
Consequently, the signature of disorder-induced critical behavior 
in Anderson localization is usually not captured by the DOS, and
one needs to resort to more complex probes such as conductance statistics, 
energy levels statistics, or multifractal spectra that are 
sensitive to a delocalization transition.

Exceptions to the rule that the DOS is smooth in Anderson localization
can occur at the band edges or at the band center.\cite{Wegner81} 
The band center is very special whenever the disorder preserves
``particle-hole'' symmetry, i.e.,
a symmetry that guarantees that energy eigenvalues always occur
in pairs of opposite signs for any given realization of the disorder.
The particle-hole symmetry can take two different forms:
(i) the sublattice
symmetry\cite{Dyson53,Theodorou76,Eggarter78,Gade93,Ludwig94,Nersesyan94}
(also called chiral symmetry) 
which is relevant in the context of
Hamiltonians defined on bipartite lattices
and is implemented by a unitary transformation
(changing the sign of the wave functions
on every site belonging to one sublattice but not on the other)
and
(ii) the Bogoliubov--de Gennes (BdG)
symmetry\cite{Senthil00,Bocquet00,Brouwer00} 
which is relevant in the context of 
disordered superconductors and is implemented by an antiunitary
transformation.\cite{Altland97}
The special role played by the band center when the disorder preserves 
particle-hole symmetry follows from the enhanced level repulsion
felt by energy eigenvalues close to the band center as a result of
their mirror images.
Naively, one would expect that level repulsion results in a depletion
of the DOS close to the band center.
This expectation is in fact always realized in finite systems
and is relevant to quantum dots or to level statistics in the core
of vortices in a superconductor.\cite{RMT}
However, it has been known since the work
of Dyson\cite{Dyson53} on disordered one-dimensional chains
that the DOS can, in spite of the enhanced level repulsion, 
diverge at the band center in the thermodynamic limit.
The band center must then represent ``some sort'' of disorder-induced
critical point characterized by a delocalization transition
of ``some kind.'' Indeed, imagine that there is no delocalization
transition, i.e., that the localization length is finite at the
band center. If so, the naive expectation of a depletion of 
the DOS as a result of enhanced level repulsion must apply 
in the thermodynamic limit as the system behaves in effect 
as a sampling of uncorrelated boxes 
of linear size of the order of the localization length
for sufficiently long length scales.

Anderson localization with particle-hole symmetry thus represents
a laboratory in which disorder-induced critical behavior can be studied
both in 1D and 2D.
In 1D a wealth of complementary approaches 
based on the nonlinear $\sigma$ model (NLSM),\cite{Bundschuh98,Altland01} 
real-space renormalization group (RG) techniques,\cite{Motrunich01}
extensions of the Dorokhov-Mello-Pereyra-Kumar (DMPK) 
equation,\cite{Brouwer98,Brouwer00,Titov01}
direct studies of critical zero modes,\cite{Shelton98,Hastings01}
and others\cite{Comtet95,Balents97,Mathur97}
are available and have been used to provide a rather detailed
description of the delocalization transition at the band center.
In 2D, the use of the NLSM was pioneered by Gade and Wegner
in the context of Anderson localization with particle-hole symmetry
implemented by a sublattice symmetry.\cite{Gade93}
They predicted that 
the disorder average {}{DOS} per
unit volume and energy diverges upon approaching the band center
$\varepsilon=0$ as 
\begin{eqnarray}
|\varepsilon|^{-\upsilon}
\exp\left(-c|\ln\varepsilon|^{\kappa}\right),
\quad
\upsilon=1,
\quad
\kappa=\frac{1}{2}.
\label{eq: DOS Gade}
\end{eqnarray}
Here, $c$ is a positive nonuniversal constant, whereas the
exponents $\upsilon$ and $\kappa$ are universal. Moreover,
neither $\upsilon$ nor $\kappa$ depends on whether
time-reversal symmetry is broken or not. The NLSM approach of
Gade and Wegner has been extended to
the random $\pi$-flux phase\cite{Fukui99,Guruswamy00}
and to the random flux model.\cite{Altland99} It also has been
refined to account for the expectation value of
the staggered {}{DOS} (Ref.\ \onlinecite{Fabrizio00})
and for the presence of Wess-Zumino-Witten (WZW)
terms.\cite{Guruswamy00} In all cases, 
the {}{DOS} (\ref{eq: DOS Gade}) was recovered.
Particle-hole symmetry is also essential to understand
the role played by disorder in a mean-field and non-self-consistent
treatment of superconductivity.\cite{Altland97}
In 2D, the disorder average {}{DOS}
was predicted\cite{Senthil00,Bocquet00} to diverge as
\begin{eqnarray}
|\ln\varepsilon|^{\kappa}
\end{eqnarray}
when spin-rotation symmetry is broken.
The universal exponent $\kappa=1/2$ in the presence of
time-reversal symmetry and $\kappa=1$
in the absence of time-reversal symmetry.

{{}}{Motrunich \textit{et al.}}\ in
Ref.\ \onlinecite{Motrunich02}
have proposed that the exponent
$\kappa$ is not given by $1/2$ 
as suggested by various field-theoretical calculations leading to
Eq.\ (\ref{eq: DOS Gade})
(see Refs.\ \onlinecite{Gade93}
and \onlinecite{Fukui99,Altland99,Fabrizio00,Guruswamy00})
but by the \textit{larger} exponent
\begin{eqnarray}
\kappa=\frac{2}{3}
\label{eq: DOS Motrunich}.
\end{eqnarray}
{{}}{
The method by which {{}}{Motrunich \textit{et al.}}\ reach their
conclusion in Ref.\ \onlinecite{Motrunich02} is based on
(i)   a real-space RG analysis 
      that is intrinsically a strong-disorder technique,\cite{Dasgupta80}
(ii)  a combination of exact variational bounds for strong random vector 
      potential disorder with the computation of the DOS in the weak-disorder 
      limit by Ludwig \textit{et al.}\ 
      in Ref.\ \onlinecite{Ludwig94} and with the 
      computation of the $\beta$ function for the variance of the 
      random vector potential by
      Guruswamy \textit{et al.}\ in Ref.\ \onlinecite{Guruswamy00}, and
(iii) a connection to the problem of monomer free energies in some problems 
      of dimer covering.
                }
In this paper we undertake the task of understanding the discrepancy
between Eqs.\ (\ref{eq: DOS Gade}) and (\ref{eq: DOS Motrunich})
for the Hatsugai-Wen-Kohmoto (HWK) 
model\cite{Hatsugai97}
in the limit of weak disorder at microscopic length scales
and within a purely field-theoretical approach.
{}{
We shall show that the RG analysis for the average DOS in the HWK model
performed in Ref.\ \onlinecite{Guruswamy00}
that leads to Eq.\ (\ref{eq: DOS Gade})
is not complete in that it fails 
to capture the broadness of 
the \textit{local} density of states (LDOS) distribution. 
A consistent RG analysis for the DOS
can, however, be constructed and yields 
Eq.\ (\ref{eq: DOS Motrunich})
in the HWK model.
               }

The HWK model is the naive continuum limit for a model of
a particle undergoing nearest-neighbor
hopping on a rectangular lattice in the
background of $\pi$-flux phase.
The microscopic disorder is introduced by allowing
real-valued random fluctuations of the nearest-neighbor
hopping amplitudes about the $\pi$-flux phase. 
The continuum limit of the clean spectrum
is approximated by two flavors of Dirac fermions.
The microscopic disorder is assumed to be sufficiently weak for
the nodal structure of the clean energy spectrum to
survive the perturbation by the disorder.
It was shown in Ref.\ \onlinecite{Hatsugai97}
that the disorder manifests itself by
a purely imaginary random vector potential and by a 
random complex valued mass in the naive continuum limit.
Furthermore, boundary conditions were chosen so as to
guarantee the existence of a zero mode for any
realization of the disorder and a numerical study
of zero modes confirmed the theoretical expectation
that zero modes are critical (multifractal).
Numerical studies of the DOS for the HWK model
are inconclusive due to severe limits on the achieved
energy resolution.\cite{Morita97,Ryu02}
{}From a purely symmetry point of view the HWK
model belongs to the chiral orthogonal universality
class since time-reversal symmetry is preserved.
If symmetries alone control the asymptotic
behavior of the {}{DOS} in the vicinity of the band center,
one would expect that the disorder average {}{DOS} 
is of the form given in
Eq.\ (\ref{eq: DOS Gade}).
This is indeed what is found in Ref.\ \onlinecite{Guruswamy00}, where
it is shown that
(i)  {{}}{the HWK model realizes a nearly conformal field 
     theory and}
{}{
(ii) the DOS (\ref{eq: DOS Gade}) follows from 
     solving a Callan-Symanzik equation that depends on three
     coupling constants.
                }

In this paper we shall build on the remarkable results (i) of
Guruswamy \textit{et al.}\ in Ref.\ \onlinecite{Guruswamy00} to
identify an infinite but countable set of relevant local operators
with negative anomalous scaling dimensions in the HWK model. 
The physical interpretation of these relevant operators
is that they govern the dependence on energy $\varepsilon$
of all moments of the LDOS provided $\varepsilon$ is
sufficiently close to the band center $\varepsilon=0$. 
Their relevance reflects the fact that the probability distribution 
for the LDOS becomes very broadly distributed 
in the thermodynamic limit as its  distribution is governed by
the (multifractal) critical state at the band center.
{}{
Correspondingly, we shall show that the disorder average DOS
$\nu(\varepsilon)$
obeys a Callan-Symanzik equation
that depends on infinitely many relevant coupling constants. 
                }
Following an \textit{approximation} 
developed by Carpentier and Le Doussal for the random phase $XY$ model
in 2D,\cite{Carpentier00}
we reduce the Callan-Symanzik equation with infinitely
many relevant coupling constants to the Callan-Symanzik equation
of Guruswamy \textit{et al.}\ with the caveat that
the anomalous scaling dimension in (ii)
{}{
that corresponds to the annealed average value of the dynamical exponent 
must be replaced by 
the anomalous scaling dimension that
corresponds to the quenched average value of the dynamical exponent.
                }
In doing so we recover Eq.\ (\ref{eq: DOS Motrunich}).

The physical origin of the infinite but countable family of operators
with negative anomalous scaling dimensions can be traced to the
multifractality of zero modes supported by the HWK model
when the random complex mass is switched off.
The explicit interplay between multifractality of zero modes
and negative scaling dimensions in a conformal field theory (CFT)
was studied in 
Refs.\ \onlinecite{Ludwig94}
and
\onlinecite{Chamon-Mudry96,Mudry96,Chamon96,Kogan96,Castillo97,Mudry99}
for a single flavor of a Dirac fermion subjected to a random vector
potential. It was shown in  Refs.\ \onlinecite{Chamon96,Kogan96,Castillo97}
that the multifractal spectrum is characterized by a freezing transition.
The multifractal spectrum probes the scaling 
of the support of a normalized zero mode 
whose squared amplitude (height) is some prescribed value
as a function of $L$ in a finite box of volume $L^2$. 
As the height is increased, 
the multifractal spectrum undergoes a freezing transition. 
{}{
Below the freezing transition, the support scales with the system size
to some finite power.
Above the freezing transition, the support does not scale 
anymore with the system size.
                }
A similar freezing transition occurs when the height is decreased
towards a minimum value of the squared amplitude of zero modes.
{}{
It was argued in 
Refs.\ \onlinecite{Motrunich02} and \onlinecite{Horovitz02}
that the freezing transition in the multifractal spectrum mirrors 
itself in the DOS for a single Dirac fermion subjected to
a random vector potential.
The relation between the multifractal spectrum and the scaling of the DOS
is embodied by the dynamical exponent that relates scaling with energy to 
scaling with system size. 
                 }
It is thus the dynamical exponent
that undergoes a freezing transition. We will argue that
the same mechanism applies, with suitable modifications, to the HWK
model.

Although the emphasis is on Anderson localization,
the models considered in this paper are also intimately connected to
disordered Coulomb gases, random $XY$, and other 2D
disordered systems from classical statistical physics.\cite
{Chamon-Mudry96,Bernard95,Chamon96,Mudry99,Carpentier00,Carpentier01,Motrunich02,Horovitz02}
Perhaps the most surprising connection between Anderson localization
and a classical stochastic model is the one conjectured
in Ref.\ \onlinecite{Chamon96} 
between the multifractal spectrum of zero modes
and the freezing transition of directed polymers on a Cayley tree
(see Refs.\ \onlinecite{Derrida88} and \onlinecite{Buffet93}). 
Carpentier and Le Doussal have convincingly argued that
a consistent RG analysis of the multifractal spectrum of zero modes
reduces to solving the so-called
Kolmogorov-Petrovsky-Piscounov (KPP)
nonlinear diffusion equation.\cite{Carpentier01}
Their approach yields new insights into the universality 
of the multifractal spectrum in terms of  
the unique properties of the asymptotic solutions of 
the KPP equation. It also gives a RG justification for the conjecture
in Ref.\ \onlinecite{Chamon96}
by building on the fact that
it is known from Ref.\ \onlinecite{Derrida88}
that the transformation law of the
generating function for the random partition function
of directed polymers on a Cayley tree 
under an infinitesimal RG step
is also governed by the KPP equation.
Hereto we shall see that the Callan-Symanzik equation obeyed by
the 
{}{DOS} 
in the HWK model is intimately related to the KPP equation.

Our paper is organized as follows.
Lattice and continuum versions of the HWK Hamiltonians are defined
in Sec.\ \ref{sec: Definitions} where we also define
the Motrunich-Damle-Huse (MDH) model.
We choose a supersymmetric (SUSY) field theory to
represent the 
{}{DOS}  
in Sec.\ \ref{sec: DOS and field theory}.
After these preliminaries, we derive the 
{}{DOS} 
for the MDH model in Sec.\ 
\ref{sec: DOS for the MDH model}. 
We compute the 
{}{DOS} 
for the HWK model
in Sec.\
\ref{sec: DOS for the HWK model}.
We conclude in Sec.\ \ref{sec: Conclusions}.
Some technical details are summarized in the Appendixes.

\section{Definitions}
\label{sec: Definitions}

\subsection{Lattice models}

We are concerned with the naive continuum limit
of a \textit{subclass} of lattice Hamiltonians belonging to the 
chiral orthogonal universality class. 
Consider some two-dimensional bipartite lattice---say, a lattice that 
is rectangular in the bulk. 
One sublattice is denoted $A$, the other $B$. 
There are $N_A$ sites on sublattice $A$ and 
$N_B$ sites on sublattice $B$. Without loss of 
generality, assume that $N_A\geq N_B$. 
The random Hamiltonian is then represented by
\begin{eqnarray}
\mathcal{H}=
\left(
\begin{array}{cc}
0 
&
T_{AB}
\\
T_{BA}
&
0 
\end{array}
\right),
\label{eq: def chiral ortho Hamil}
\end{eqnarray}
where $T_{AB}$ is a $N_A\times N_B$ real-valued matrix 
with the random matrix elements
$
(T_{AB})_{ij}=
t_{ij}
$
if $ij$ are nearest-neighbor sites with $i\in A$
and
$
(T_{AB})_{ij}=
0
$
otherwise,
while the $N_B\times N_A$ matrix $T_{BA}$ 
is the transpose of $T_{AB}$:
$
T_{BA}=(T_{AB})^{\hbox{t}}
$.
Random Hamiltonians of the form (\ref{eq: def chiral ortho Hamil})
are generic representatives of the chiral orthogonal universality class.
Wave functions in the $N$-dimensional Hilbert space
$
(N:=N_A+N_B)
$
are represented by
\begin{eqnarray}
\Psi=
\left(
\begin{array}{cc}
\Psi_A
\\
\Psi_B
\end{array}
\right).
\end{eqnarray}
By construction, $\mathcal{H}$ changes by a sign
under the unitary transformation
\begin{eqnarray}
\Psi_A\to+\Psi_A,
\qquad
\Psi_B\to-\Psi_B.
\label{eq: def chiral trsf on lattice}
\end{eqnarray}
Zero modes have the form
\begin{eqnarray}
\Psi=
\left(
\begin{array}{cc}
\Psi_A
\\
0 
\end{array}
\right),
\end{eqnarray}
where
\begin{eqnarray}
0= 
T_{BA}\Psi_A=
(T_{AB})^{\hbox{t}}\Psi_A.
\label{eq: condition on zero modes}
\end{eqnarray}
Zero modes are thus eigenstates of the unitary
transformation (\ref{eq: def chiral trsf on lattice}).
There are $N_A-N_B$ zero modes.

The HWK lattice model is given
by the Hamiltonian $\mathcal{H}_{\hbox{\tiny HWK}}$
of the form (\ref{eq: def chiral ortho Hamil})
whereby it is assumed that
(i)
a $\pi$-flux phase threads every 
elementary plaquette $P$ 
of the rectangular lattice and
(ii) 
there are small random fluctuations 
$\delta t_{ij}$ of the 
nearest-neighbor hopping amplitude:
\begin{equation}
t_{ij}^{\vphantom{(0)}}=
t_{ij}^{(0)}
+
\delta    
t_{ij}^{\vphantom{(0)}},
\qquad
\prod_{ij\in P}t_{ij}^{(0)}= 
-t^4.
\label{eq: def random hopping HWK}
\end{equation}
Here, $t$ is the characteristic energy scale 
($\propto$ bandwidth) for the clean system.
In Ref.\ \onlinecite{Hatsugai97},
the multifractal spectrum of zero modes
of $\mathcal{H}_{\hbox{\tiny HWK}}$
was studied numerically and shown to agree within
numerical error bars with a theoretical prediction
made in Refs.\ \onlinecite{Ludwig94} and
\onlinecite{Mudry96,Chamon96,Kogan96,Castillo97}.
In subsequent papers the DOS was also studied 
numerically.\cite{Morita97,Ryu02}
Unfortunately, the lack of energy resolution was so severe
that the observation of a diverging DOS could only
be established in Ref.\ \onlinecite{Ryu02}
through a tiny upturn of the DOS away from an
algebraic decaying crossover regime. 

{{}}{Motrunich \textit{et al.}}\ in Ref.\ \onlinecite{Motrunich02}
proposed a different parametrization of the random 
fluctuations of the nearest-neighbor hopping amplitudes.
Their lattice Hamiltonian $\mathcal{H}_{\hbox{\tiny MDH}}$
is of the form (\ref{eq: def chiral ortho Hamil})
but with
\begin{equation}
t_{ij}^{\vphantom{(0)}}=
e^{+\phi_i}\,
t_{ij}^{(0)}\,
e^{-\phi_j},
\qquad
\prod_{ij\in P}t_{ij}^{(0)}= 
-t^4,
\end{equation}
instead of Eq.\ (\ref{eq: def random hopping HWK}).
The $\phi$'s are real-valued, white-noise-correlated,
Gaussian random variables of vanishing means.
With the help of the 
$N_A\times N_A$ diagonal matrix
\begin{equation}
U_A:=
\left(
\begin{array}{ccc}
e^{+\phi_1} & & \\
 & \ddots &  \\
 & & e^{+\phi_{N_A}}\\
\end{array}
\right)
\label{eq: def UA}
\end{equation}
and of the $N_B\times N_B$ diagonal matrix
\begin{eqnarray}
U_B:=
\left(
\begin{array}{ccc}
e^{+\phi_{N_A+1}} & & \\
 & \ddots & \\
 & & e^{+\phi_{N_A+N_B}} \\
\end{array}
\right),
\label{eq: def UB}
\end{eqnarray}
the MDH random Hamiltonian becomes
\begin{eqnarray}
&&
\mathcal{H}_{\hbox{\tiny MDH}}=
\left(
\begin{array}{cc}
0 
&
U_A\,
T_{AB}^{(0)}\,
(U_B)^{-1}
\\
(U_B)^{-1}\,
\left(T_{AB}^{(0)}\right)^{\hbox{t}}\,
U_A
&
0 
\end{array}
\right),
\nonumber\\
&&
\label{eq: lattice MDH after gauge trsf}
\end{eqnarray}
where $T_{AB}^{(0)}$ is a $N_A\times N_B$ matrix made of
$t_{ij}^{(0)}$.
In this model, it is straightforward to construct 
explicitly zero modes.
To each solution
\begin{eqnarray}
\Psi'=
\left(
\begin{array}{cc}
\Psi_A'
\\
0 
\end{array}
\right)
\end{eqnarray}
of 
\begin{eqnarray}
0= 
T_{BA}^{(0)}\,\Psi_A'=
\left(T_{AB}^{(0)}\right)^{\hbox{t}}\,\Psi_A'
\end{eqnarray}
[i.e., Eq.\ (\ref{eq: condition on zero modes})
 without disorder],
there corresponds a zero mode of 
$\mathcal{H}_{\hbox{\tiny MDH}}$
given by
\begin{eqnarray}
\Psi=
\left(
\begin{array}{cc}
\left(U_A^{\ }\right)^{-1}
&
0 
\\
0 
&
U_B^{\ }
\end{array}
\right)
\Psi'=
\left(
\begin{array}{cc}
\left(U_A^{\ }\right)^{-1}\Psi_A'
\\
0 
\end{array}
\right).
\label{eq: generic zero mode in MDH}
\end{eqnarray}
This explicit construction of zero modes will have a
counterpart in the field theory later on
and plays a very special role. 
The {imaginary gauge chiral transformation} 
(\ref{eq: generic zero mode in MDH}) turns 
the Schr\"odinger equation at any finite $\varepsilon$ into
\begin{eqnarray}
&&
T_{AB}^{(0)}\,
\Psi_{B}'=
\varepsilon\,
\left(U_A^{\ }\right)^{-2}\,
\Psi_A',
\nonumber\\
&&
\left(T_{AB}^{(0)}\right)^{\hbox{t}}\,
\Psi_{A}'=
\varepsilon\,
\left(U_B^{\ }\right)^{+2}\,
\Psi_B'.
\end{eqnarray}

\smallskip

\subsection{Naive continuum limit}

Common to the HWK and MDH lattice Hamiltonians
in the clean limit is the property that the Fermi 
surface collapses to four Fermi points at 
$(\pm\pi/2,\pm\pi/2)$ in the
Brillouin zone at half filling.
Correspondingly,
the clean DOS has a V-shaped singularity at the band
center. Linearization of the spectrum around those four
Fermi points describes two copies of ``relativistic''
particles moving with the Fermi velocity $v_f:=2\mathfrak{a}t$,
$\mathfrak{a}$ being the lattice spacing.
These are the two flavors of Dirac fermions.
Unless specified, we will set $v_f$ to 1.

The difference in the parametrization of the disorder
in the HWK and MDH lattice Hamiltonians survives the continuum
limit. For sufficiently weak disorder, the naive continuum limit
for the HWK lattice Hamiltonian reads\cite{Hatsugai97}
\begin{subequations}
\begin{equation}
H_{\hbox{\tiny HWK}}=
\left(
\begin{array}{cc}
0 & D_{\hbox{\tiny HWK}} \\
D^{\dag}_{\hbox{\tiny HWK}} & 0 \\
\end{array}
\right),
\label{eq: HWK Hamiltonian}
\end{equation}
with
\begin{equation}
D_{\hbox{\tiny HWK}}=
i\sigma_\mu\partial_\mu
+i\sigma_\mu A_\mu
+i\sigma_0A_0
+\sigma_3A_3
\label{eq: HWK upper right block}
\end{equation}
and
\begin{equation}
D^\dag_{\hbox{\tiny HWK}}=
i\sigma_\mu\partial_\mu
-i\sigma_\mu A_\mu
-i\sigma_0A_0
+\sigma_3A_3.
\label{eq: HWK lower left block}
\end{equation}
\label{eq: naive cont limit HWK}
\end{subequations}

\par\noindent Pauli matrices 
$\boldsymbol{\sigma}\equiv(\sigma_1,\sigma_2)$,
together with the unit $2\times2$ matrix $\sigma_0$ and the third
Pauli matrices
$\sigma_3$ have been introduced.
Summation over the repeated indices ($\mu=1,2$) is assumed throughout
this paper.
This Hamiltonian is invariant under time reversal, i.e.,
\begin{equation}
\mathcal{T}
\left(H_{\hbox{\tiny HWK}}\right)^*
\mathcal{T}=
H_{\hbox{\tiny HWK}},
\quad
\mathcal{T}:=
\left(
\begin{array}{cc}
\sigma_1 & 0 \\
0 & -\sigma_1 \\
\end{array}
\right).
\end{equation}
It also changes by a sign under charge conjugation 
\begin{equation}
\mathcal{C}
H_{\hbox{\tiny HWK}}
\mathcal{C}=
-H_{\hbox{\tiny HWK}},
\quad
\mathcal{C}:=
\left(
\begin{array}{cc}
\sigma_0 & 0 \\
0 & -\sigma_0 \\
\end{array}
\right),
\end{equation}
which is nothing but the continuum counterpart of
the transformation that implements the
transformation $T_{AB}\to-T_{AB}$ 
in Eq.\ (\ref{eq: def chiral trsf on lattice}).
The $\sigma$'s encode the spinor grading of a single flavor
of Dirac fermions.
The (real-valued) disorder is represented by
(i)
the ``{purely imaginary}'' random vector potential 
$\boldsymbol{A}=(A_1,A_2)\in\mathbb{R}^2$,
(ii)
the ``{purely imaginary}'' random scalar potential 
(chemical potential) $A_0\in\mathbb{R}$, and
(iii)
the random mass potential $A_3\in\mathbb{R}$.

The terminology ``{purely imaginary}'' comes from the factor of 
$i$ multiplying $\boldsymbol{A}$ and $A_0$ in the 
operator $D^{\vphantom{\dag}}_{\hbox{\tiny HWK}}$; 
i.e., $\boldsymbol{A}$ and $A_0$ parametrize the
anti-Hermitian contributions to 
$D^{\vphantom{\dag}}_{\hbox{\tiny HWK}}$. 
The Hermitian contributions to
$D^{\vphantom{\dag}}_{\hbox{\tiny HWK}}$ 
are the kinetic energy 
$i\boldsymbol{\sigma}\cdot\boldsymbol{\partial}$
and the random mass contributions $\sigma_3 A_3$.
With a prejudice towards a field-theoretical approach,
it is standard practice to assume that $\boldsymbol{A}$,
$A_0$, and $A_3$ are all 
independent white-noise-correlated Gaussian random variables
of vanishing means.\cite{Fradkin80's}
The implicit assumption made here is that 
the detailed shape of the disorder should not matter as long as it is
weak and short-range correlated in space.
So far the discussion is generic to the chiral orthogonal universality
class.
We now specialize to a subclass of the chiral orthogonal universality
class by, following Hatsugai \textit{et al.}, making a further important
simplification.
Motivated by analytical continuation
of a one-loop RG calculation performed in Ref.\ \onlinecite{Bernard95}
for a single flavor of Dirac fermions subjected to 
``{purely real}''
random scalar, vector, and mass potentials,\cite{footnote purely real A} 
we assume that, on the one hand,
the two components of the vector potential have the same variance 
$g^{\ }_A$
and, on the other hand, the random scalar potential and the random mass 
have the same variance 
$g^{\ }_M$ (disorder averaging is denoted by an overbar):
\begin{eqnarray}
\overline{
A^{\vphantom{*}}_\mu(\bm{x})A^{\vphantom{*}}_\nu(\bm{y})
         }
=
\left\{
\begin{array}{ll}
g^{\vphantom{*}}_A
\delta^{\vphantom{*}}_{\mu\nu}\delta(\bm{x}-\bm{y}), 
&
\mu,\nu=1,2, 
\\
&\\
g^{\vphantom{*}}_M
\delta^{\vphantom{*}}_{\mu\nu}\delta(\bm{x}-\bm{y}), 
&
\mu,\nu=0,3, 
\\
&\\
0, & \mbox{otherwise}.
\end{array}
\right.
\end{eqnarray}
This completes the definition of the HWK model.

For sufficiently weak disorder,
the naive continuum limit for the MDH Hamiltonian is given by
Eq.\ (\ref{eq: HWK Hamiltonian})
without the purely imaginary random scalar potential and random mass. 
To see this, we demand a counterpart to 
Eqs.\ (\ref{eq: lattice MDH after gauge trsf}) 
and   (\ref{eq: generic zero mode in MDH})
in the continuum limit.
First, we introduce the parametrization
\begin{eqnarray}
A_\mu=
\widetilde\partial_\mu
\Phi_1
+
\partial_\mu
\Phi_2
\quad
(\widetilde\partial_\mu:=
\epsilon_{\mu\nu}\partial_\nu,
\quad
\mu=1,2)
\end{eqnarray}
of the random vector potential in terms of two 
independent scalar fields $\Phi_1$ and $\Phi_2$,
where $\epsilon_{\mu\nu}$ is the antisymmetric tensor with
$\epsilon_{12}=1$. 
Second, we observe that
\begin{eqnarray}
&&
i\sigma_\mu(\partial_\mu+A_\mu)=
e^{+i\sigma_3\Phi_1-\sigma_0\Phi_2}\,
i\sigma_\mu\partial_\mu\,
e^{-i\sigma_3\Phi_1+\sigma_0\Phi_2},
\nonumber\\
&&
i\sigma_\mu(\partial_\mu-A_\mu)=
e^{-i\sigma_3\Phi_1+\sigma_0\Phi_2}\,
i\sigma_\mu\partial_\mu\,
e^{+i\sigma_3\Phi_1-\sigma_0\Phi_2}.
\nonumber\\
&&
\label{eq: decoupling trsf in continuum}
\end{eqnarray}
We thus identify Eqs.\ (\ref{eq: def UA}) and (\ref{eq: def UB}) with
\begin{eqnarray}
&&
U_A,
U_B\to
e^{+i\sigma_3\Phi_1-\sigma_0\Phi_2}.
\end{eqnarray}
To sum up,
in the naive continuum limit the MDH model is given by
\begin{equation}
H^{\vphantom{\dag}}_{\hbox{\tiny MDH}}=
\left(
\begin{array}{cc}
0
&
D^{\vphantom{\dag}}_{\hbox{\tiny MDH}}
\\
D^{          \dag} _{\hbox{\tiny MDH}}
&
0
\end{array}
\right),
\quad
D^{\vphantom{\dag}}_{\hbox{\tiny MDH}}=
i\sigma_\mu\left(\partial_\mu+A_\mu\right),
\label{eq: MDH Hamiltonian}
\end{equation}
where the vector potential $\bf{A}$ is randomly distributed according
to
\begin{equation}
\overline{
A_\mu(\bm{x})
A_\nu(\bm{y})
         }=
g_{A}
\delta_{\mu\nu}\delta(\bm{x}-\bm{y}),
\qquad
\mu,\nu=1,2.
\end{equation}

\section{DOS and field theory}
\label{sec: DOS and field theory}

\subsection{Generalities}

The DOS $\nu(\varepsilon)$
per unit energy and per unit volume $L^2$
for the single-particle Hamiltonian $H$ 
with eigenvalues $\varepsilon_n$
is defined by 
\begin{eqnarray}
\nu(\varepsilon)&:=&
\frac{1}{L^2}\sum_n
\delta(\varepsilon-\varepsilon_n)
\nonumber\\
&=&
\frac{1}{L^2}\sum_n
\lim_{\eta\to0}
\frac{1}{\pi}
\mathrm{Im}
\left(
\frac{1}{\varepsilon-\varepsilon_n-i\eta}
\right)
\nonumber\\
&=&
\lim_{\eta\to0}
\frac{1}{\pi L^2}
\sum_n
\frac{{}{\eta}}
     {(\varepsilon-\varepsilon_n)^2+\eta^2}.
\label{eq: smeared DOS}
\end{eqnarray}
With $\eta$ kept small but finite, Eq.\ (\ref{eq: smeared DOS})
defines the smeared DOS.
The LDOS $\nu(\varepsilon;\bm{x})$
is defined in terms of the exact normalized eigenfunctions
$\Psi_n$ of $H$ by
\begin{eqnarray}
&&
\nu(\varepsilon;\bm{x}):=
\lim_{\eta\to0}
\frac{1}{\pi}
\sum_n
\frac{\eta|\Psi_n(\bm{x})|^2}
     {(\varepsilon-\varepsilon_n)^2+\eta^2}.
\label{eq: Lorentzian local DOS}
\end{eqnarray}
When the single-particle Hamiltonian $H$ is one member of  
a statistical ensemble, it is commonly held that
{}{
the smeared DOS 
is self-averaging in the thermodynamic limit.
               }
On the other hand, 
critical statistical fluctuations of the wave function amplitudes 
at $\bm{x}$ 
open up the possibility that the (smeared) LDOS is broadly
distributed in the thermodynamic limit.

The computation of the (smeared) LDOS in field theory is
very natural as it reduces to the calculation of the expectation
value of a {local composite operator}.
For example, the smeared LDOS can be equivalently well
represented by fermionic or bosonic path integrals
\begin{widetext}
\begin{eqnarray}
\nu(\varepsilon-i\eta;\bm{x})&=&
-
\frac{1}{\pi}
\mathrm{Im}\,
\frac{
\int \mathcal{D}[\bar\psi,\psi]\,
    \psi(\bm{x})
\bar\psi(\bm{x})\,
\exp
\left[
-\int d^2\bm{r}\,
\bar\psi(\bm{r})
\left(
H-\varepsilon+i\eta
\right)
    \psi(\bm{r})
\right]
     }
     {
\int \mathcal{D}[\bar\psi,\psi]\,
\hphantom{
\bar\psi(\bm{x})
    \psi(\bm{x})\,
         }
\exp
\left[
-\int d^2\bm{r}\,
\bar\psi(\bm{r})
\left(
H-\varepsilon+i\eta
\right)
    \psi(\bm{r})
\right]
     }
\nonumber\\
&=&
-
\frac{1}{\pi}
\mathrm{Im}\,
\frac{
\int \mathcal{D}[\bar\beta,\beta]\,
    \beta(\bm{x})
\bar\beta(\bm{x})\,
\exp
\left[
-\int d^2\bm{r}\,
\bar\beta(\bm{r})
\left(
H-\varepsilon+i\eta
\right)
    \beta(\bm{r})
\right]
     }
     {
\int \mathcal{D}[\bar\beta,\beta]\,
\hphantom{
\bar\beta(\bm{x})
    \beta(\bm{x})\,
         }
\exp
\left[
-\int d^2\bm{r}\,
\bar\beta(\bm{r})
\left(
H-\varepsilon+i\eta
\right)
    \beta(\bm{r})
\right]
     },
\label{eq: fermion vs boson rep of local dos}
\end{eqnarray}
respectively. Here, $\bar\psi$ and $\psi$ denote two independent
Grassmann fields whereas $\bar\beta$ denotes the complex conjugate
of the complex-valued field $\beta$.

A SUSY representation for 
the smeared DOS can be achieved
by combining the first with the second line of 
Eq.\ (\ref{eq: fermion vs boson rep of local dos}):
\begin{subequations}
\label{eq: def SUSY nu(varepsilon-ieta) for fixed disorder}
\begin{eqnarray}
\nu(\varepsilon-i\eta)&=&
-\frac{1}{\pi L^2}{\rm Im}
 \int \mathcal{D}[\bar\psi,\psi,\bar\beta,\beta]
 \int d^2\bm{x}\,
 \psi(\bm{x})\bar\psi(\bm{x})\,
e^{
 -\int d^2\bm{r}
  \left[
  \bar\psi (\bm{r})(H-\varepsilon+i\eta)\psi (\bm{r})
 +\bar\beta(\bm{r})(H-\varepsilon+i\eta)\beta(\bm{r})
  \right]
  }
\nonumber\\
&=&
-
\frac{1}{2\pi L^2}
\mathrm{Im}\,
\left\langle
 \int d^2\bm{x}
 \left[
 \psi(\bm{x})\bar\psi(\bm{x})
 +
 \beta(\bm{x})\bar\beta(\bm{x})
 \right]
\right\rangle^{\vphantom{A^A}}_{Z_{\hbox{\tiny SUSY}}(\varepsilon-i\eta)},
\label{eq: def SUSY DOS without disorder}
\end{eqnarray}
where 
$
\left\langle
(\cdots)
\right\rangle^{\vphantom{A^A}}_{Z_{\hbox{\tiny SUSY}}(\varepsilon-i\eta)}
$
denotes averaging with the SUSY partition functions defined by
\begin{eqnarray}
&&
Z_{\hbox{\tiny SUSY}}(\varepsilon-i\eta):=
\int \mathcal{D}[\bar\psi,\psi,\bar\beta,\beta]\,
\exp\left\{
-\int d^2\bm{r}\,
\left[
\bar\psi(\bm{r})
\left(
H-\varepsilon+i\eta
\right)
     \psi(\bm{r})
+
\bar\beta(\bm{r})
\left(
H-\varepsilon+i\eta
\right)
    \beta(\bm{r})
\right]
\right\}.
\label{eq: def SUSY Z without disorder}
\end{eqnarray}
\end{subequations}
\end{widetext}
Alternatively, one can replicate either line of
Eq.\ (\ref{eq: fermion vs boson rep of local dos})
to obtain a  fermion (boson) replica representation of the DOS.

By assuming that disorder averaging restores translation invariance,
we infer that the smeared LDOS
$
\nu(\varepsilon-i\eta;\bm{x})
$
and the smeared DOS
$
\nu(\varepsilon-i\eta)
$
share the same mean value as statistical variables.
{}{
However, their distributions are likely to be very different in the
thermodynamic limit $L\to\infty$ at some fixed energy 
$\varepsilon-i\eta$
as the former is expected to be broadly distributed whereas the latter 
is expected to be self-averaging.
From now on, we will refer to the disorder average DOS as the DOS
having in mind the expectation that the DOS is self-averaging.
                }

\subsection{SUSY partition function for the LDOS
in the HWK model}

Needed is a SUSY partition function from which the
LDOS in the HWK model can be represented by the expectation value of 
some local composite operator. 
To construct  a convenient SUSY partition function we choose 
a realization of the disorder $A_0$, $A_1$, $A_2$, and $A_3$ in 
Eq.\ (\ref{eq: naive cont limit HWK})
and define the fermionic partition function
\begin{eqnarray}
&&
Z^{(f)}:=
\int\!\mathcal{D}
[\bar\psi_A,\psi_A,
 \bar\psi_B,\psi_B]
e^{-\int d^2\bm{r}\,
\left[
\mathcal{L}^{(f)}_{AB}
+
\mathcal{L}^{(f)}_{BA}
+
\mathcal{L}^{(f)}_{\varepsilon}
\right]
  },
\nonumber\\&&\nonumber\\
&&
\mathcal{L}^{(f)}_{AB}=
\bar\psi_A D^{\vphantom{\dag}}_{\hbox{\tiny HWK}}\psi_B,
\nonumber\\&&\nonumber\\
&&
\mathcal{L}^{(f)}_{BA}=
\bar\psi_B D^{             \dag }_{\hbox{\tiny HWK}}\psi_A,
\nonumber\\&&\nonumber\\
&&
\mathcal{L}^{(f)}_{\varepsilon}=
-\varepsilon
\left(
\bar\psi_A 
    \psi_A 
+
\bar\psi_B 
    \psi_B 
\right).
\end{eqnarray}
In view of the particle-hole symmetry, we will assume 
that $\varepsilon>0$ from now on and without loss of generality.
The SUSY partition function is obtained by first, 
multiplying $Z^{(f)}$ by $Z^{(b)}$, whereby
$Z^{(b)}$ follows from $Z^{(f)}$ after
replacing all Grassmann fields 
$\bar\psi$ and $\psi$
by the complex-valued fields
$\bar\beta$ and $\beta$,
respectively, and second, by integrating over disorder.
We are now going to perform three successive
changes of path integral variables in the 
``matter'' ($\bar\psi$'s, $\psi$'s, $\bar\beta$'s, and $\beta$'s) sector.
We will only specify
the first two changes of variables in the fermion ($\psi$) sector
as it is understood that the same changes of variables 
apply to the boson ($\beta$) sector. We will treat the fermion and
boson sectors in an asymmetric way in the last step.
Although these are symbolic manipulations, 
they can be justified by choosing appropriate 
unitary equivalent representations
of 
$H^{\vphantom{\dag}}_{\hbox{\tiny HWK}}$ 
for the fermion and boson partition functions,
respectively.\cite{Guruswamy00}

First, introduce
\begin{eqnarray}
\bar\psi_A^{\ }=:\psi_A^{* }\sigma_1,
\qquad
\bar\psi_B^{\ }=:\psi_B^{* }\sigma_1,
\end{eqnarray}
whereby $\psi_A$ and $\psi_B$ are left unchanged.
Denote eigenstates of $\sigma_3$ according to
\begin{eqnarray}
\sigma_3\psi_{A\pm}=\mp\psi_{A\pm},
\qquad
\sigma_3\psi_{B\pm}=\mp\psi_{B\pm},
\end{eqnarray}
say, and rewrite the disorder as
\begin{eqnarray}
A_{\pm}:=A_1\pm iA_2,
\qquad
M_{\pm}:=A_0\pm iA_3.
\end{eqnarray}
This gives the fermion Lagrangians
\begin{eqnarray}
\mathcal{L}^{(f)}_{AB}&=&
\!i\!
\left[
\psi_{A-}^{* }\, 
\left(
\partial_+^{\vphantom{*}}
+
A_+^{\vphantom{*}}
\right)
\psi_{B-}^{\vphantom{*}}
\!+\!
\psi_{A+}^{* }\, 
\left(
\partial_-^{\vphantom{*}}
+
A_-^{\vphantom{*}}
\right)
\psi_{B+}^{\vphantom{*}}
\right]
\nonumber\\
&&
\!+i\!
\left(
\psi_{A-}^{* }\, 
M_+^{\vphantom{*}}
\psi_{B+}^{\vphantom{*}}
+
\psi_{A+}^{* }\, 
M_-^{\vphantom{*}}
\psi_{B-}^{\vphantom{*}}
\right),
\nonumber\\&&\nonumber\\
\mathcal{L}^{(f)}_{BA}&=&
\!
i
\left[
\psi_{B-}^{* }\, 
\left(
\partial_+^{\vphantom{*}}
-
A_+^{\vphantom{*}}
\right)
\psi_{A-}^{\vphantom{*}}
\!+\!
\psi_{B+}^{* }\, 
\left(
\partial_-^{\vphantom{*}}
-
A_-^{\vphantom{*}}
\right)
\psi_{A+}^{\vphantom{*}}
\right]
\nonumber\\
&&
-
i
\left(
\psi_{B-}^{* }\, 
M_-^{\vphantom{*}}\,
\psi_{A+}^{\vphantom{*}}
+
\psi_{B+}^{* }\, 
M_+^{\vphantom{*}}\,
\psi_{A-}^{\vphantom{*}}
\right),
\nonumber\\&&\nonumber\\
\mathcal{L}^{(f)}_{\varepsilon}&=&
-\varepsilon
\left[
\psi_{A-}^{* }
\psi_{A+}^{\vphantom{*}}
+
\psi_{B-}^{* }
\psi_{B+}^{\vphantom{*}}
+
(+\longleftrightarrow-)
\right],
\nonumber\\
&&
\end{eqnarray}
whereby $\partial_\pm:=\partial_1\pm i\partial_2$. 
Their boson counterparts follow from substitution of the
$\psi^*$'s by the $\beta^*$'s and of the
$\psi  $'s by the $\beta  $'s.

Second, define 
\begin{eqnarray} 
\psi^{* }_A=:\frac{1}{\sqrt{2\pi} i} \psi^{* }_1,
\quad 
\psi^{* }_B=:\frac{1}{\sqrt{2\pi} }\psi^{* }_2,
\nonumber\\
\psi^{\vphantom{*}}_A=:\frac{1}{\sqrt{2\pi} i}\psi^{\vphantom{*}}_2,
\quad
\psi^{\vphantom{*}}_B=:\frac{1}{\sqrt{2\pi}}\psi^{\vphantom{*}}_1. 
\end{eqnarray}
The rescaling of the fields is done for later convenience when computing 
operator product expansion (OPE).
We also rescale the energy $\varepsilon\rightarrow 2\pi \varepsilon$.
This gives the fermion Lagrangians
\begin{eqnarray}
\mathcal{L}^{(f)}_{AB}&=&
\!\!
\frac{1}{2\pi}
\left[
\psi_{1-}^{* }\, 
\left(
\partial_+^{\vphantom{*}}
+
A_+^{\vphantom{*}}
\right)
\psi_{1-}^{\vphantom{*}}
\!+\!
\psi_{1+}^{* }\, 
\left(
\partial_-^{\vphantom{*}}
+
A_-^{\vphantom{*}}
\right)
\psi_{1+}^{\vphantom{*}}
\right]
\nonumber\\
&&+
\frac{1}{2\pi}
\left(
\psi_{1-}^{* }\, 
M_+^{\vphantom{*}}
\psi_{1+}^{\vphantom{*}}
+
\psi_{1+}^{* }\, 
M_-^{\vphantom{*}}
\psi_{1-}^{\vphantom{*}}
\right),
\nonumber\\&&\nonumber\\
\mathcal{L}^{(f)}_{BA}&=&
\!\!
\frac{1}{2\pi}
\left[
\psi_{2-}^{* }\, 
\left(
\partial_+^{\vphantom{*}}
-
A_+^{\vphantom{*}}
\right)
\psi_{2-}^{\vphantom{*}}
\!+\!
\psi_{2+}^{* }\, 
\left(
\partial_-^{\vphantom{*}}
-
A_-^{\vphantom{*}}
\right)
\psi_{2+}^{\vphantom{*}}
\right]
\nonumber\\
&&-
\frac{1}{2\pi}
\left(
\psi_{2-}^{* }\, 
M_-^{\vphantom{*}}\,
\psi_{2+}^{\vphantom{*}}
+
\psi_{2+}^{* }\, 
M_+^{\vphantom{*}}\,
\psi_{2-}^{\vphantom{*}}
\right),
\nonumber\\&&\nonumber\\
\mathcal{L}^{(f)}_{\varepsilon}&=&
\varepsilon
\left(
\psi_{1-}^{* }
\psi_{2+}^{\vphantom{*}}
-
\psi_{2-}^{* }
\psi_{1+}^{\vphantom{*}}
+
\psi_{1+}^{* }
\psi_{2-}^{\vphantom{*}}
-
\psi_{2+}^{* }
\psi_{1-}^{\vphantom{*}}
\right).
\nonumber\\
&&
\label{eq: fermionic Hermitian Lagrangian with sublattice asymmetry}
\end{eqnarray}
Their boson counterparts follow from substitution of the
$\psi^*$'s by the $\beta^*$'s and of the
$\psi  $'s by the $\beta  $'s.
Observe that to go from 
$\mathcal{L}^{(f)}_{AB}+\mathcal{L}^{(b)}_{AB}$
to
$\mathcal{L}^{(f)}_{BA}+\mathcal{L}^{(b)}_{BA}$
it suffices to change by a sign the vector potential 
$\boldsymbol{A}$ and the scalar potential $A_0$, 
while leaving the sign of the mass potential $A_3$ unchanged.

This {asymmetry} between the two sublattices can be removed
by performing the final change of path integral variables
according to\cite{Guruswamy00}
\begin{eqnarray}
&&
\psi^{* }_2\longrightarrow
+\psi^{\vphantom{*}}_2,
\qquad
\psi^{\vphantom{*} }_2\longrightarrow
+\psi^{*}_2,
\nonumber\\
&&
\beta^{* }_2\longrightarrow
+\beta^{\vphantom{*}}_2,
\qquad
\beta^{\vphantom{*} }_2\longrightarrow
-\beta^{*}_2,
\end{eqnarray}
with all other fields unchanged. This gives the fermion Lagrangian
\begin{subequations}
\label{eq: final SUSY Copper lagrangians}
\begin{eqnarray}
\mathcal{L}^{(f)}_{AB}+\mathcal{L}_{BA}^{(f)}&=&
\sum_{a=1,2}
\frac{1}{2\pi}
\left[
\psi_{a-}^{* }\, 
\left(
\partial_+^{\vphantom{*}}
+
A_+^{\vphantom{*}}
\right)
\psi_{a-}^{\vphantom{*}}
\right.
\nonumber\\
&&
\hphantom{\sum_{a=1,2}}
+
\psi_{a+}^{* }\, 
\left(
\partial_-^{\vphantom{*}}
+
A_-^{\vphantom{*}}
\right)
\psi_{a+}^{\vphantom{*}}
\nonumber\\
&&
\hphantom{\sum_{a=1,2}}
\left.
+
\psi_{a-}^{* }\, 
M_+^{\vphantom{*}}
\psi_{a+}^{\vphantom{*}}
+
\psi_{a+}^{* }\, 
M_-^{\vphantom{*}}
\psi_{a-}^{\vphantom{*}}
\right],
\nonumber\\
&&
\label{eq: final fermionic lagrangian for HWK}
\end{eqnarray}
together with its boson counterparts obtained by substitution of the
$\psi^*$'s by the $\beta^*$'s and of the
$\psi  $'s by the $\beta  $'s:
\begin{eqnarray}
\mathcal{L}^{(b)}_{AB}+\mathcal{L}_{BA}^{(b)}&=&
\sum_{a=1,2}
\frac{1}{2\pi}
\left[
\beta_{a-}^{* }\, 
\left(
\partial_+^{\vphantom{*}}
+
A_+^{\vphantom{*}}
\right)
\beta_{a-}^{\vphantom{*}}
\right.
\nonumber\\
&&
\hphantom{\sum_{a=1,2}}
+
\beta_{a+}^{* }\, 
\left(
\partial_-^{\vphantom{*}}
+
A_-^{\vphantom{*}}
\right)
\beta_{a+}^{\vphantom{*}}
\nonumber\\
&&
\hphantom{\sum_{a=1,2}}
\left.
+
\beta_{a-}^{* }\, 
M_+^{\vphantom{*}}
\beta_{a+}^{\vphantom{*}}
+
\beta_{a+}^{* }\, 
M_-^{\vphantom{*}}
\beta_{a-}^{\vphantom{*}}
\right].
\nonumber\\
&&
\label{eq: final bosonic lagrangian for HWK}
\end{eqnarray}
{{}}{The asymmetry between sublattices is not completely
removed, however, since it is disguised as a 
{{}}{perturbation for the DOS
with {non-Hermitian} appearence}:}\cite{Mistake-Guruswamy00}
\begin{eqnarray}
\mathcal{L}^{(f)}_{\varepsilon}&=&
\varepsilon
\left(
\psi_{1-}^{* }
\psi_{2+}^{* }
-
\psi_{2-}^{\vphantom{*}}
\psi_{1+}^{\vphantom{*}}
+
\psi_{1+}^{* }
\psi_{2-}^{* }
-
\psi_{2+}^{\vphantom{*}}
\psi_{1-}^{\vphantom{*}}
\right),
\nonumber\\
\mathcal{L}^{(b)}_{\varepsilon}&=&
-
\varepsilon
\left(
\beta_{1-}^{* }
\beta_{2+}^{* }
+
\beta_{2-}^{\vphantom{*}}
\beta_{1+}^{\vphantom{*}}
+
\beta_{1+}^{* }
\beta_{2-}^{* }
+
\beta_{2+}^{\vphantom{*}}
\beta_{1-}^{\vphantom{*}}
\right).
\nonumber\\
&&
\label{eq: non-Hermitian source term}
\end{eqnarray}
\label{eq: def SUSY representation HWK no chiral trsf}
\end{subequations}

The ``decoupling transformation'' 
(\ref{eq: decoupling trsf in continuum})
now reads (the flavor index runs over $a=1,2$)
\begin{subequations}
\label{eq: chiral rep of dec trsf}
\begin{eqnarray}
&&
\psi^{* }_{a-}=
\psi^{\prime*}_{a-}\,
e^{-i\Phi_1+\Phi_2},
\qquad
\psi^{\ }_{a-}=
e^{+i\Phi_1-\Phi_2}\,
\psi^{\prime}_{a-},
\nonumber\\
&&
\psi^{* }_{a+}=
\psi^{\prime*}_{a+}\,
e^{+i\Phi_1+\Phi_2},
\qquad
\psi^{\ }_{a+}=
e^{-i\Phi_1-\Phi_2}\,
\psi^{\prime}_{a+},
\nonumber\\
&&
\label{eq: chiral rep of dec trsf for psi}
\end{eqnarray}
in the fermion sector, and
\begin{eqnarray}
&&
\beta^{* }_{a-}=
\beta^{\prime*}_{a-}\,
e^{-i\Phi_1+\Phi_2},
\qquad
\beta^{\ }_{a-}=
e^{+i\Phi_1-\Phi_2}\,
\beta^{\prime}_{a-},
\nonumber\\
&&
\beta^{* }_{a+}=
\beta^{\prime*}_{a+}\,
e^{+i\Phi_1+\Phi_2},
\qquad
\beta^{\ }_{a+}=
e^{-i\Phi_1-\Phi_2}\,
\beta^{\prime}_{a+},
\nonumber\\
&&
\label{eq: chiral rep of dec trsf for beta}
\end{eqnarray}
\end{subequations}

\noindent in the boson sector.
Under the chiral transformation
(\ref{eq: chiral rep of dec trsf})
we note that the fermion Lagrangians become
\begin{subequations}
\label{eq: decoupling trsf on SUSY HWK}
\begin{eqnarray}
&&\mathcal{L}^{(f)}_{AB}+\mathcal{L}^{(f)}_{BA}
=
\frac{1}{2\pi}\sum_{a=1,2}
\left(\vphantom{E^{\Phi^{*}_1}}
\psi_{a-}^{\prime* }\, 
\partial_+^{\vphantom{*}}
\psi_{a-}^{\prime\vphantom{*}}
+
\psi_{a+}^{\prime* }\, 
\partial_-^{\vphantom{*}}
\psi_{a+}^{\prime\vphantom{*}}
\right.
\nonumber\\
&&
\qquad
\left.
+
\psi_{a-}^{\prime* }\, 
M_+^{\vphantom{*}}
e^{-2i\Phi_1^{\vphantom{*}}}
\psi_{a+}^{\prime\vphantom{*}}
+
\psi_{a+}^{\prime* }\, 
M_-^{\vphantom{*}}
e^{+2i\Phi_1^{\vphantom{*}}}
\psi_{a-}^{\prime\vphantom{*}}
\right),
\nonumber\\&&\nonumber\\
&&\mathcal{L}^{(f)}_{\varepsilon}=
+\varepsilon
e^{+2\Phi_2^{\vphantom{*}}}\,
\left(
\psi_{1-}^{\prime* }\,
\psi_{2+}^{\prime* }
+
\psi_{1+}^{\prime* }\,
\psi_{2-}^{\prime* }
\right)
\nonumber\\
&&
\qquad\qquad
-\varepsilon
e^{-2\Phi_2^{\vphantom{*}}}\,
\left(
\psi_{2-}^{\prime\vphantom{*}}\,
\psi_{1+}^{\prime\vphantom{*}}
+
\psi_{2+}^{\prime\vphantom{*}}\,
\psi_{1-}^{\prime\vphantom{*}}
\right),
\label{eq: decoupling trsf on fermion HWK}
\end{eqnarray}
whereas the boson Lagrangians become
\begin{eqnarray}
&&\mathcal{L}^{(b)}_{AB}+\mathcal{L}^{(b)}_{BA}
=
\frac{1}{2\pi}
\sum_{a=1,2}
\left(\vphantom{E^{\Phi^{*}_1}}
\beta_{a-}^{\prime* }\, 
\partial_+^{\vphantom{*}}
\beta_{a-}^{\prime\vphantom{*}}
+
\beta_{a+}^{\prime* }\, 
\partial_-^{\vphantom{*}}
\beta_{a+}^{\prime\vphantom{*}}
\right.
\nonumber\\
&&
\qquad
\left.
+
\beta_{a-}^{\prime* }\, 
M_+^{\vphantom{*}}
e^{-2i\Phi_1^{\vphantom{*}}}
\beta_{a+}^{\prime\vphantom{*}}
+
\beta_{a+}^{\prime* }\, 
M_-^{\vphantom{*}}
e^{+2i\Phi_1^{\vphantom{*}}}
\beta_{a-}^{\prime\vphantom{*}}
\right),
\nonumber\\ &&\nonumber\\
&&\mathcal{L}^{(b)}_{\varepsilon}=
-\varepsilon
e^{+2\Phi_2^{\vphantom{*}}}\,
\left(
\beta_{1-}^{\prime* }\,
\beta_{2+}^{\prime* }
+
\beta_{1+}^{\prime* }\,
\beta_{2-}^{\prime* }
\right)
\nonumber\\
&&
\qquad\qquad
-\varepsilon
e^{-2\Phi_2^{\vphantom{*}}}\,
\left(
\beta_{2-}^{\prime\vphantom{*}}\,
\beta_{1+}^{\prime\vphantom{*}}
+
\beta_{2+}^{\prime\vphantom{*}}\,
\beta_{1-}^{\prime\vphantom{*}}
\right).
\label{eq: decoupling trsf on boson HWK}
\end{eqnarray}
\end{subequations}

\noindent The complex-valued random masses
$M_\pm$ couple to
the transversal component $\Phi_1$ of the random vector
potential through a {bounded phase factor} $\exp(\pm2i\Phi_1)$
whereas they decouple from the longitudinal 
component $\Phi_2$ of the random vector potential.
Conversely, the energy term couples to the longitudinal 
component $\Phi_2$ of the random vector
potential through an {unbounded exponential factor}
$\exp(\pm2\Phi_2)$
whereas it decouples from the transversal
component $\Phi_1$ of the random vector potential.
We conclude,
with the help of the decoupling transformation
(\ref{eq: chiral rep of dec trsf}),
that the random vector potential enters in the
energy perturbation
$
\mathcal{L}^{(f)}_{\varepsilon}
+
\mathcal{L}^{(b)}_{\varepsilon}
$
as it does in the squared amplitude of the zero mode
of the operator $D^{\dag}_{\hbox{\tiny MDH}}$
in Eq.\ (\ref{eq: MDH Hamiltonian})---namely, 
through the unbounded exponential factors 
$\exp(\pm2\Phi_2)$. We shall see that  
these exponential factors dictate the
dependence on $\varepsilon$ of the DOS in the close vicinity
of the band center.\cite{Mistake-Hatsugai97}

The SUSY partition function for the HWK model that we will choose
to work with treats quantum mechanical 
(through the SUSY matter fields $\psi$ and $\beta$)
and disorder 
(through the random vector and mass potentials)
averaging on an equal footing:
\begin{eqnarray}
&&
Z=
\int\mathcal{D}
[
A^{\vphantom{*}}_{\pm},
M^{\vphantom{*}}_{\pm},
\psi^{          * }_{\mp},
\psi^{\vphantom{*}}_{\mp},
\beta^{          * }_{\mp},
\beta^{\vphantom{*}}_{\mp}]\,
e^{-\int d^2\bm{r}\,\mathcal{L}},
\nonumber\\
&&
\mathcal{L}=
\frac{1}{2 g^{\ }_A}  A_+A_-
+
\frac{1}{2 g^{\ }_M}  M_+M_-
\nonumber\\
&&
\hphantom{\mathcal{L}=}
{}{
+
\mathcal{L}_{AB}^{(f)}
+
\mathcal{L}_{BA}^{(f)}
+
\mathcal{L}_{\varepsilon}^{(f)}
+
\mathcal{L}_{AB}^{(b)}
+
\mathcal{L}_{BA}^{(b)}
+
\mathcal{L}_{\varepsilon}^{(b)},
                 }
\nonumber\\
&&
\label{eq: final SUSY theory for HWK}
\end{eqnarray}
where
$
\mathcal{L}_{AB}^{(f/b)}
+
\mathcal{L}_{BA}^{(f/b)}
+
\mathcal{L}_{\varepsilon}^{(f/b)}
$
are given by 
Eq.\ (\ref{eq: final SUSY Copper lagrangians}).
Correspondingly,
the SUSY partition function for the MDH model is obtained
from Eq.\ (\ref{eq: final SUSY theory for HWK})
by switching off the random masses $M_\pm$, i.e., setting the
variance $g^{\ }_M=0$. This SUSY partition function is
critical at the band center $\varepsilon=0$
as follows from the decoupling transformation
(\ref{eq: chiral rep of dec trsf}). Indeed, it is then
given by
\begin{eqnarray}
&&
Z^{\vphantom{*}}_{*}=
\int\mathcal{D}
[
\bar\alpha,
\alpha,
\Phi^{\vphantom{*}}_1,
\Phi^{\vphantom{*}}_2,
\psi^{\prime          * }_{\mp},
\psi^{\prime\vphantom{*}}_{\mp},
\beta^{\prime          * }_{\mp},
\beta^{\prime\vphantom{*}}_{\mp}]\,
e^{-\int d^2\bm{r}\,\mathcal{L}^{\vphantom{*}}_{*}},
\nonumber\\
&&
\mathcal{L}^{\vphantom{*}}_{*}=
\bar\alpha\boldsymbol{\partial}^2 \alpha
+
\frac{1}{2 g^{\ }_A}  
(\partial^{\vphantom{*}}_\mu\Phi^{\vphantom{*}}_1)^2
+
\frac{1}{2 g^{\ }_A}
(\partial^{\vphantom{*}}_\mu\Phi^{\vphantom{*}}_2)^2
\nonumber\\
&&
\hphantom{\mathcal{L}^{\vphantom{*}}_{*}=}
+
\mathcal{L}_{AB}^{(f)}
+
\mathcal{L}_{BA}^{(f)}
+
\mathcal{L}_{AB}^{(b)}
+
\mathcal{L}_{BA}^{(b)},
\label{eq: final SUSY theory for MDH}
\end{eqnarray}
where
$
\mathcal{L}_{AB}^{(f)}
+
\mathcal{L}_{BA}^{(f)}
$
and
$
\mathcal{L}_{AB}^{(b)}
+
\mathcal{L}_{BA}^{(b)}
$
follow from
Eqs.\ (\ref{eq: decoupling trsf on fermion HWK})
and
      (\ref{eq: decoupling trsf on boson HWK})
with the random masses $M_\pm$ switched off,
respectively.
We have exponentiated the Jacobian that arises
from expressing the vector potential in terms
of its longitudinal and transversal components
with the help of fermionic ghosts $\bar\alpha$
and $\alpha$. In doing so we ensure that the SUSY
partition function remains unity. The Fujikawa
Jacobian arising from the decoupling transformation
in the fermionic sector cancels with that from the bosonic
sector. 

It was shown in Ref.\ \onlinecite{Mudry96}
that Eq.\ (\ref{eq: final SUSY theory for MDH})
defines a CFT for each
value of the marginal coupling constant $g^{\ }_A$. 
The full operator content was constructed explicitly;
i.e., anomalous scaling dimensions of all local operators
that can be built out of the $\psi$'s and $\beta$'s
were calculated. Furthermore, it was established that
this CFT realizes a gl($2|2$) current algebra.
This property allows the explicit construction of primary
fields realizing one- and two-dimensional representations, say,
of the current algebra. The anomalous scaling dimensions of primary
fields carrying multiples of the U(1) gauge charge of 2 
were shown to be unbounded from below, related to the
``Cooper-like'' terms entering $\mathcal{L}^{(b)}_{\varepsilon}$,
and responsible for the multifractality of the zero mode.
We are going to combine these results with a method
pioneered by Carpentier and Le Doussal
in Ref.\ \onlinecite{Carpentier00}
to compute the 
{}{DOS}
for the MDH model.
We will then show that the anomalous scaling dimensions of the 
family of Cooper-like terms that govern both
multifractality of the zero mode and the 
{}{DOS}
in the
MDH model also control the 
{}{DOS}
in the HWK model. 
To this end, we will make use of the fact, demonstrated 
by Guruswamy \textit{et al.}\ in Ref.\ \onlinecite{Guruswamy00},
that Eq.\ (\ref{eq: final SUSY theory for HWK})
defines a nearly conformal field theory in which the coupling
constant $g^{\ }_A$ is marginally relevant 
{{}}{(in the sense that it grows logarithmically with length)}
and the coupling constant
$g^{\ }_M$ is exactly marginal.

\section{DOS for $H_{\hbox{\tiny MDH}}$ }
\label{sec: DOS for the MDH model}

In this section, we are going to compute the 
{}{DOS} 
in the MDH model.
We shall develop an approximation that, presumably, becomes exact in the limit
$\varepsilon\to0$ and in which the 
{}{DOS}  for the MDH model reduces to
the {}{DOS} 
for a single species of Dirac fermion subjected to  
a real-valued random vector potential.
{}{
The cornerstone of our calculation is to distinguish between
the annealed and quenched average values of the dynamical exponent
in very much the same way as one must distinguish 
between the scaling with system size of the average and typical values
of the normalization of zero modes.
                }
In this way we will have established
a one-to-one correspondence between the power-law dependence of the 
{}{DOS}  
and the multifractal spectrum of zero modes.

Our starting point is the critical theory
(\ref{eq: final SUSY theory for MDH})
from which we extract the nonvanishing two-point functions
\begin{eqnarray}
&&
\left\langle
\Phi^{\vphantom{*}}_2(z,\bar z)
\Phi^{\vphantom{*}}_2(0)
\right\rangle_{*}
=
-\frac{g^{\ }_A}{2\pi}\ln|z|,
\nonumber\\
&&
\left\langle
\psi^{\prime\vphantom{*}}_{a-}(z)
\psi^{\prime         * }_{b-}(0)
\right\rangle_{*}
=+
\left\langle
\psi^{\prime          * }_{b-}(z)
\psi^{\prime\vphantom{*}}_{a-}(0)
\right\rangle_{*}
=
\frac{\delta_{ab}}{z},
\nonumber\\
&&
\left\langle
\psi^{\prime\vphantom{*}}_{a+}(\bar z)
\psi^{\prime         * }_{b+}(0)
\right\rangle_{*}
=+
\left\langle
\psi^{\prime         *  }_{b+}(\bar z)
\psi^{\prime\vphantom{*}}_{a+}(0)
\right\rangle_{*}
=
\frac{\delta_{ab}}{\bar z},
\nonumber\\
&&
\left\langle
\beta^{\prime\vphantom{*}}_{a-}(z)
\beta^{\prime         * }_{b-}(0)
\right\rangle_{*}
=
-
\left\langle
\beta^{\prime          * }_{b-}(z)
\beta^{\prime\vphantom{*}}_{a-}(0)
\right\rangle_{*}
=
\frac{\delta_{ab}}{z},
\nonumber\\
&&
\left\langle
\beta^{\prime\vphantom{*}}_{a+}(\bar z)
\beta^{\prime         * }_{b+}(0)
\right\rangle_{*}
=-
\left\langle
\beta^{\prime          * }_{b+}(\bar z)
\beta^{\prime\vphantom{*}}_{a+}(0)
\right\rangle_{*}
=
\frac{\delta_{ab}}{\bar z},
\nonumber\\
&&
a,b=1,2.
\end{eqnarray}
A point $\bm{r}=(r_1,r_2)\in\mathbb{R}^2$ in the two-dimensional
Euclidean plane is hereby parametrized by the complex variables
$z:=r_{1}+ir_{2}$ and $\bar{z}:=r_{1}-ir_{2}$,
respectively.

The disorder-average-smeared DOS in the MDH model is to be calculated by
adding to the critical theory
(\ref{eq: final SUSY theory for MDH})
the perturbation
$
\mathcal{L}^{\vphantom{*}}_{\varepsilon-i\eta}:=
-(\varepsilon-i\eta)
\mathcal{O}^{\vphantom{*}}_{\varepsilon-i\eta}
$
where
\begin{subequations}
\begin{eqnarray}
\mathcal{O}^{\vphantom{*}}_{\varepsilon-i\eta}&:=&
+e^{+2\Phi^{\vphantom{*}}_2}
\left(
\mathcal{B}^{\vphantom{*}}_{12+}
+
\mathcal{B}^{\vphantom{*}}_{21+}
-
\mathcal{F}^{\vphantom{*}}_{12+}
-
\mathcal{F}^{\vphantom{*}}_{21+}   
\right)
\nonumber\\
&&+
e^{-2\Phi^{\vphantom{*}}_2}
\left(
\mathcal{B}^{\vphantom{*}}_{12-}
+
\mathcal{B}^{\vphantom{*}}_{21-}
+
\mathcal{F}^{\vphantom{*}}_{12-}
+
\mathcal{F}^{\vphantom{*}}_{21-}
\right)
\nonumber\\
&&
\end{eqnarray}
and we have defined the composite operators
\begin{eqnarray}
\mathcal{F}^{\vphantom{*}}_{12+}&:=&
\psi^{\prime          * }_{1+}
\psi^{\prime          * }_{2-},
\qquad
\mathcal{F}^{\vphantom{*}}_{21+} :=
\psi^{\prime          * }_{1-}
\psi^{\prime          * }_{2+},
\nonumber\\
\mathcal{F}^{\vphantom{*}}_{12-}&:=&
\psi^{\prime\vphantom{*}}_{2-}
\psi^{\prime\vphantom{*}}_{1+},
\qquad
\mathcal{F}^{\vphantom{*}}_{21-} :=
\psi^{\prime\vphantom{*}}_{2+}
\psi^{\prime\vphantom{*}}_{1-}, 
\end{eqnarray}
in the fermionic sector 
[see Eq.\ (\ref{eq: decoupling trsf on fermion HWK})], 
and the composite operators
\begin{eqnarray}
\mathcal{B}^{\vphantom{*}}_{12+}&:=&
\beta^{\prime          * }_{1+}
\beta^{\prime          * }_{2-},
\qquad
\mathcal{B}^{\vphantom{*}}_{21+} :=
\beta^{\prime          * }_{1-}
\beta^{\prime          * }_{2+},
\nonumber\\
\mathcal{B}^{\vphantom{*}}_{12-}&:=&
\beta^{\prime\vphantom{*}}_{2-}
\beta^{\prime\vphantom{*}}_{1+},
\qquad
\mathcal{B}^{\vphantom{*}}_{21-} :=
\beta^{\prime\vphantom{*}}_{2+}
\beta^{\prime\vphantom{*}}_{1-},
\end{eqnarray}
\label{eq: smeared energy perturbation}
\end{subequations}

\noindent in the bosonic sector 
[see Eq.\ (\ref{eq: decoupling trsf on boson HWK})].
Furthermore, in the definition of the DOS
(\ref{eq: def SUSY DOS without disorder}) 
the operator to be averaged over is now given by
\begin{eqnarray}
\mathcal{O}_{\nu}&=&
+e^{+2\Phi_2}\left(
\mathcal{B}^{}_{12+}+\mathcal{B}^{}_{21+}
+\mathcal{F}^{}_{12+}+\mathcal{F}^{}_{21+}
\right)
\nonumber\\
&&
+e^{-2\Phi_2}\left(
\mathcal{B}^{}_{12-}+\mathcal{B}^{}_{21-}
-\mathcal{F}^{}_{12-}-\mathcal{F}^{}_{21-}
\right).
\label{eq: def cal Onu}
\end{eqnarray}
Observe that $\mathcal{O}_{\nu}$
has the same anomalous scaling dimension as the energy perturbation
$\mathcal{O}^{}_{\varepsilon-i\eta}$.

Our strategy is, first, to derive a differential
(Callan-Symanzik) equation
obeyed by the {}{DOS} that encodes its change
under the infinitesimal rescaling of the lattice constant
\begin{eqnarray}
\mathfrak{a}
\longrightarrow
\mathfrak{a}^{\prime}=
(1+dl)\mathfrak{a}.
\label{eq: rescaling lattice spacing}
\end{eqnarray}
Second, we must solve 
the Callan-Symanzik equation obeyed by the {}{DOS}.

In this spirit, 
we modify the critical action $S_{*}$ defined by
Eq.\ (\ref{eq: final SUSY theory for MDH})
by adding to it the energy perturbation
and any other operator 
$\mathcal{O}^{\vphantom{*}}_{\iota}$
needed to ensure 
invariance under infinitesimal rescaling of the lattice spacing.
In other words, we seek a family of local composite operators
$\{\mathcal{O}^{\vphantom{*}}_{\iota}\}$
such that
\begin{eqnarray}
&&
S:=
S^{\vphantom{*}}_{*}
-
\int d^2\bm{r}
\left(
\lambda^{\vphantom{*}}_{\varepsilon}
\mathfrak{a}^{x^{\vphantom{*}}_{\varepsilon}-2}
\mathcal{O}^{\vphantom{*}}_{\varepsilon-i\eta}
+
\sum_{\iota}
\lambda^{\vphantom{*}}_{\iota}
\mathfrak{a}^{x^{\vphantom{*}}_{\iota}-2}
\mathcal{O}^{\vphantom{*}}_{\iota}
\right)
\nonumber\\
&&
\label{eq: action with source invariant under RG}
\end{eqnarray}
is invariant under infinitesimal rescaling of the lattice spacing.
The dimensionless coupling constants $\lambda^{\vphantom{*}}_{\iota}$
are scale ($l$) dependent and the anomalous scaling dimensions $x_\iota$
of the composite operators $\mathcal{O}^{\vphantom{*}}_{\iota}$
are to be calculated at criticality. 
Invariance of the right- hand side under
infinitesimal rescaling of the lattice spacing is equivalent to
demanding that the family of composite operators
labeled by $\varepsilon$ and $\iota$ 
is closed under the OPE.
Although the range of $\iota$ is always infinite (for one,
composite operators can always be point split and Taylor expanded), 
perturbation of a unitary CFT simplifies greatly in that the family 
$\{\mathcal{O}^{\vphantom{*}}_{\iota}\}$
contains no more than a 
{finite number} of relevant ($0<x_\iota<2$) operators.
In practice, it is then sufficient to truncate the family
$\{\mathcal{O}^{\vphantom{*}}_{\iota}\}$
to this finite subset of relevant operators
if one is only interested in the effect of an arbitrary small perturbation
to criticality. For a nonunitary CFT, as is 
Eq.\ (\ref{eq: final SUSY theory for MDH}),
the situation is dramatically different, however, 
since $\mathcal{O}^{\vphantom{*}}_{\varepsilon-i\eta}$ 
induces under repeated OPE infinitely many 
relevant ($x_\iota<2$) operators.
After selection of the family 
$\{\mathcal{O}^{\vphantom{*}}_{\iota}\}$,
the 
{}{
DOS $\nu(\varepsilon)$
                }        
obeys the Callan-Symanzik equation
\begin{eqnarray}
&&
0=
\left(
\beta^{\vphantom{*}}_{\lambda^{\vphantom{*}}_{\varepsilon}}
\frac{\partial}{\partial\lambda^{\vphantom{*}}_{\varepsilon}}
+
\sum_\iota
\beta^{\vphantom{*}}_{\lambda^{\vphantom{*}}_{\iota}}
\frac{\partial}{\partial\lambda^{\vphantom{*}}_{\iota}}
-
x^{\vphantom{*}}_{\varepsilon}
\right)
{}{\nu(\varepsilon)},
\nonumber\\
&&
\beta^{\vphantom{*}}_{\lambda^{\vphantom{*}}_{\varepsilon}}=
\frac{d\lambda^{\vphantom{*}}_{\varepsilon}}{dl},
\qquad
\beta^{\vphantom{*}}_{\lambda^{\vphantom{*}}_{\iota}}=
\frac{d\lambda^{\vphantom{*}}_{\iota}}{dl}.
\label{eq: Callan-Symanzik eq in all generality}
\end{eqnarray}
We then go on to solve
Eq.\ (\ref{eq: Callan-Symanzik eq in all generality})
in some approximate scheme.

\subsection{Operator content at criticality and RG equations}
\label{subsec: Operator content at criticality and RG equations}

At criticality and through OPE with itself, 
the energy perturbation 
$\mathcal{O}^{\vphantom{*}}_{\varepsilon-i\eta}$
induces local composite operators of the form
\begin{widetext}
\begin{subequations}
\label{eq: most general relavant operator without gradient}
\begin{eqnarray}
\mathcal{O}^{\vphantom{*}}_{
{}{
\boldsymbol{m},\boldsymbol{n}
                }
                           }&=&
\left(\mathcal{F}^{\vphantom{*}}_{12+}\right)^{m^{\vphantom{*}}_{12+}}
\left(\mathcal{F}^{\vphantom{*}}_{21+}\right)^{m^{\vphantom{*}}_{21+}}
\left(\mathcal{B}^{\vphantom{*}}_{12+}\right)^{n^{\vphantom{*}}_{12+}}
\left(\mathcal{B}^{\vphantom{*}}_{21+}\right)^{n^{\vphantom{*}}_{21+}}
\left(\mathcal{F}^{\vphantom{*}}_{12-}\right)^{m^{\vphantom{*}}_{12-}}
\left(\mathcal{F}^{\vphantom{*}}_{21-}\right)^{m^{\vphantom{*}}_{21-}}
\left(\mathcal{B}^{\vphantom{*}}_{12-}\right)^{n^{\vphantom{*}}_{12-}}
\left(\mathcal{B}^{\vphantom{*}}_{21-}\right)^{n^{\vphantom{*}}_{21-}}
\nonumber\\
&&\times
\exp
\left[
2
\left(
m^{\vphantom{*}}_{12+}
+
m^{\vphantom{*}}_{21+}
+
n^{\vphantom{*}}_{12+}
+
n^{\vphantom{*}}_{21+}
-
m^{\vphantom{*}}_{12-}
-
m^{\vphantom{*}}_{21-}
-
n^{\vphantom{*}}_{12-}
-
n^{\vphantom{*}}_{21-}
\right)\Phi^{\vphantom{*}}_2
\right],
\qquad
{}{
\forall\boldsymbol{m},\boldsymbol{n}
                }
\in\mathbb{N}^4.
\label{eq: most general local comp opera}
\end{eqnarray}
It can be shown along the lines of
Refs.\ \onlinecite{Chamon-Mudry96} and \onlinecite{Mudry96} that 
{}{
$\mathcal{O}^{\vphantom{*}}_{\boldsymbol{m},\boldsymbol{n}}$
                }
carries the anomalous scaling dimension
\begin{eqnarray}
x^{\vphantom{*}}_{
{}{
\boldsymbol{m},\boldsymbol{n}
                }
                 }&=&
(m^{\vphantom{*}}_{12+} - m^{\vphantom{*}}_{12-})^2
+
(m^{\vphantom{*}}_{21+} - m^{\vphantom{*}}_{21-})^2
+
|n^{\vphantom{*}}_{12+}-n^{\vphantom{*}}_{12-}|
+
|n^{\vphantom{*}}_{21+}-n^{\vphantom{*}}_{21-}|
\nonumber\\
&&-
\frac{g^{\vphantom{*}}_A}{\pi}
\left[
\left(
m^{\vphantom{*}}_{12+}
-
m^{\vphantom{*}}_{12-}
\right)
+
\left(
m^{\vphantom{*}}_{21+}
-
m^{\vphantom{*}}_{21-}
\right)
+
\left(
n^{\vphantom{*}}_{12+}
-
n^{\vphantom{*}}_{12-}
\right)
+
\left(
n^{\vphantom{*}}_{21+}
-
n^{\vphantom{*}}_{21-}
\right)
\right]^2,
\label{eq: scaling dim for most general local comp opera}
\end{eqnarray}
\end{subequations}
\end{widetext}

\noindent which should be contrasted with its engineering dimension
$
m^{\vphantom{*}}_{12+}
+
m^{\vphantom{*}}_{12-}
+
m^{\vphantom{*}}_{21+}
+
m^{\vphantom{*}}_{21-}
+
n^{\vphantom{*}}_{12+}
+
n^{\vphantom{*}}_{12-}
+
n^{\vphantom{*}}_{21+}
+
n^{\vphantom{*}}_{21-}.
$
Observe the difference in the contributions to the 
anomalous scaling dimensions between the fermion and boson sectors. 
Fermi statistics demands 
point splitting when taking the product of Grassmann variables.
This leads to the quadratic dependence on components of $\boldsymbol{m}$
in Eq.\ (\ref{eq: scaling dim for most general local comp opera})
when $g^{\vphantom{*}}_A=0$.
Boson statistics accommodates arbitrary positive integer powers of 
complex-valued fields.
This leads to the linear dependence on components of $\boldsymbol{n}$
in Eq.\ (\ref{eq: scaling dim for most general local comp opera})
when $g^{\vphantom{*}}_A=0$.
The effect of disorder is thus much more potent in the boson sector
than it is in the fermion sector since the positive term, 
which is linear in $\boldsymbol{n}$ when $g^{\vphantom{*}}_A=0$,
is always dominated by the negative term, 
which is quadratic in $\boldsymbol{n}$ when $g^{\vphantom{*}}_A>0$,
for sufficiently large $|\boldsymbol{n}|$.
The boson sector of the SUSY theory conspires with the unboundedness of
the exponential factor $\exp(2\Phi^{\vphantom{*}}_2)$ that enters in
zero modes to produce
infinitely many operators with negative anomalous scaling dimensions.

Alternatively, the difference between fermion and boson sectors
can be deduced by noting that the boson sector is responsible
for the normalization factor in correlation functions.
Suppose that we perform an expansion of some correlation function in
powers of $\exp(\pm 2\Phi^{\vphantom{*}}_2)$ without using the replica or
SUSY method.
There are contributions to this expansion that originate from the
numerator and from the denominator of the correlation function.
Perform term-by-term disorder averaging.
It was shown in the context of the 2D random phase $XY$ model
in Ref.\ \onlinecite{Mudry99} that
(i)
the counterpart to the contribution from the 
boson (fermion) sector to the anomalous scaling dimension
(\ref{eq: scaling dim for most general local comp opera})
arises from disorder averaging over the contributions
from the denominator (numerator) to the perturbative expansion
in powers of $\exp(\pm 2\Phi^{\vphantom{*}}_2)$ and
(ii)
disorder averaging over the perturbative expansion in powers
of $\exp(\pm 2\Phi^{\vphantom{*}}_2)$
leads to the breakdown of perturbation theory.
The lesson from the 2D random phase $XY$ model
is that we cannot truncate the family 
$\{\mathcal{O}_{\boldsymbol{m},\boldsymbol{n}}\}$
to some \textit{finite} subset.
We shall nevertheless argue that the family
$\{\mathcal{O}_{\boldsymbol{m},\boldsymbol{n}}\}$
can be truncated to an infinite but countable subset
when computing the {}{DOS} for the MDH model.

To this end, we will again rely on the lessons to be drawn from 
a perturbative expansion of correlation functions
in powers of $\exp(\pm 2\Phi^{\vphantom{*}}_2)$.
To $N$th order in powers of $\exp(\pm 2\Phi^{\vphantom{*}}_2)$
the most singular contributions to the disorder average are
associated with terms proportional to 
$\exp(\pm 2N\Phi^{\vphantom{*}}_2)$.
For given 
\begin{eqnarray}
N&=&
m^{\vphantom{*}}_{12+}
+
m^{\vphantom{*}}_{21+}
+
n^{\vphantom{*}}_{12+}
+
n^{\vphantom{*}}_{21+}
\nonumber\\
&&-
m^{\vphantom{*}}_{12-}
-
m^{\vphantom{*}}_{21-}
-
n^{\vphantom{*}}_{12-}
-
n^{\vphantom{*}}_{21-},
\label{eq: def N in terms m's and n's}
\end{eqnarray}
the most relevant operators of the form
(\ref{eq: most general local comp opera})
are given by
\begin{eqnarray}
&&
\left(
\mathcal{B}^{\vphantom{*}}_{12\pm}
\right)^{n}
\left(
\mathcal{B}^{\vphantom{*}}_{21\pm}
\right)^{N-n}
e^{\pm2N\Phi^{\vphantom{*}}_{2}},
\nonumber\\
&&
\hphantom{AAAAAAAAAAAAAAAAAAAAAA-1}
N\geq n\geq0,
\nonumber\\
&&
\left(
\mathcal{F}^{\vphantom{*}}_{12\pm}
\right)
\left(
\mathcal{B}^{\vphantom{*}}_{12\pm}
\right)^{n}
\left(
\mathcal{B}^{\vphantom{*}}_{21\pm}
\right)^{N-1-n}
e^{\pm2N\Phi^{\vphantom{*}}_{2}},
\nonumber\\
&&
\hphantom{AAAAAAAAAAAAAAAAAAAAAA}
N-1\geq n\geq0,
\nonumber\\
&&
\left(
\mathcal{F}^{\vphantom{*}}_{21\pm}
\right)
\left(
\mathcal{B}^{\vphantom{*}}_{12\pm}
\right)^{n}
\left(
\mathcal{B}^{\vphantom{*}}_{21\pm}
\right)^{N-1-n}
e^{\pm2N\Phi^{\vphantom{*}}_{2}},
\nonumber\\
&&
\hphantom{AAAAAAAAAAAAAAAAAAAAAA}
N-1\geq n\geq0,
\nonumber\\
&&
\left(
\mathcal{F}^{\vphantom{*}}_{12\pm}
\right)
\left(
\mathcal{F}^{\vphantom{*}}_{21\pm}
\right)
\left(
\mathcal{B}^{\vphantom{*}}_{12\pm}
\right)^{n}
\left(
\mathcal{B}^{\vphantom{*}}_{21\pm}
\right)^{N-2-n}
e^{\pm2N\Phi^{\vphantom{*}}_{2}},
\nonumber\\
&&
\hphantom{AAAAAAAAAAAAAAAAAAAAAA}
N-2\geq n\geq0,
\nonumber\\
&&
\label{eq: selection most relevant operators}
\end{eqnarray}
with the ($2\times4N$)-fold degenerate anomalous scaling dimension
\begin{eqnarray}
x^{\vphantom{*}}_{N}=N-\frac{g^{\vphantom{*}}_A}{\pi}N^2.
\label{eq: most important scaling dimension within perturbation theory}
\end{eqnarray}
The ($2\times4N$)-fold degeneracy of operators 
with the anomalous scaling dimension
(\ref{eq: most important scaling dimension within perturbation theory})
suggests that we truncate the family
(\ref{eq: most general local comp opera})
to the subset
\begin{subequations}
\label{eq: choice for A_Npm}
\begin{eqnarray}
\mathcal{A}^{\vphantom{*}}_{+;N}
&:=&
\frac{e^{+2N\Phi^{\vphantom{*}}_{2}}}{N!}
\left[
(\mathcal{B}^{\vphantom{*}}_{12+}+\mathcal{B}^{\vphantom{*}}_{21+})^{N}
\right.
\nonumber\\
&&
\left.
-
N(\mathcal{F}^{\vphantom{*}}_{12+}+\mathcal{F}^{\vphantom{*}}_{21+})
(\mathcal{B}^{\vphantom{*}}_{12+}+\mathcal{B}^{\vphantom{*}}_{21+})^{N-1}
\right.
\nonumber\\
&&
\left.
+N(N-1)
\mathcal{F}^{\vphantom{*}}_{12+}\mathcal{F}^{\vphantom{*}}_{21+}
(\mathcal{B}^{\vphantom{*}}_{12+}+\mathcal{B}^{\vphantom{*}}_{21+})^{N-2}
\right],
\nonumber\\
&&
\label{eq: def A_N+}
\\
\mathcal{A}^{\vphantom{*}}_{-;N}
&:=&
\frac{e^{-2N\Phi^{\vphantom{*}}_{2}}}{N!}
\left[
(\mathcal{B}^{\vphantom{*}}_{12-}+\mathcal{B}^{\vphantom{*}}_{21-})^{N}
\right.
\nonumber\\
&&
\left.
+
N(\mathcal{F}^{\vphantom{*}}_{12-}+\mathcal{F}^{\vphantom{*}}_{21-})
(\mathcal{B}^{\vphantom{*}}_{12-}+\mathcal{B}^{\vphantom{*}}_{21-})^{N-1}
\right.
\nonumber\\
&&
\left.
+N(N-1)
\mathcal{F}^{\vphantom{*}}_{12-}\mathcal{F}^{\vphantom{*}}_{21-}
(\mathcal{B}^{\vphantom{*}}_{12-}+\mathcal{B}^{\vphantom{*}}_{21-})^{N-2}
\right],
\nonumber\\
&&
\label{eq: def A_N-}
\end{eqnarray}
\end{subequations}

\noindent labeled by $0<N\in\mathbb{N}$ and $\pm$.
Be aware of the sign difference on the second lines of
Eqs.\ (\ref{eq: def A_N+}) 
and
      (\ref{eq: def A_N-}),
respectively. The choice 
$
 \{\mathcal{A}^{\vphantom{*}}_{\pm;N}\}
 \subset
 \{\mathcal{O}^{\vphantom{*}}_{\boldsymbol{m},\boldsymbol{n}}\}
$
is dictated by the algebraic structure gl($2|2$)
of the critical theory
(\ref{eq: final SUSY theory for MDH}).
Indeed, by construction, the {}{left-hand side} of
{}{
\begin{eqnarray}
&&
\frac{e^{\pm2N\Phi^{\vphantom{*}}_{2}}}{N!}
\left(
 \mathcal{B}^{\vphantom{*}}_{12\pm}+\mathcal{B}^{\vphantom{*}}_{21\pm}
\mp
 \mathcal{F}^{\vphantom{*}}_{12\pm}\mp \mathcal{F}^{\vphantom{*}}_{21\pm}
\right)^{N}
\nonumber\\
&&
\
=
+
\frac{e^{\pm2N\Phi^{\vphantom{*}}_{2}}}{N!}
\left(
\mathcal{B}^{\vphantom{*}}_{12\pm}+\mathcal{B}^{\vphantom{*}}_{21\pm}
\right)^N
\nonumber\\
&&
\
\hphantom{=}
\mp
N
\frac{e^{\pm2N\Phi^{\vphantom{*}}_{2}}}{N!}
\left(
\mathcal{F}^{\vphantom{*}}_{12\pm}+\mathcal{F}^{\vphantom{*}}_{21\pm}
\right)
\left(
\mathcal{B}^{\vphantom{*}}_{12\pm}+\mathcal{B}^{\vphantom{*}}_{21\pm}
\right)^{N-1}
\nonumber\\
&&
\
\hphantom{=}
+
N(N-1)
\frac{e^{\pm2N\Phi^{\vphantom{*}}_{2}}}{N!}
\mathcal{F}^{\vphantom{*}}_{12\pm}\mathcal{F}^{\vphantom{*}}_{21\pm}
\left(
\mathcal{B}^{\vphantom{*}}_{12\pm}+\mathcal{B}^{\vphantom{*}}_{21\pm}
\right)^{N-2}
\nonumber\\
&&
\
\hphantom{=}
+\cdots
\label{eq: gl(2|2) singlet}
\end{eqnarray}
                 }
is a SUSY singlet [though not a gl($2|2$) singlet].
Contributions to the binomial expansion of the left-hand side 
are related to higher-dimensional
irreducible representations of gl($2|2$).\cite{Mudry96}
Similarly, by taking the $N$th power of the operator 
$\mathcal{O}_{\nu}$ in Eq.\ (\ref{eq: def cal Onu})
we can construct the counterparts
$\mathcal{A}_{\nu\,\pm;N}$
to
$\mathcal{A}_{     \pm;N}$
that obey the {same closed OPE's}
and share the {same anomalous scaling dimensions}.
The consistency of the choice (\ref{eq: choice for A_Npm})
is guaranteed by the fact that the OPE's
(see Appendix \ref{app: sec The process of complete annihilation})
\begin{widetext}
\begin{subequations}
\begin{eqnarray}
{\cal A}^{\vphantom{*}}_{+;N}(z,\bar{z})\,
{\cal A}^{\vphantom{*}}_{+;N^{\prime}}(w,\bar{w})
&=&
|z-w|^{x_{N+N^{\prime}}-x_{N^{\vphantom{\prime}}}-x_{N^{\prime}}}
\left(\begin{array}{c}
N+N^{\prime}
\\
N 
\end{array}\right)
{\cal A}^{\vphantom{*}}_{+;N+N^{\prime}}(w,\bar{w})+\cdots,
\qquad
0<N,N^{\prime}\in\mathbb{N},
\nonumber\\
&&
\label{eq: OPE fusion + to +}\\
{\cal A}^{\vphantom{*}}_{-;N}(z,\bar{z})\,
{\cal A}^{\vphantom{*}}_{-;N^{\prime}}(w,\bar{w})
&=&
|z-w|^{x_{N+N^{\prime}}-x_{N^{\vphantom{\prime}}}-x_{N^{\prime}}}
\left(\begin{array}{c}
N+N^{\prime}
\\
N 
\end{array}\right)
{\cal A}^{\vphantom{*}}_{-;N+N^{\prime}}(w,\bar{w})+\cdots,
\qquad
0<N,N^{\prime}\in\mathbb{N},
\nonumber\\
&&
\label{eq: OPE fusion - to -}\\
{\cal A}^{\vphantom{*}}_{+;N^{\vphantom{*}}_{+}}(z,\bar{z})\,  
{\cal A}^{\vphantom{*}}_{-;N^{\vphantom{*}}_{-}}(w,\bar{w})
&=&
|z-w|^{{}{x^{\vphantom{\prime}}_{N_{+}-N_{-}}
   -x^{\vphantom{\prime}}_{N_{+}}-x^{\vphantom{\prime}}_{N_{-}}}}
\left(\begin{array}{c}
N^{\vphantom{*}}_{+}-1  
\\
N^{\vphantom{*}}_{-} 
\end{array}\right)
{\cal A}^{\vphantom{*}}_{+;N^{\vphantom{*}}_{+}-N^{\vphantom{*}}_{-}}(w,\bar{w})
+\cdots,
\qquad
0<N^{\vphantom{*}}_{-}<N^{\vphantom{*}}_{+}\in\mathbb{N},
\nonumber\\
&&
\label{eq: OPE annihi + to -}\\
{\cal A}^{\vphantom{*}}_{-;N^{\vphantom{*}}_{-}}(z,\bar{z})\,  
{\cal A}^{\vphantom{*}}_{+;N^{\vphantom{*}}_{+}}(w,\bar{w})
&=&
|z-w|^{x^{\vphantom{\prime}}_{N_{-}-N_{+}}
     -x^{\vphantom{\prime}}_{N_{-}}-x^{\vphantom{\prime}}_{N_{+}}}
\left(\begin{array}{c}
N^{\vphantom{*}}_{-}-1  
\\
N^{\vphantom{*}}_{+} 
\end{array}\right)
{\cal A}^{\vphantom{*}}_{-;N^{\vphantom{*}}_{-}-N^{\vphantom{*}}_{+}}(w,\bar{w})
+\cdots,
\qquad
0<N^{\vphantom{*}}_{+}<N^{\vphantom{*}}_{-}\in\mathbb{N},
\nonumber\\
&&
\label{eq: OPE annihi - to +}
\end{eqnarray}
\label{eq: closed subOPE's}
\end{subequations}
\end{widetext}

\noindent are closed within the subset
$
 \{\mathcal{A}^{\vphantom{*}}_{\pm;N}\}
 \subset
 \{\mathcal{O}^{\vphantom{*}}_{\boldsymbol{m},\boldsymbol{n}}\}.
$
Equation (\ref{eq: closed subOPE's})
is the most important technical result of this section.
With the choice (\ref{eq: choice for A_Npm}),
Eq.\ (\ref{eq: action with source invariant under RG})
is approximated by
\begin{eqnarray}
&&
S=
S^{\vphantom{*}}_{*}
-
\sum_{s=\pm}
\sum_{N^{\vphantom{*}}_s=1}^{\infty}
Y^{\vphantom{*}}_{s;N^{\vphantom{*}}_s}
\mathfrak{a}^{x^{\vphantom{*}}_{s;N^{\vphantom{*}}_s}-2}
\int d^2\bm{r}\,
\mathcal{A}^{\vphantom{*}}_{s;N^{\vphantom{*}}_s},
\nonumber\\
&&
x^{\vphantom{*}}_{s;N^{\vphantom{*}}_s}=
x^{\vphantom{*}}_{N^{\vphantom{*}}_s}=
N^{\vphantom{*}}_s
-
\frac{g^{\vphantom{*}}_A}{\pi}
(N^{\vphantom{*}}_s)^2.
\label{eq: approximated action with source invariant under RG}
\end{eqnarray}
Furthermore, 
by neglecting processes of ``charge annihilation''
as encoded by 
Eqs.\ (\ref{eq: OPE annihi + to -})
and   (\ref{eq: OPE annihi - to +})
and their extensions when 
$N^{\vphantom{*}}_+=N^{\vphantom{*}}_-$
in favor of ``charge fusion''
as encoded by
Eqs.\ (\ref{eq: OPE fusion + to +})
and   (\ref{eq: OPE fusion - to -})
we obtain the counterpart
\begin{eqnarray}
&&
0=
\left(
\sum_{s=\pm}
\sum_{N^{\vphantom{*}}_s=1}^{\infty}
\frac{d Y^{\vphantom{*}}_{s;N^{\vphantom{*}}_s}}
     {d l}
\frac{\partial}
     {\partial Y^{\vphantom{*}}_{s;N^{\vphantom{*}}_s}}
-
x^{\vphantom{*}}_{+;1}
\right)
{}{\nu(\varepsilon)},
\nonumber\\
&&
\frac{d Y^{\vphantom{*}}_{s;N^{\vphantom{*}}_s}}{d l}=
\left(
2
-
x^{\vphantom{*}}_{s;N^{\vphantom{*}}_s}
\right)
Y^{\vphantom{*}}_{s;N^{\vphantom{*}}_s}
\nonumber\\
&&
\hphantom{
\frac{d Y^{\vphantom{*}}_{s;N^{\vphantom{*}}_s}}{d l}
         }
+
\pi
\sum_{ N^{\prime      }_{s}=1}^{N^{\vphantom{*}}_s-1}
\left(\begin{array}{c}
N^{\vphantom{*}}_s
\\
N^{\prime}_s
\end{array}\right)
Y^{\vphantom{*}}_{s;N^{\prime      }_s}
Y^{\vphantom{*}}_{s;N^{\vphantom{*}}_s-N^{\prime      }_s},
\nonumber\\
&&
\label{eq: CS eq for DOS MDG: interm}
\end{eqnarray}
to the Callan-Symanzik equation  
(\ref{eq: Callan-Symanzik eq in all generality}).
At last, initial conditions on the coupling constants (fugacities)
$Y^{\vphantom{*}}_{s;N^{\vphantom{*}}_s}$
are independent of $s=\pm$ and given by
\begin{eqnarray}
Y^{\vphantom{*}}_{s;N^{\vphantom{*}}_s}(l=0)=
\frac{
\varepsilon-i\eta
     }
     {
\mathfrak{a}^{x^{\vphantom{*}}_{s;1}-2}
     }
\delta^{\vphantom{*}}_{1,N^{\vphantom{*}}_s}.
\end{eqnarray}
The justification for neglecting all annihilation processes
is that this approximation preserves the freezing transition
that characterizes both the multifractal spectrum
of zero modes\cite{Chamon96} and, as we shall see below, the DOS.
This approximation is expected to  
capture the leading singularity of the DOS close to the band center.
Furthermore, this approximation is consistent with
neglecting renormalization effects on $g^{\vphantom{*}}_A$
induced by the energy perturbation
(see Appendix \ref{app: sec The process of complete annihilation}).
At this level of approximation, the distinction between positive
and negative charges is irrelevant to the Callan-Symanzik equation
obeyed by the {}{DOS}. 
Equation (\ref{eq: CS eq for DOS MDG: interm}) thus simplifies to
\begin{eqnarray}
&&
0=
\left(
\sum_{N=1}^{\infty}
\beta^{\vphantom{*}}_{Y^{\vphantom{*}}_{N}}
\frac{\partial}{\partial Y^{\vphantom{*}}_{N}}
-
x^{\vphantom{*}}_1
\right)
{}{\nu(\varepsilon)},
\nonumber\\
&&
\beta^{\vphantom{*}}_{Y^{\vphantom{*}}_N}=
(2-x^{\vphantom{*}}_N) Y^{\vphantom{*}}_N
+
\pi
\sum_{N^{\prime}=1}^{N-1}
\left(\begin{array}{c}
N^{\vphantom{*}}
\\
N^{\prime}
\end{array}\right)
Y^{\vphantom{*}}_{N^{\prime}}
Y^{\vphantom{*}}_{N^{\vphantom{*}}-N^{\prime}},
\nonumber\\
&&
x^{\vphantom{*}}_N=
N-\frac{g^{\vphantom{*}}_A}{\pi}N^2,
\nonumber\\
&&
Y^{\vphantom{*}}_{N}(l=0)=
\frac{
\varepsilon-i\eta
     }
     {
\mathfrak{a}^{x^{\vphantom{*}}_{1}-2}
     }
\delta^{\vphantom{*}}_{1,N}.
\label{eq: final CZ eq for MDH model}
\end{eqnarray}

It can be shown that
Eq.\ (\ref{eq: final CZ eq for MDH model})
is also obeyed by the {}{DOS} for a single flavor
of Dirac fermions coupled to a real-valued random vector potential
once all annihilation processes in the RG analysis have been neglected.
{}{
Furthermore, 
essentially the same $\beta$ functions for the fugacities in
Eq.\ (\ref{eq: final CZ eq for MDH model})
encode a RG analysis of the probability distribution for the normalization
                }
\begin{eqnarray}
\mathcal{Z}(L):=
\int_L d^2\bm{r}\, e^{-2\Phi^{\vphantom{*}}_2(\bm{r})}
\label{eq: def normalization zero mode}
\end{eqnarray}
of the squared zero mode amplitude in the MDH model
provided one replaces the anomalous scaling dimension
$
x^{\vphantom{*}}_N=
N-({g^{\vphantom{*}}_A}/{\pi})N^2
$
by
$
 -({g^{\vphantom{*}}_A}/{\pi})N^2.
$
This is most easily seen after
(i)
performing the decoupling transformation
(\ref{eq: chiral rep of dec trsf})
on
Eq.\ (\ref{eq: def SUSY DOS without disorder})
applied to $H^{\vphantom{*}}_{\hbox{\tiny MDH}}$,
(ii)
isolating in the source term all terms that couple to
$\exp(-2\Phi^{\vphantom{*}}_2)$,
(iii)
ignoring the RG contributions from the SUSY matter fields, and
(iv)
identifying the energy $\varepsilon$ with 
$L^{-z^{\vphantom{*}}_A}$ 
where $L$ is the system size and $z^{\vphantom{*}}_A$ 
is the dynamical scaling exponent that relates scaling with respect to
$L$ to scaling with respect to $\varepsilon$.
We have thus explicitly related the transformation law obeyed 
by the DOS under rescaling to the multifractal properties of zero modes.
Similar RG equations were derived by Carpentier and Le Doussal
within a replicated approach in Refs.\ 
\onlinecite{Carpentier00}
and
\onlinecite{Carpentier01}
in the context of the 2D random $XY$ model and the freezing
transition of the multifractal spectrum of zero modes
in the random vector potential problem, respectively.
We now turn to the analysis of the Callan-Symanzik equation
(\ref{eq: final CZ eq for MDH model}).

\subsection{From the Callan-Symanzik equation to the KPP equation}

The Callan-Symanzik equation
(\ref{eq: final CZ eq for MDH model})
reflects the transformation law under infinitesimal rescaling of the
ultraviolet cutoff
(lattice spacing) of the SUSY partition function
\begin{eqnarray}
Z&:=&
\int\mathcal{D}
[
\bar\alpha,
\alpha,
\Phi ^{\vphantom{*}}_1,
\Phi ^{\vphantom{*}}_2,
\psi ^{\prime          * },
\psi ^{\prime\vphantom{*}},
\beta^{\prime          * },
\beta^{\prime\vphantom{*}}]\,
e^{-\int d^2 \bm{r}\mathcal{L}},
\nonumber\\
\mathcal{L}&:=&
\mathcal{L}^{\vphantom{*}}_{*}
-\sum_{n=1}^{\infty}
Y^{\vphantom{*}}_n \mathfrak{a}^{x_n-2}
\mathcal{A}^{\vphantom{*}}_n,
\end{eqnarray}
where 
$\mathcal{L}^{\vphantom{*}}_{*}$
is given by Eq.\ (\ref{eq: final SUSY theory for MDH})
and 
$\mathcal{A}^{\vphantom{*}}_n$
is, say, defined by the right-hand side of
Eq.\ (\ref{eq: def A_N-}).
As noted earlier 
[see Eq.\ (\ref{eq: gl(2|2) singlet})],
\begin{eqnarray}
\mathcal{A}^{\vphantom{*}}_n=
\frac{1}{n!}(\mathcal{A}^{\vphantom{*}}_1)^n
+
\cdots
\label{eq: a useful property of cal A's}
\end{eqnarray}
to leading order in an expansion organized by the relevance of
the operators occurring in OPE's. It is then tempting to think of
the coupling constants (fugacities) $Y^{\vphantom{*}}_n$ 
as being related to
the cumulants of some probability distribution.\cite{Mudry99} 
Initially, this probability distribution is very narrow since only the first 
cumulant is nonvanishing. However, as the lattice rescaling parameter
$l$ is increased, the cumulants grow; i.e., the probability
distribution becomes broader. In the infrared limit $l\to\infty$
the ratios $Y^{\vphantom{*}}_{n}/(Y^{\vphantom{*}}_1)^n$ diverge.
The asymptotic RG flow of the fugacities thus 
offers little information on the bulk of the underlying
probability distribution. Hence, it would be desirable to recast
the RG equations obeyed by the fugacities $Y^{\vphantom{*}}_n$
in terms of their generating function and to follow the RG flow
of the generating function instead of that of the cumulants 
it generates so as to obtain
some information on the bulk of the generating function for the
fugacities. Equation (\ref{eq: a useful property of cal A's})
suggests that the generating function for the fugacities
takes the form
\begin{subequations}
\begin{eqnarray}
\Theta(\xi,l)=
\sum_{n=1}^{\infty}
\frac{\xi^n}{n!}Y^{\vphantom{n}}_n(l),
\qquad \xi\in\mathbb{C},
\end{eqnarray}
with the initial condition
\begin{eqnarray}
\Theta(\xi,0)=
\frac{
\varepsilon-i\eta
     }
     {
\mathfrak{a}^{x^{\vphantom{*}}_{1}-2}
     }
\xi
\end{eqnarray}
by Eq.\ (\ref{eq: final CZ eq for MDH model}).
Differentiation with respect to the lattice rescaling parameter $l$
gives, with the help of Eq.\ (\ref{eq: final CZ eq for MDH model}),
the second-order nonlinear partial differential equation
\begin{eqnarray}
\partial^{\vphantom{n}}_l\Theta=
\left[
2
-
\xi\partial^{\vphantom{n}}_{\xi}
+
\frac{g^{\vphantom{n}}_A}{\pi}
\left(\xi\partial^{\vphantom{n}}_{\xi}\right)^2
\right]\Theta
+\pi\Theta^2.
\end{eqnarray}
\end{subequations}

\noindent In turn, reparametrization of $\Theta$ and $\xi$ 
\begin{subequations}
\begin{eqnarray}
\Theta(\xi,l)=:
\frac{2}{\pi}
\left[
\widetilde G(y,l)-1
\right],
\qquad
\xi=:-e^{-y},
\end{eqnarray}
gives
\begin{eqnarray}
&&
\partial^{\vphantom{n}}_l \widetilde G=
\left(
\partial^{\vphantom{n}}_y
+
\frac{g^{\vphantom{n}}_A}{\pi}
\partial^{          2}_{y}
\right)\widetilde G
+2\widetilde G(\widetilde G-1),
\\
&&
\widetilde G(y,0)=
1
-
\frac{\pi}{2}
\frac{
\varepsilon-i\eta
     }
     {
\mathfrak{a}^{x^{\vphantom{*}}_{1}-2}
     }
e^{-y}.
\label{eq: Initial condition for widetilde G}
\end{eqnarray}
\label{eq: RG for widetilde G}
\end{subequations}

\noindent The convection term on the right-hand side is induced by the
RG flow of the matter fields in the bosonic sector. This term
is absent from the RG analysis of the normalization 
(\ref{eq: def normalization zero mode})
of the zero mode. 
At the level of the generating function,
the convection term can be disposed of by
the change of variable consisting of a Galilean boost and of a rescaling
\begin{subequations}
\begin{equation}
G(x,t):=
\widetilde G(\sqrt{g_A/\pi}\,x-t/2,t/2),
\label{eq: def G canonical for KPP}
\end{equation}
after which
\begin{eqnarray}
&&
\partial^{\vphantom{n}}_t G=
\frac{1}{2}
\partial^{          2}_{x}
G
+G(G-1),
\label{eq: KPP equation with f=G(G-1)}
\\
&&
G(x,0)=
1
-
\frac{\pi}{2}
\frac{
\varepsilon-i\eta
     }
     {
\mathfrak{a}^{x^{\vphantom{*}}_{1}-2}
     }
\exp\left(-\sqrt{\frac{g^{\vphantom{*}}_A}{\pi}}x\right)
\nonumber\\
&&\hphantom{G(x,0)}
\approx
\exp\left[
-
\frac{\pi}{2}
\frac{
\varepsilon-i\eta
     }
     {
\mathfrak{a}^{x^{\vphantom{*}}_{1}-2}
     }
\exp\left(-\sqrt{\frac{g^{\vphantom{*}}_A}{\pi}}x\right)
  \right].
\nonumber\\
&&
\label{eq: initial conditions on G(x,l)}
\end{eqnarray}
\label{eq: RG for G}
\end{subequations}

\noindent The error caused by exponentiation in the last step should be
comparable to the one caused by neglecting the
renormalization of $g^{\vphantom{*}}_A$ induced by
the energy perturbation.
Equation (\ref{eq: KPP equation with f=G(G-1)}) 
is known in the mathematical literature as the 
KPP equation
with the forcing term 
\begin{eqnarray}
F(G):=G(G-1)
\label{eq: forcing term}
\end{eqnarray}
when $x\in\mathbb{R}$ and $0\leq G\leq 1$.\cite{Kolmogorov37,Bramson83}

Trading $\Theta(\xi,l)$ for $\widetilde{G}(y,l)$ [or $G(x,t)$]
offers the following advantage. It is possible to interpret
$\widetilde{G}(y,l)$ 
as the generating function of a random variable $Z$
with the (scale-dependent) probability distribution $P(Z,l)$
through the definition
\begin{subequations}
\begin{eqnarray}
\widetilde G(y,l)&=:&
\left\langle
\exp
\left(
-
e^{-y}\, Z
\right)
\right\rangle^{\vphantom{*}}_{P(Z,l)}
\end{eqnarray}
with the initial condition
\begin{eqnarray}
P(Z,0)=
\delta
\biglb(
Z-\pi Y^{\vphantom{*}}_1(0)/2
\bigrb).
\end{eqnarray}
This definition is consistent since 
the probability distribution $P(Z,l)$ is normalized to 1 initially 
and since its normalization is invariant under the RG flow.
In turn, we can reinterpret the fugacities $Y_{n}$
as the moments of the (scale-dependent) probability distribution $P(Z,l)$
with the initial conditions
\begin{eqnarray}
Y_{n}(l=0)=
\frac{2}{\pi}
\left(
\frac{\pi}{2}
\frac{\varepsilon-i\eta}{\mathfrak{a}^{x^{\vphantom{*}}_{1}-2}}
\right)^n.
\label{eq: new initial conditions on fugacities}
\end{eqnarray}
\end{subequations}
Equation (\ref{eq: new initial conditions on fugacities})
should be compared with Eq.\ (\ref{eq: final CZ eq for MDH model}).
This immediately leads to the identification of the 
(scale-dependent) random variable $Z$ with the 
{scale-dependent random} energy
$
({\pi}/{2})
[{\varepsilon(l)-i\eta(l)}]/{\mathfrak{a}^{x^{\vphantom{*}}_{1}-2}}
$,
i.e.,
\begin{subequations}
\begin{eqnarray}
\widetilde{G}(y,l)&=&
1+\frac{\pi}{2}\Theta(\xi,l)
\nonumber\\
&=:&
\left\langle
\exp
\left[
-e^{-y}
\frac{\pi}{2}
\frac{\varepsilon(l)-i\eta(l)}{\mathfrak{a}^{x^{\vphantom{*}}_{1}-2}}
\right]
\right\rangle_{P(Z,l)}
\end{eqnarray}
and, more importantly,
\begin{eqnarray}
&&
\left\langle
\ln\frac{Z(l)}{Z(0)}
\right\rangle_{P(Z,l)}=
\left\langle
\ln\frac{\varepsilon(l)}{\varepsilon(0)}
\right\rangle_{P(Z,l)},
\label{eq: <ratio ln Z>}
\\
&&
\frac{\varepsilon^{\vphantom{*}}_\mathrm{typ}(l)}
     {\varepsilon^{\vphantom{*}}_\mathrm{typ}(0)}:=
\exp\left[\left\langle\ln\frac{Z(l)}{Z(0)}\right\rangle_{P(Z,l)}\right].
\label{eq: def typical ratio energies}
\end{eqnarray}
\end{subequations}
As we shall see in the next two subsections, it is possible
to extract from Eq.\ (\ref{eq: def typical ratio energies})
a scaling exponent $z^{\vphantom{*}}_A$
called the dynamical scaling exponent that encodes the scaling
of the typical energy $\varepsilon^{\vphantom{*}}_\mathrm{typ}(l)$
in the infrared limit $l\to\infty$
through
\begin{eqnarray}
z^{\vphantom{*}}_A:=
\lim_{l\to\infty}
\frac{1}{l}
\ln
\frac{\varepsilon^{\vphantom{*}}_\mathrm{typ}(l)}
     {\varepsilon^{\vphantom{*}}_\mathrm{typ}(0)}.
\label{eq: def zA}
\end{eqnarray}
Knowledge of the dynamical scaling exponent $z^{\vphantom{*}}_A$
suffices to compute the 
{}{DOS} 
in the MDH model.

\subsection{Analysis of the KPP equation}

In the limit $t\to\infty$, the KPP equation
(\ref{eq: KPP equation with f=G(G-1)})
possesses a solution in the form of a traveling wave 
$w^{\vphantom{*}}_A(x-v^{\vphantom{*}}_At)$ 
that is unique up to translation
and whose velocity $v^{\vphantom{*}}_A$
is solely controlled by the exponential decay of the initial data
for large $x$, provided the initial data obey the conditions
\begin{eqnarray}
G(x,0)=
\left\{
\begin{array}{ll}
1,
&
\hbox{ for }
x\to+\infty,\\
&\\
0,
&
\hbox{ for }
x\to-\infty,\\
\end{array}
\right.
\label{eq: initial conditions for x to pm infty}
\end{eqnarray}
which they do according to 
Eq.\ (\ref{eq: initial conditions on G(x,l)}).
More precisely, if we choose the origin of $x$ such that
\begin{eqnarray}
\lim_{x\to\infty}
\left[
1-G(x,0)
\right]=
e^{
{}{
-\sqrt{
{g^{\vphantom{*}}_A}/{\pi}
      }
\,
\,x
                } 
  },
\label{eq: relating asymptotics of initial data to velocity}
\end{eqnarray} 
then
\begin{eqnarray}
v^{\vphantom{*}}_A=
\left\{
\begin{array}{ll}
\frac{1}{2}
\sqrt{\frac{g^{\vphantom{*}}_{A}}{\pi}}
+
\sqrt{\frac{\pi}{g^{\vphantom{*}}_{A}}},
&
\hbox{ for }
\frac{g^{\vphantom{*}}_{A}}{\pi}<2,
\\
&\\
\sqrt{2},
&
\hbox{ for }
\frac{g^{\vphantom{*}}_{A}}{\pi}\geq 2,
\end{array}
\right.
\label{eq: vA for canonical KPP}
\end{eqnarray}
and
\begin{eqnarray}
\lim_{x\to\infty}[1-w^{\vphantom{*}}_A(x)]=
\left\{
\begin{array}{ll}
e^{
{}{
-
\sqrt{{g^{\vphantom{*}}_A}/{\pi}
      }
\,
\,x
                }
  },
&
\hbox{ for }
v^{\vphantom{*}}_A>\sqrt{2},
\\
&\\
xe^{-\sqrt{2}x
  },
&
\hbox{ for }
v^{\vphantom{*}}_A=\sqrt{2}.
\end{array}
\right.
\label{eq: wA as x infty}
\end{eqnarray}
It is a remarkable property of the KPP
equation\cite{Kolmogorov37,Bramson83} that
the form of the forcing term $F(G)$ in Eq.\ (\ref{eq: forcing term})
does not affect the velocity of the traveling wave
as long as
\begin{eqnarray} 
&&
F(0)=F(1)=0, 
\qquad
F(G)<0,
\nonumber\\
&&
F'(0)=-1, 
\qquad
F'(G)\geq-1,
\qquad
0<G<1.
\nonumber\\
&&
\label{eq: universality in KPP equation}
\end{eqnarray}
As noted in 
Refs.\ \onlinecite{Carpentier00}
and    \onlinecite{Carpentier01},
had we chosen a different RG scheme such as smoothened
the hard-core condition in real space used to obtain
Eq.\ (\ref{eq: CS eq for DOS MDG: interm}), say,
we would have obtained another forcing term in the KPP
equation. Universality, i.e., the independence of the choice
of RG scheme, is thus equivalent to finding a forcing term that
obeys Eq.\ (\ref{eq: universality in KPP equation})
from the point of view of the KPP equation.

Equations 
(\ref{eq: vA for canonical KPP}) 
and 
(\ref{eq: wA as x infty})
allow us to perform the summation over all fugacities
in the generating function 
(\ref{eq: def G canonical for KPP})
when $l\to\infty$:
\begin{eqnarray}
1-G(x,2l)&=&
-
\frac{\pi}{2}
\sum_{n=1}^{\infty}
\frac{
\left(-e^{{}{-\sqrt{{g_A}/{\pi}}\,}\,x+l}\right)^n
     }
     {
n!
     } Y_n(l)
\nonumber\\
&=&
\left\{
\begin{array}{ll}
e^{
{}{-\sqrt{{g^{\vphantom{*}}_A}/{\pi}}\,}\,
(x-2v^{\vphantom{*}}_Al)},
&
\hbox{ for }
v^{\vphantom{*}}_A>\sqrt{2},
\\
&\\
x\,
e^{-\sqrt{2}(x-2v^{\vphantom{*}}_Al)},
&
\hbox{ for }
v^{\vphantom{*}}_A=\sqrt{2},
\\
\end{array}
\right.
\nonumber\\
&&
\label{eq: travelling wave asymptotics}
\end{eqnarray}
as $x\to\infty$.
Although individual fugacities are relevant and grow without
bounds in the infrared limit, a finite limiting value of the
generating function can nevertheless be extracted if $x$ 
scales linearly with $l$ as a result of the traveling-wave
asymptotic of $G(x,2l)$.

\subsection{KPP equation and {}{DOS} }

We are ready to make an estimate for 
Eq.\ (\ref{eq: <ratio ln Z>})
that justifies the definition of the 
dynamical exponent (\ref{eq: def zA}).
To this end, we take advantage of the facts that,
by 
Eqs.\ (\ref{eq: def G canonical for KPP}),
      (\ref{eq: initial conditions on G(x,l)}),
and   (\ref{eq: travelling wave asymptotics}),
the generating function $G(x,2l)$ can be thought of as a wave front
which, initially, was chosen to coincide with the origin $x=0$
and is located at $x=2v^{\vphantom{*}}_A l$ 
in the long-time limit $t=2l\to\infty$.
In the same way,
$\widetilde G(y,l)=G\biglb(\sqrt{\pi/g_A}(y+l),2l\bigrb)$
is a wave front located at
$y=0$ and $y=(2\sqrt{g^{\vphantom{*}}_A/\pi}\,v^{\vphantom{*}}_A-1)l$
initially and in the infrared limit $l\to\infty$, respectively.
This gives the estimate 
\begin{eqnarray}
\left\langle
\ln\frac{Z(l)}{Z(0)}
\right\rangle^{\vphantom{*}}_{P(Z,l)}
&=&
\int_{-\infty}^{+\infty}dy
\left[
\widetilde G(y,0)
-
\widetilde G(y,l)
\right]
\nonumber\\
&=&
\sqrt{\frac{g^{\vphantom{*}}_A}{\pi}}
\int_{-\infty}^{+\infty}dx
\biggl[
G(x,0)
\nonumber\\
&&\qquad\qquad
-
G(x+l\sqrt{\pi/g_A},2l)
\biggr]
\nonumber\\
&=&
\left(
2\sqrt{\frac{g^{\vphantom{*}}_A}{\pi}}\,
v^{\vphantom{*}}_A\,
-
1
\right)
l
+\mathcal{O}(\ln l),
\nonumber\\
&&
\label{eq: position of the wave front in MDH}
\end{eqnarray}
where the logarithmic corrections are universal
(see Ref.\ \onlinecite{Bramson83}).
It is then tempting to identify
\begin{equation}
\frac{\varepsilon^{\vphantom{*}}_{\mathrm{typ}}(l)}
     {\varepsilon^{\vphantom{*}}_{\mathrm{typ}}(0)}
=
\exp
\left(
\left\langle
\ln \frac{Z(l)}{Z(0)}
\right\rangle^{\vphantom{*}}_{P(Z,l)}
\right)
\propto
e^{z^{\vphantom{*}}_A l},
\end{equation}
whereby
\begin{eqnarray}
z^{\vphantom{*}}_{A}&:=&
\lim_{l\to\infty}
\frac{d \ln \varepsilon^{\vphantom{*}}_{\mathrm{typ}}}{dl}
\nonumber\\
&=&
\left\{
\begin{array}{ll}
1
+
\frac{g^{\vphantom{*}}_{A}}{\pi},
&
\hbox{ for }
g^{\vphantom{*}}_{A}<2\pi,
\\
&\\
4\sqrt{\frac{g^{\vphantom{*}}_{A}}{2\pi}}
-
1,
&
\hbox{ for }
g^{\vphantom{*}}_{A}\geq 2\pi,
\end{array}
\right.
\label{eq: dynamical exponent of MDH}
\end{eqnarray}

\noindent{defines} the quenched average value $z^{\vphantom{*}}_{A}$ 
of the dynamical exponent,
as a coupling constant that enters the Callan-Symanzik equation obeyed
by the DOS, which we here {assume} to be self-averaging and to obey some
sort of scaling ansatz.
This conjecture is bolstered by the fact 
(see Refs.\ \onlinecite{Derrida88} and \onlinecite{Buffet93})
that the distribution of
$\ln Z(l)$ is self-averaging in the infrared limit 
$l\to\infty$ as we had hoped for.
{}{
Consequently, it is plausible that
$Y^{\vphantom{*}}_{\mathrm{typ}}\sim\varepsilon^{\vphantom{*}}_{\mathrm{typ}}$
is the only relevant coupling constant that enters 
the Callan-Symanzik equation for the (self-averaging) DOS
                }
\begin{eqnarray}
&&
0=
\left[
z^{\vphantom{*}}_A
\varepsilon
\frac{\partial}{\partial\varepsilon}
-
\left(
2-z^{\vphantom{*}}_A
\right)
\right]
{}{\nu(\varepsilon)}.
\label{eq: Callan-Symanzik MDH typical DOS}
\end{eqnarray}

\noindent(Observe that 
$2-z^{\vphantom{*}}_A=1-g^{\vphantom{*}}_A/\pi=x^{\vphantom{*}}_1$
for $g^{\vphantom{*}}_A<2\pi$.)
A rigorous proof that the solutions to
Eqs.\ (\ref{eq: Callan-Symanzik MDH typical DOS}) 
and   (\ref{eq: final CZ eq for MDH model})
are identical in the thermodynamic limit $L\to\infty$
at some fixed energy $\varepsilon$ is beyond the scope of this paper.

The mechanism by which solutions to the KPP equations are
traveling waves causes the 
dynamical exponent $z^{\vphantom{*}}_A$ to undergo a
freezing transition when the disorder strength 
$g^{\vphantom{*}}_A$
reaches the critical value $2\pi$.
The freezing transition of the dynamical exponent
reflects itself by a nonanalyticity of the multifractal spectrum
of zero modes as was shown in 
Refs.\ \onlinecite{Chamon96} and \onlinecite{Castillo97}.
It also reveals itself through a nonanalytic dependence on
$g^{\vphantom{*}}_A$ in the power-law behavior of the 
{}{DOS}
since integration of Eq.\ (\ref{eq: Callan-Symanzik MDH typical DOS})
yields
\begin{eqnarray}
\frac{
{}{\nu(\varepsilon)}
     }
     {
{}{\nu(\varepsilon^{\vphantom{*}}_{\hbox{\tiny ren}})}
     }&=&
\left(
\frac{
\varepsilon
     }
     {
\varepsilon^{\vphantom{*}}_{\hbox{\tiny ren}}
     }
\right)^{
\frac{
2-z^{\vphantom{*}}_{A}
     }
     {
  z^{\vphantom{*}}_{A}
     }}
\nonumber\\
&=&
\left\{
\begin{array}{ll}
\left(
\frac{
\varepsilon
     }
     {
\varepsilon^{\vphantom{*}}_{\hbox{\tiny ren}}
     }
\right)^{
\frac{
1-g^{\vphantom{*}}_{A}/\pi
     }
     {
1+g^{\vphantom{*}}_{A}/\pi
     }  },
&
\hbox{ for }
g^{\vphantom{*}}_{A}<2\pi,
\\&\\
\left(
\frac{
\varepsilon
     }
     {
\varepsilon^{\vphantom{*}}_{\hbox{\tiny ren}}
     }
\right)^{
\frac{
3-4\sqrt{g^{\vphantom{*}}_{A}/(2\pi)}
     }
     {
  4\sqrt{g^{\vphantom{*}}_{A}/(2\pi)}-1
     }  },
&
\hbox{ for }
g^{\vphantom{*}}_{A}\geq2\pi.
\end{array}
\right.
\nonumber\\
\label{eq: final typical DOS for MDH}
\end{eqnarray}
The energy scale $\varepsilon^{\vphantom{*}}_{\hbox{\tiny ren}}$
is the one for which errors caused by
neglecting renormalization effects on
$g^{\vphantom{*}}_{A}$
are of order 1. 
{}{
For weak disorder ($g^{\vphantom{*}}_{A}<2\pi$), 
quenched and annealed average values of the dynamical exponent agree
since $2-z^{\vphantom{*}}_{A}=x^{\vphantom{*}}_1$.
For strong disorder ($g^{\vphantom{*}}_{A}\geq2\pi$), 
quenched and annealed average values of the dynamical exponent differ
since $2-z^{\vphantom{*}}_{A}\neq x^{\vphantom{*}}_1$.
                }
Bounds on the dynamical exponent that are consistent
with Eq.\ (\ref{eq: dynamical exponent of MDH})
were established in Ref.\ \onlinecite{Motrunich02}.
The same power-law dependence for the {}{DOS}
of a single species of Dirac fermions subjected to a real random vector
potential was derived in Ref.\ \onlinecite{Ludwig94}
for weak disorder. In this simpler context,
the transition from weak to strong disorder for the 
{}{DOS}
was established in
Ref.\ \onlinecite{Horovitz02}
using a replica variational method.
We now turn our attention to the computation of the 
{}{DOS}
in the HWK model for which a field-theoretical approach is lacking.

\section{DOS for $H_{\hbox{\tiny HWK}}$ }
\label{sec: DOS for the HWK model}

The HWK model differs from the MDH model by the presence of 
a complex-valued random mass 
$M^{\vphantom{*}}_{-}(\bm{r})=[M^{\vphantom{*}}_{+}(\bm{r})]^*$. 
It was pointed out in Ref.\ \onlinecite{Hatsugai97} 
that, although $g^{\vphantom{*}}_A$ is exactly marginal
in the MDH model (\ref{eq: final SUSY theory for MDH}),
the presence of a random mass renders
$g^{\vphantom{*}}_A$ marginally relevant
{{}}{(in the sense that it grows logarithmically with length)}
whereas $g^{\vphantom{*}}_M$ remains marginal
in a one-loop RG analysis.
Guruswamy \textit{et al.}\ computed the exact $\beta$ functions for 
$g^{\vphantom{*}}_A$
and
$g^{\vphantom{*}}_M$,
thereby confirming the one-loop RG analysis.\cite{Guruswamy00}
Furthermore, they found the anomalous scaling dimension
\begin{eqnarray}
&&
x^{\vphantom{*}}_{\varepsilon}=
{{{}}{\zeta}^{\vphantom{2}}_{\vphantom{+}}}
\left[1+\mathcal{O}(g^{\vphantom{*}}_M)\right]
-\frac{g^{\vphantom{*}}_A}{\pi}{{{}}{\zeta}^{2}_{\vphantom{+}}},
\nonumber\\
&&
{{{}}{\zeta}^{\vphantom{2}}_{\vphantom{+}}}:=
\frac{1}{1+g^{\vphantom{*}}_M/\pi},
\label{eq: anomalous scaling dimension GLF}
\end{eqnarray}
for the energy perturbation 
in Eq.\ (\ref{eq: decoupling trsf on SUSY HWK})
by taking advantage of enhanced symmetries of the
HWK model at the {band center}. Remarkably, 
the dependence 
on $g^{\vphantom{*}}_A$
in Eq.\ (\ref{eq: anomalous scaling dimension GLF})
is exact while that on $g^{\vphantom{*}}_M$
can be obtained by perturbation theory.
Finally, they wrote down a Callan-Symanzik equation
for the {}{DOS} of the form 
(\ref{eq: Callan-Symanzik eq in all generality})
with a {finite} number (3) of coupling constants.
The three coupling constants are
$\varepsilon$,
$g^{\vphantom{*}}_A$,
and 
$g^{\vphantom{*}}_M$.
The corresponding $\beta$ functions are
\begin{eqnarray}
&&
\beta^{\vphantom{*}}_{\varepsilon}=
(2-x^{\vphantom{*}}_{\varepsilon})\varepsilon,
\qquad
\beta^{\vphantom{*}}_{g^{\vphantom{*}}_A}=
\frac{(g^{\vphantom{*}}_M{{{}}{\zeta}^{\vphantom{2}}_{\vphantom{-}}})^2}{2\pi^2},
\qquad
\beta^{\vphantom{*}}_{g^{\vphantom{*}}_M}=0,
\nonumber\\
\label{eq: beta functions in GLL}
\end{eqnarray}
respectively. Additionally,
the anomalous scaling dimension $x^{\vphantom{*}}_\varepsilon$
entering the Callan-Symanzik
equation is given in
Eq.\ (\ref{eq: anomalous scaling dimension GLF}).
Solving this Callan-Symanzik equation gives 
the diverging {}{DOS}
(\ref{eq: DOS Gade})
in the vicinity of the band center.

One can tune the HWK model to the MDH model by
switching off the random mass perturbation, i.e., by
taking the limit $g^{\vphantom{*}}_M=0$. In this limit,
the Callan-Symanzik equation used by Guruswamy \textit{et al.}\ 
only agrees with the Callan-Symanzik equation 
(\ref{eq: Callan-Symanzik MDH typical DOS})
for the 
{}{DOS}
when $g^{\vphantom{*}}_A<2\pi$, 
i.e., below the freezing transition.
As we have seen in Sec.\ \ref{sec: DOS for the MDH model}, 
this discrepancy is caused by a failure to account
for all the relevant coupling constants associated 
{}{with powers of the energy perturbation.}
In this section, we are going to show that infinitely many
relevant operators were overlooked by Guruswamy \textit{et al.}\ 
in their RG analysis of Eq.\ (\ref{eq: decoupling trsf on SUSY HWK})
at the band center.
We will then derive a generalized KPP equation for the generating
function of these coupling constants. This generalized
KPP equation differs from the KPP equation
(\ref{eq: RG for widetilde G})
by a scale-dependent diffusion ``constant'' driven by the scale dependence
of $g^{\vphantom{*}}_A$.
Within an adiabatic approximation, we shall 
extract from this generalized KPP equation a Callan-Symanzik
equation for the 
{}{DOS}
of the form (\ref{eq: Callan-Symanzik MDH typical DOS})
except for the presence of the running coupling constant 
$g^{\vphantom{*}}_A$ whose $\beta$ function is given by 
Eq.\ (\ref{eq: beta functions in GLL}).
However small the bare value of
$g^{\vphantom{*}}_A$, 
it always flows to infinitely strong disorder.
The freezing transition must then
necessarily affect the 
{}{DOS}
 in the HWK model.
The solution for the 
{}{DOS}
in the HWK model is in fact given by
Eq.\ (\ref{eq: DOS Gade})
with the exponent $\kappa=2/3$ instead of $\kappa=1/2$.
It is through the larger exponent $\kappa=2/3$ that
the freezing transition reveals itself
as was noted by {{}}{Motrunich \textit{et al.}}\cite{Motrunich02}
A more detailed analysis of the Callan-Symanzik equation shows
that the asymptotic regime
(\ref{eq: DOS Gade}) 
with $\kappa=2/3$
can be preceded by crossover regimes characterized by
the Gade singularity or by the power-law behavior of the MDH model
provided bare coupling constants are fine-tuned.

\begin{table*}[t]
\caption{
\label{table: bosonization rules: diag in flavor}
Abelian bosonization rules for gl($2|2$) matter fields and currents:
flavor-diagonal components of the currents.
The scalar fields $\phi_a$, 
$\varphi_{a-}$, and $\varphi_{a+}$
are related by
$\phi_a(z,\bar z)= 
 \varphi_{a-}(z)+\varphi_{a+}(\bar z)$.
The scalar fields $\phi^{\prime}_a$,
$\varphi^{\prime}_{a-}$, and $\varphi^{\prime}_{a+}$
are related by
$\phi^{\prime}_a(z,\bar z)= 
 \varphi^{\prime}_{a-}(z)+\varphi^{\prime}_{a+}(\bar z)$.
The lattice spacing $\mathfrak{a}$ is used as a short-distance cutoff.
The magnitude of the complex number ${{}}{u}$
is arbitrary as it reflects 
the nonuniversality of the short-distance regularization. 
The phase of the complex number ${{}}{u}$ is
fixed by the boundary condition obeyed by the scalar fields at
infinity, which, in turn, is uniquely fixed by the SUSY of the theory.
Demanding that the scalar fields vanish at infinity amounts to
fixing the phase of ${{}}{u}$ to be $-\pi/2$.
We will choose ${{}}{u}=-i$ for convenience.}
\begin{ruledtabular}
\begin{tabular}{lcclc}
\hbox{SUSY matter }
&
\hbox{Bosonized matter}
&
&
\hbox{SUSY currents}
&
\hbox{Bosonized currents}
\\\hline\\
$\psi ^{* }_{a-}$
& 
$
 {{}}{u^{+1/2}}\hphantom{()^*}
 \mathfrak{a}^{-1/2}\, e^{-{i}\varphi_{a-}}
$
&
&
$     J^{\ }_{aa}:=\psi ^{\ }_{a-}\psi ^{* }_{a-}$
&
$(+{i}    \partial\phi_a)$
\\
$\psi ^{\ }_{a-}$
& 
$
 {{}}{u^{-1/2}}\hphantom{()^*}
 \mathfrak{a}^{-1/2}\, e^{+{i}\varphi_{a-}}
$
&
&
$
 \bar J^{\ }_{aa}:=\psi ^{\ }_{a+}\psi ^{* }_{a+}
$
&
$(-{i}\bar\partial\phi_a)$
\\
$\psi ^{* }_{a+}$
&
$
 {{}}{(u^*)^{+1/2}}
 \mathfrak{a}^{-1/2}\, e^{+{i}\varphi_{a+}}
$
&
&
$     
 J^{\prime}_{aa}:=\beta^{\ }_{a-}\beta^{* }_{a-}
$
&
$(+{i}    \partial\phi'_a)$
\\
$\psi ^{\ }_{a+}$
& 
$
 {{}}{(u^*)^{-1/2}}
 \mathfrak{a}^{-1/2}\, e^{-{i}\varphi_{a+}}
$
&
&
$
 \bar J^{\prime}_{aa}:=\beta^{\ }_{a+}\beta^{* }_{a+}
$
&
$(-{i}\bar\partial\phi'_a)$
\\
$\beta^{* }_{a-}$
& 
$
 {{}}{u^{-1/2}}\hphantom{()^*}
 \mathfrak{a}^{+1/2} \,e^{-{i}\varphi^{\prime}_{a-}}
 ({i}   \partial\chi_{a})
$
&
&
$     
 G^{\ }_{aa+}:=\beta^{* }_{a-}\psi ^{\ }_{a-}
$
&
$
 {{}}{\hphantom{(}u^{-1}\hphantom{)^*}}
 e^{-{i}(\varphi_{a-}^{\prime}-\varphi_{a-}^{\     })}
 ({i}    \partial\chi^{\ }_{a})
$
\\
$\beta^{\ }_{a-}$
&
$
 {{}}{u^{+1/2}}\hphantom{()^*}
 \mathfrak{a}^{+1/2} \,e^{+{i}\varphi^{\prime}_{a-}}
 ({i}    \partial\chi^{* }_{a})
$
&
&
$
 \bar G^{\ }_{aa+}:=\beta^{* }_{a+}\psi ^{\ }_{a+}
$
&
$
 {{}}{(u^{*})^{-1}}
 e^{+{i}(\varphi_{a+}^{\prime}-\varphi_{a+}^{\     })}
 ({i}\bar\partial\chi^{\ }_{a})
$
\\
$\beta^{* }_{a+}$
&
$
 {{}}{(u^*)^{-1/2}}
 \mathfrak{a}^{+1/2} \,e^{+{i}\varphi^{\prime}_{a+}}
 ({i}\bar\partial\chi^{\ }_{a})
$
&
&
$     
 G^{\ }_{aa-}:=\beta^{\ }_{a-}\psi^{* }_{a-}
$
&
$
 {{}}{\hphantom{(}u\hphantom{^{-1})^*}}
 e^{+{i}(\varphi_{a-}^{\prime}-\varphi_{a-}^{\     })}
 ({i}    \partial\chi^{* }_{a})
$
\\
$\beta^{\ }_{a+}$
&
$
 {{}}{(u^*)^{+1/2}}
 \mathfrak{a}^{+1/2} \,e^{-{i}\varphi^{\prime}_{a+}}
 ({i}\bar\partial\chi^{* }_{a})
$
&
&
$
 \bar G^{\ }_{aa-}:=\beta^{\ }_{a+}\psi^{* }_{a+}
$
&
$
 {{}}{\hphantom{(}u^*\hphantom{^{-1})}}
 e^{-{i}(\varphi_{a+}^{\prime}-\varphi_{a+}^{\     })}
 ({i}\bar\partial\chi^{* }_{a})
$
\end{tabular}
\end{ruledtabular}
\end{table*}

\begin{table*}[t]
\caption{
\label{table: bosonization rules: off-diag in flavor}
Abelian bosonization rules for gl($2|2$) matter fields and currents:
flavor--off-diagonal components of the currents, i.e., $a\neq b$.
The scalar fields $\phi_a$, 
$\varphi_{a-}$, and $\varphi_{a+}$
are related by
$\phi_a(z,\bar z)= 
 \varphi_{a-}(z)+\varphi_{a+}(\bar z)$.
The scalar fields $\phi^{\prime}_a$,
$\varphi^{\prime}_{a-}$, and $\varphi^{\prime}_{a+}$
are related by
$\phi^{\prime}_a(z,\bar z)= 
 \varphi^{\prime}_{a-}(z)+\varphi^{\prime}_{a+}(\bar z)$.
The lattice spacing $\mathfrak{a}$ is used as a short-distance cutoff.
The magnitude of the complex number ${{}}{u}$
is arbitrary as it reflects 
the nonuniversality of the short-distance regularization. 
The phase of the complex number ${{}}{u}$ is
fixed by the boundary condition obeyed by the scalar fields at
infinity, which, in turn, is uniquely fixed by the SUSY of the theory.
Demanding that the scalar fields vanish at infinity amounts to
fixing the phase of ${{}}{u}$ to be $-\pi/2$.
We will choose ${{}}{u}=-i$ for convenience.}
\begin{ruledtabular}
\begin{tabular}{lcclc}
\hbox{SUSY matter }
&
\hbox{Bosonized matter}
&
&
\hbox{SUSY currents}
&
\hbox{Bosonized currents}
\\\hline\\
$\psi ^{* }_{a-}$
& 
$
 {{}}{u^{+1/2}}\hphantom{()^*}
 \mathfrak{a}^{-1/2}\, e^{-{i}\varphi_{a-}}
$
&
&
$     J^{\ }_{ba}:=\psi ^{\ }_{a-}\psi ^{* }_{b-}$
&
$\mathfrak{a}^{-1}e^{+{i}(\varphi_{a-}-\varphi_{b-})}$
\\
$\psi ^{\ }_{a-}$
& 
$
 {{}}{u^{-1/2}}\hphantom{()^*}
 \mathfrak{a}^{-1/2}\, e^{+{i}\varphi_{a-}}
$
&
&
$\bar J^{\ }_{ba}:=\psi ^{\ }_{a+}\psi ^{* }_{b+}$
&
$\mathfrak{a}^{-1}e^{-{i}(\varphi_{a+}-\varphi_{b+})}$
\\
$\psi ^{* }_{a+}$
&
$
 {{}}{(u^*)^{+1/2}}
 \mathfrak{a}^{-1/2}\, e^{+{i}\varphi_{a+}}
$
&
&
$     J^{\prime}_{ba}:=\beta^{\ }_{a-}\beta^{* }_{b-}$
&
$
\mathfrak{a}e^{+{i}(\varphi'_{a-}-\varphi'_{b-})}
({i}\partial\chi^{* }_a)
({i}\partial\chi^{\ }_b)
$
\\
$\psi ^{\ }_{a+}$
& 
$
 {{}}{(u^*)^{-1/2}}
 \mathfrak{a}^{-1/2}\, e^{-{i}\varphi_{a+}}
$
&
&
$\bar J^{\prime}_{ba}:=\beta^{\ }_{a+}\beta^{* }_{b+}$
&
$
\mathfrak{a}e^{-{i}(\varphi'_{a+}-\varphi'_{b+})}
({i}\bar\partial\chi^{* }_a)
({i}\bar\partial\chi^{\ }_b)
$
\\
$\beta^{* }_{a-}$
& 
$
 {{}}{u^{-1/2}}\hphantom{()^*}
 \mathfrak{a}^{+1/2} \,e^{-{i}\varphi^{\prime}_{a-}}
 ({i}   \partial\chi^{\ }_{a})
$
&
&
$     G^{\ }_{ba+}:=\beta^{* }_{a-}\psi ^{\ }_{b-}$
&
$
 {{}}{\hphantom{(}u^{-1}\hphantom{)^*}}
 e^{-{i}(\varphi_{a-}^{\prime}-\varphi_{b-}^{\     })}
 ({i}    \partial\chi^{\ }_{a})
$
\\
$\beta^{\ }_{a-}$
&
$
 {{}}{u^{+1/2}}\hphantom{()^*}
 \mathfrak{a}^{+1/2} \,e^{+{i}\varphi^{\prime}_{a-}}
 ({i}    \partial\chi^{* }_{a})
$
&
&
$\bar G^{\ }_{ba+}:=\beta^{* }_{a+}\psi ^{\ }_{b+}$
&
$
 {{}}{(u^{*})^{-1}}
 e^{+{i}(\varphi_{a+}^{\prime}-\varphi_{b+}^{\     })}
 ({i}\bar\partial\chi^{\ }_{a})
$
\\
$\beta^{* }_{a+}$
&
$
 {{}}{(u^*)^{-1/2}}
 \mathfrak{a}^{+1/2} \,e^{+{i}\varphi^{\prime}_{a+}}
 ({i}\bar\partial\chi^{\ }_{a})
$
&
&
$     G^{\ }_{ba-}:=\beta^{\ }_{a-}\psi^{* }_{b-}$
&
$
 {{}}{\hphantom{(}u\hphantom{^{-1})^*}}
 e^{+{i}(\varphi_{a-}^{\prime}-\varphi_{b-}^{\     })}
 ({i}    \partial\chi^{* }_{a})
$
\\
$\beta^{\ }_{a+}$
&
$
 {{}}{(u^*)^{+1/2}}
 \mathfrak{a}^{+1/2} \,e^{-{i}\varphi^{\prime}_{a+}}
 ({i}\bar\partial\chi^{* }_{a})
$
&
&
$\bar G_{ba-}:=\beta^{\ }_{a+}\psi^{* }_{b+}$
&
$
 {{}}{\hphantom{(}u^*\hphantom{^{-1})}}
 e^{-{i}(\varphi_{a+}^{\prime}-\varphi_{b+}^{\     })}
 ({i}\bar\partial\chi^{* }_{a})
$
\end{tabular}
\end{ruledtabular}
\end{table*}

\subsection{Nearly conformal field theory}
\label{subsec: Nearly conformal field theory}

The chiral decoupling transformation
(\ref{eq: chiral rep of dec trsf})
on the HWK model 
in the SUSY representation 
(\ref{eq: def SUSY representation HWK no chiral trsf})
is not as useful when $g^{\vphantom{*}}_M>0$
as it is when $g^{\vphantom{*}}_M=0$
since rotated SUSY matter fields remain coupled to the disorder
through the random mass. Our starting point in this section will
thus be the effective theory obtained from the HWK model 
{at the band center}
after integration over the Gaussian-distributed random vector  
and mass potentials
\begin{eqnarray}
&&
Z:=
\int\mathcal{D}
[
\psi ^{* },\psi ^{\vphantom{*}},
\beta^{* },\beta^{\vphantom{*}}
]\,
e^{
-\int \frac{d^2\bm{r}}{\pi}
\left(
\mathcal{L}^{\vphantom{*}}_0
+
\frac{g^{\vphantom{*}}_A}{2\pi}
\mathcal{O}^{\vphantom{*}}_A
+
\frac{g^{\vphantom{*}}_M}{2\pi}
\mathcal{O}^{\vphantom{*}}_M
\right)
  },
\nonumber\\
&&
\mathcal{L}^{\vphantom{*}}_0:=
\sum_{a=1}^{2}
\left(
\psi ^{* }_{a-}\bar\partial\psi ^{\vphantom{*}}_{a-}
+
\psi ^{* }_{a+}    \partial\psi ^{\vphantom{*}}_{a+}
\right.
\nonumber\\
&&\hphantom{
\mathcal{L}^{\vphantom{*}}_0:=
\sum_{a=1}^{2}\big(
           }
+
\left.
\beta^{* }_{a-}\bar\partial\beta^{\vphantom{*}}_{a-}
+
\beta^{* }_{a+}    \partial\beta^{\vphantom{*}}_{a+}
\right),
\nonumber\\
&&
\mathcal{O}^{\vphantom{*}}_A=
\!-\!
\sum_{a,b=1}^{2}\!
\left(
\psi ^{* }_{a-}\psi ^{\vphantom{*}}_{a-}
\!+\!
\beta^{* }_{a-}\beta^{\vphantom{*}}_{a-}
\right)
\left(
\psi ^{* }_{b+}\psi ^{\vphantom{*}}_{b+}
\!+\!
\beta^{* }_{b+}\beta^{\vphantom{*}}_{b+}
\right),
\nonumber\\
&&
{}{
\mathcal{O}^{\vphantom{*}}_M=
\!-\!
\sum_{a,b=1}^{2}\!
\left(
\psi ^{* }_{a+}\psi ^{\vphantom{*}}_{a-}
\!+\!
\beta^{* }_{a+}\beta^{\vphantom{*}}_{a-}
\right)
\left(
\psi ^{* }_{b-}\psi ^{\vphantom{*}}_{b+}
\!+\!
\beta^{* }_{b-}\beta^{\vphantom{*}}_{b+}
\right).
                }
\nonumber\\
&&
\label{eq: def HWK model after integration over disorder}
\end{eqnarray}

\noindent
Here
$
\bar\partial:=\frac{1}{2}\partial^{\vphantom{*}}_+
$,
and
$
    \partial:=\frac{1}{2}\partial^{\vphantom{*}}_-
$.
Observe that the effective theory
(\ref{eq: def HWK model after integration over disorder})
is invariant under any permutation of the flavor indices. 
Flavor permutation symmetry is broken by the energy perturbation
(\ref{eq: non-Hermitian source term}); i.e.,
the SUSY symmetry group in the presence of 
an energy perturbation is {enlarged} to the group GL($2|2$)
at the band center.
Since we are after the behavior of the 
{}{DOS}
in the immediate vicinity of
the band center, it is essential to account for
the enlarged SUSY symmetry group GL($2|2$) at the band center. 

Fermionic interacting field theories can be considerably simplified
by the method of bosonization. This is also the case of the 
SUSY representation 
(\ref{eq: def HWK model after integration over disorder})
of the HWK model at the band center
as shown by Guruswamy \textit{et al.},\cite{Guruswamy00}
{{}}{and we now summarize their results.
We thus adopt} Abelian bosonization rules
that are summarized in Tables
\ref{table: bosonization rules: diag in flavor}\!
--\ref{table: bosonization rules: off-diagonal interactions}.\cite{errorguru}
Abelian bosonization in the fermionic sector is performed by trading
left, $\psi^{*}_{a-}${}{, say,} 
and 
right, $\psi^{* }_{a+}${}{, say,} 
moving fermions 
for the exponential of holomorphic 
$\varphi^{\vphantom{*}}_{a-}$
and antiholomorphic
$\varphi^{\vphantom{*}}_{a+}$ 
scalar fields, 
respectively.
Abelian bosonization in the bosonic sector was introduced in
string theory.\cite{Friedan86,Gaberdiel96}
It is performed by trading
left, $\beta^{*}_{a-}$, say, 
and 
right, $\beta^{* }_{a+}$, say, 
moving bosons 
for the exponential of holomorphic
$\varphi^{\prime}_{a-}$
and antiholomorphic
$\varphi^{\prime}_{a+}$
scalar fields, 
respectively. 
Additional fermionic fields 
$\chi^{*}_{a}$
and 
$\chi^{\vphantom{*}}_{a}$
must also be introduced when bosonizing bosonic spinors.
A simplifying feature of the two interactions induced by the disorder
in Eq.\ (\ref{eq: def HWK model after integration over disorder})
is that they can be written as bilinears in the gl($2|2$) currents
that generate the GL($2|2$) symmetry group. These currents are of two
kinds.
The first kind is made of bosonic currents denoted
$J^{\vphantom{\prime}}_{ab}$, 
$\bar J^{\vphantom{\prime}}_{ab}$, 
$J^{\prime}_{ab}$, 
and 
$\bar J^{\prime}_{ab}$,
which are defined in 
tables \ref{table: bosonization rules: diag in flavor}
and    \ref{table: bosonization rules: off-diag in flavor}.
The second kind is made of fermionic currents denoted
$G^{\vphantom{\prime}}_{ab+}$, 
$\bar G^{\vphantom{\prime}}_{ab+}$, 
$G^{\vphantom{\prime}}_{ab-}$, 
and 
$\bar G^{\vphantom{\prime}}_{ab-}$,
which are also defined in
Tables \ref{table: bosonization rules: diag in flavor}
and    \ref{table: bosonization rules: off-diag in flavor}.
Currents of the bosonic kind permute the flavors of right 
and left movers but do not change their statistics.
Currents of the fermionic kind exchange fermionic spinors
of one flavor for bosonic spinors of another flavor, and conversely.
The importance of the gl($N|N$), $N\in\mathbb{N}$ 
algebra for Dirac fermions coupled to random fields was emphasized in 
Refs.\ \onlinecite{Chamon-Mudry96},
       \onlinecite{Mudry96},
and    \onlinecite{Bernard95}.
The central role of the symmetry group GL($N|N$) 
also appears in the quantum field theory description of the 
multivariable Alexander polynomial.\cite{Rozansky92}

After Abelian bosonization,
the holomorphic and antiholomorphic scalar fields 
$\varphi^{\vphantom{*}}_{a\pm}$ and $\varphi^{\prime}_{a\pm}$
enter Eq.\ (\ref{eq: def HWK model after integration over disorder})
through the combinations
\begin{eqnarray}
\phi^{\vphantom{*}}_{a}:=
\varphi^{\vphantom{*}}_{a-}+\varphi^{\vphantom{*}}_{a+},
\quad
\phi^{\prime}_{a}:=
\varphi^{\prime}_{a-}+\varphi^{\prime}_{a+},
\quad
{}{
a=1,2
}
\nonumber\\
&&
\label{eq: def fields phi and phi'}
\end{eqnarray}
only
{}{
(see 
Tables \ref{table: bosonization rules: diagonal interactions}
and    \ref{table: bosonization rules: off-diagonal interactions}),
                }
\begin{widetext}
\begin{subequations}
\begin{eqnarray}
&&
Z=
\int\mathcal{D}
[
\phi^{\vphantom{*}}_a,
\phi^{\prime}_a,
\chi^{*}_a,
\chi^{\vphantom{*}}_a
]
\exp
\left(
-
\int \frac{d^2\bm{r}}{2\pi}
\mathcal{L}^{\vphantom{*}}
\right),
\qquad
\mathcal{L}^{\vphantom{*}}:=
\mathcal{L}^{\vphantom{*}}_1
+
\mathcal{L}^{\vphantom{*}}_2,
\label{eq: reorganised def bosonic int field theory N>1 a}\\
&&
\mathcal{L}^{\vphantom{*}}_1=
\left(1+\frac{g^{\vphantom{*}}_M}{\pi}\right)
\sum_{a=1}^{2}
\left[
(    \partial\phi^{\vphantom{*}}_{a})
(\bar\partial\phi^{\vphantom{*}}_{a})
-
(    \partial\phi^{\prime}_{a})
(\bar\partial\phi^{\prime}_{a})
\right]
-
\frac{g^{\vphantom{*}}_A}{\pi}
(    \partial\Sigma^{\vphantom{*}}_-)
(\bar\partial\Sigma^{\vphantom{*}}_-)
\nonumber\\
&&
\hphantom{\mathcal{L}^{\vphantom{*}}_1=}
-
\sum_{a=1}^{2}
\left\{
\left[
1
{{}}{+}
\frac{g^{\vphantom{*}}_M}{\pi}
e^{-{i}(\phi^{\vphantom{*}}_a-\phi^{\prime}_a)}
\right]
(     \partial\chi^{* }_{a})
(\bar \partial\chi^{\vphantom{*}}_{a})
+
\left[
1
{{}}{+}
\frac{g^{\vphantom{*}}_M}{\pi}
e^{+{i}(\phi^{\vphantom{*}}_a-\phi^{\prime}_a)}
\right]
(\bar\partial\chi^{* }_{a})
(    \partial\chi^{\vphantom{*}}_{a})
\right\},
\label{eq: reorganised def bosonic int field theory N>1 b}\\
&&
\mathcal{L}^{\vphantom{*}}_2=
\frac{g^{\vphantom{*}}_M}{\pi}
\sum_{a\neq b}^{2}
\left[
\frac{1}{\mathfrak{a}^2}
e^{+{i}(\phi^{\vphantom{*}}_a-\phi^{\vphantom{*}}_b )}
\!-\!
         \mathfrak{a}^2
e^{+{i}(\phi^{\prime}_a-\phi^{\prime}_b )}
(    \partial\chi^{* }_a)
(    \partial\chi^{\vphantom{*}}_b)
(\bar\partial\chi^{* }_b)
(\bar\partial\chi^{\vphantom{*}}_a)
\!{{}}{-}\!
 e^{-{i}(\phi^{\vphantom{*}}_{b}-\phi^{\prime}_{a})}
 (    \partial\chi^{* }_{a})
 (\bar\partial\chi^{\vphantom{*}}_{a})
\!{{}}{-}\!
 e^{+{i}(\phi^{\vphantom{*}}_{a}-\phi^{\prime}_{b})}
 (\bar\partial\chi^{* }_{b})
 (    \partial\chi^{\vphantom{*}}_{b})
\right].
\nonumber\\
&&
\label{eq: reorganised def bosonic int field theory N>1 c}
\end{eqnarray}
\label{eq: reorganised def bosonic int field theory N>1}
\end{subequations}

\noindent
Here, we have introduced the ``center of mass'' 
$\Sigma^{\vphantom{*}}_{+}$ 
and the ``relative coordinate''
$\Sigma^{\vphantom{*}}_{-}$ 
for the scalar fields
\begin{eqnarray}
\Sigma^{\vphantom{*}}_{+}:=
\Sigma
+
\Sigma^{\prime},
\qquad
\Sigma^{\vphantom{*}}_{-}:=
\Sigma
-
\Sigma^{\prime},
\qquad
\Sigma:=
\sum_{a=1}^{2}
\phi^{\vphantom{*}}_a,
\qquad
\Sigma^{\prime}:=
\sum_{a=1}^{2}
\phi^{\prime}_a.
\end{eqnarray}
After Abelian bosonization, the decoupling transformation
(\ref{eq: chiral rep of dec trsf})
takes the very simple and useful form
\begin{eqnarray}
\mathcal{L}
[\phi^{\vphantom{*}}_a,
 \phi^{\prime}_a,
 \chi^{*     }_a,
 \chi^{\vphantom{*}}_a;g^{\vphantom{*}}_A,g^{\vphantom{*}}_M]
=
\mathcal{L}
\left[
\phi^{\vphantom{*}}_a
-\frac{{g}_A}{2\pi}{{{}}{\zeta}^{\vphantom{2}}_{\vphantom{+}}}
 \Sigma^{\vphantom{*}}_-,
 \phi^{\prime}_a
-\frac{{g}_A}{2\pi}{{{}}{\zeta}^{\vphantom{2}}_{\vphantom{+}}}
 \Sigma^{\vphantom{*}}_-,
 \chi^{*     }_a,
 \chi^{\vphantom{*}}_a
;g^{}_A=0,g^{}_M\right]
.
\label{eq: bosonized version of chiral trsf}
\end{eqnarray}

\begin{table*}[t]
\caption{
\label{table: bosonization rules: diagonal interactions}
Abelian bosonization rules for mass and vector potential
current-current interactions:
flavor-diagonal contributions.
Columns 1 and 2 are the contributions to the mass perturbation
and
columns 3 and 4 are the contributions to the vector perturbation.
With the choice ${{}}{u}=-i$ in Tables
\ref{table: bosonization rules: diag in flavor}
and
\ref{table: bosonization rules: off-diag in flavor}
${{}}{u^{-1}u^*=u(u^*)^{-1}=-1}$.}
\begin{ruledtabular}
\begin{tabular}{lcclc}
SUSY matter
&
Bosonized matter
&
\hphantom{AAAA}
&
SUSY matter
&
Bosonized matter
\\\hline\\
$
+
     J^{\vphantom{\prime}}_{aa}
\bar J^{\vphantom{\prime}}_{aa}
$
&
$
+
(    \partial\phi_a)
(\bar\partial\phi_a)
$
&
&
$
-
     J^{\vphantom{\prime}}_{aa}
\bar J^{\vphantom{\prime}}_{aa}
$
&
$
-
(    \partial\phi_a)
(\bar\partial\phi_a)
$
\\
$
-
     J^{         {\prime}}_{aa}
\bar J^{         {\prime}}_{aa}
$
&
$
-
(    \partial\phi^{\prime}_a)
(\bar\partial\phi^{\prime}_a)
$
&
&
$
-
     J^{         {\prime}}_{aa}
\bar J^{         {\prime}}_{aa}
$
&
$
-
(    \partial\phi^{\prime}_a)
(\bar\partial\phi^{\prime}_a)
$
\\
$
+
     G^{\vphantom{\prime}}_{aa+}
\bar G^{\vphantom{\prime}}_{aa-}
$
&
$
+
{{}}{u^{-1}u^*}
e^{-{i}(\phi^{\prime}_a-\phi^{\ }_a)}
({i}    \partial\chi^{\ }_a)
({i}\bar\partial\chi^{* }_a)
$
&
&
$
+
     J^{\vphantom{\prime}}_{aa}
\bar J^{         {\prime}}_{aa}
$
&
$
+
(    \partial\phi^{\vphantom{\prime}}_a)
(\bar\partial\phi^{         {\prime}}_a)
$
\\
$
-
     G^{\vphantom{\prime}}_{aa-}
\bar G^{\vphantom{\prime}}_{aa+}
$
&
$
-{{}}{u(u^*)^{-1}}
e^{+{i}(\phi^{\prime}_a-\phi^{\ }_a)}
({i}    \partial\chi^{* }_a)
({i}\bar\partial\chi^{\ }_a)
$
&
&
$
+
     J^{         {\prime}}_{aa}
\bar J^{\vphantom{\prime}}_{aa}
$
&
$
+
(    \partial\phi^{         {\prime}}_a)
(\bar\partial\phi^{\vphantom{\prime}}_a)
$
\end{tabular}
\end{ruledtabular}
\end{table*}

\begin{table*}[t]
\caption{
\label{table: bosonization rules: off-diagonal interactions}
Abelian bosonization rules for mass and vector potential
current-current interactions:
flavor--off-diagonal contributions, i.e., $a\neq b$.
The lattice spacing $\mathfrak{a}$ is used as a short-distance cutoff.
Columns 1 and 2 are the contributions to the mass perturbation,
and
columns 3 and 4 are the contributions to the vector perturbation.
With the choice ${{}}{u}=-i$ in Tables
\ref{table: bosonization rules: diag in flavor}
and
\ref{table: bosonization rules: off-diag in flavor}
${{}}{u^{-1}u^*=u(u^*)^{-1}=-1}$.}
\begin{ruledtabular}
\begin{tabular}{lcclc}
SUSY matter
&
Bosonized matter
&
\hphantom{AAAA}
&
SUSY matter
&
Bosonized matter
\\\hline\\
$
+
     J^{\vphantom{\prime}}_{ba}
\bar J^{\vphantom{\prime}}_{ab}
$
&
$
+
\mathfrak{a}^{-2} e^{+{i}(\phi_a-\phi_b)}
$
&
&
$
-
     J^{\vphantom{\prime}}_{aa}
\bar J^{\vphantom{\prime}}_{bb}
$
&
$
-
(    \partial\phi_a)
(\bar\partial\phi_b)
$
\\
$
-
     J^{         {\prime}}_{ba}
\bar J^{         {\prime}}_{ab}
$
&
$
-
\mathfrak{a}^{+2} e^{+{i}(\phi^{\prime}_a-\phi^{\prime}_b)}
(    \partial\chi^{* }_a)
(    \partial\chi^{\vphantom{*}}_b)
(\bar\partial\chi^{* }_b)
(\bar\partial\chi^{\vphantom{*}}_a)
$
&
&
$
-
     J^{         {\prime}}_{aa}
\bar J^{         {\prime}}_{bb}
$
&
$
-
(    \partial\phi^{\prime}_a)
(\bar\partial\phi^{\prime}_b)
$
\\
$
+
     G^{\vphantom{\prime}}_{ab+}
\bar G^{\vphantom{\prime}}_{ab-}
$
&
$
+{{}}{u^{-1}u^*}
e^{-{i}(\phi^{\prime}_b-\phi^{\ }_a)}
({i}    \partial\chi^{\ }_b)
({i}\bar\partial\chi^{* }_b)
$
&
&
$
+
     J^{\vphantom{\prime}}_{aa}
\bar J^{         {\prime}}_{bb}
$
&
$
+
(    \partial\phi^{\vphantom{\prime}}_a)
(\bar\partial\phi^{         {\prime}}_b)
$
\\
$
-
     G^{\vphantom{\prime}}_{ba-}
\bar G^{\vphantom{\prime}}_{ba+}
$
&
$
-{{}}{u(u^*)^{-1}}
e^{+{i}(\phi^{\prime}_a-\phi^{\ }_b)}
({i}    \partial\chi^{* }_a)
({i}\bar\partial\chi^{\ }_a)
$
&
&
$
+
     J^{         {\prime}}_{aa}
\bar J^{\vphantom{\prime}}_{bb}
$
&
$
+
(    \partial\phi^{         {\prime}}_a)
(\bar\partial\phi^{\vphantom{\prime}}_b)
$
\end{tabular}
\end{ruledtabular}
\end{table*}

\end{widetext}

{{}}{As pointed out by Guruswamy 
\textit{et al}.,\cite{Guruswamy00}}
Eq.\ (\ref{eq: bosonized version of chiral trsf}) is
crucial to establish that the HWK model is,
in essence, a nearly conformal field theory at the band center. 
Equation (\ref{eq: bosonized version of chiral trsf})
follows from the peculiarity of the Lagrangian $\mathcal{L}$ 
by which the center of mass $\Sigma^{\vphantom{*}}_+$
only enters (in $\mathcal{L}^{\vphantom{*}}_1$)
through the linear term
\begin{eqnarray}
\frac{1}{2{{{}}{\zeta}^{\vphantom{2}}_{\vphantom{+}}}}
(\partial\Sigma^{\vphantom{*}}_+)(\bar\partial\Sigma^{\vphantom{*}}_-).
\label{eq: coupling between Sigma+ and Sigma-}
\end{eqnarray}
To put it differently, the term 
(\ref{eq: coupling between Sigma+ and Sigma-})
is the only one in $\mathcal{L}$ affected by
the chiral transformation
\begin{eqnarray}
&&
\phi^{\vphantom{*}}_a\to
\phi^{\vphantom{*}}_a
+\frac{{g}^{\vphantom{*}}_A}{2\pi}{{{}}{\zeta}^{\vphantom{2}}_{\vphantom{+}}}
 \Sigma^{\vphantom{*}}_-,
\nonumber\\
&&
\phi^{\prime}_a\to
\phi^{\prime}_a
+\frac{{g}^{\vphantom{*}}_A}{2\pi}{{{}}{\zeta}^{\vphantom{2}}_{\vphantom{+}}}
 \Sigma^{\vphantom{*}}_-,
\qquad
a=1,2.
\label{eq: chiral transf for phi's}
\end{eqnarray}
All other terms in $\mathcal{L}$ and, in particular,
all terms in $\mathcal{L}^{\vphantom{*}}_2$
are left unchanged by the chiral transformation
(\ref{eq: chiral transf for phi's})
as is best seen after trading in $\mathcal{L}$
the four coordinates 
$\phi^{\vphantom{*}}_a$ 
and 
$\phi^{\prime}_a$
($a=1,2$)
for $\Sigma^{\vphantom{*}}_+$, $\Sigma^{\vphantom{*}}_-$ and the
relative coordinates
$\sigma^{\vphantom{*}}_{a+}:=\sigma^{\vphantom{*}}_a+\sigma^{\prime}_a$ 
and 
$\sigma^{\prime}_{a-}      :=\sigma^{\vphantom{*}}_a-\sigma^{\prime}_a$ 
whereby
$\phi^{\vphantom{*}}_a=:(\Sigma/2)+\sigma^{\vphantom{*}}_a$
and
$\phi^{\prime}_a=:(\Sigma^{\prime}/2)+\sigma^{\prime}_a$.
The role of the center of mass $\Sigma^{\vphantom{*}}_+$ is 
thus to enforce a global constraint on the coordinates 
$\phi^{\ }_a$
and
$\phi^{\prime}_a$,
$a=1,2$
of the gl($2|2$)-graded Lie algebra. 

We are mostly (if not only) concerned with
the evaluation of the expectation value of the product of vertex
operators (up to Klein factors that do not matter for 
the calculation of anomalous scaling dimensions)
\begin{eqnarray}
\mathcal{F}
[
\phi^{\vphantom{*}}_a,
\phi^{\prime}_a,
\chi^{* }_a,
\chi^{\vphantom{*}}_a
]:=
\prod_{\iota,\iota'}
e^{
{i}
\alpha^{\     }_\iota
\phi  ^{\     }_\iota
+
{i}
\alpha^{\prime}_{\iota'}
\phi  ^{\prime}_{\iota'}
}
({i}\partial\chi^{\vphantom{*}}_{\iota'}),
\label{eq: def prod vertex}
\end{eqnarray}
where $\iota$ and $\iota'$ are collective labels for flavor and
space coordinates,
$({i}\partial\chi^{\vphantom{*}}_{\iota'})$
should be understood as any of the four options
$({i}    \partial\chi^{\vphantom{*}}_{\iota'})$,
$({i}\bar\partial\chi^{\vphantom{*}}_{\iota'})$,
$({i}    \partial\chi^{* }_{\iota'})$,
and 
$({i}\bar\partial\chi^{* }_{\iota'})$,
and
$\alpha^{\vphantom{*}}_\iota,\alpha^{\prime}_{\iota^{\prime}}\in\mathbb{R}$.
Under the chiral (additive) transformation law
(\ref{eq: chiral transf for phi's})
the transformation law of a product of vertex operators is multiplicative
\begin{eqnarray}
\mathcal{F}
[
\phi^{\vphantom{*}}_a,
\phi^{\prime}_a,
\chi^{* }_a,
\chi^{\vphantom{*}}_a
]
&\to&
e^{
{i}(g^{\vphantom{*}}_A/2\pi)
{{{}}{\zeta}^{\vphantom{2}}_{\vphantom{+}}}\Sigma^{\vphantom{*}}_{-}
\left(
\sum_{\iota }
\alpha^{\vphantom{*}}_{\iota }
+
\sum_{\iota'}
\alpha^{\prime      }_{\iota'}
\right)
  }
\nonumber\\
&&
\times
\mathcal{F}
[
\phi^{\vphantom{*}}_a,
\phi^{\prime}_a,
\chi^{* }_a,
\chi^{\vphantom{*}}_a
].
\label{eq: trsf law vertex operators under chiral}
\end{eqnarray}
The usefulness of 
Eqs.\ 
(\ref{eq: bosonized version of chiral trsf}),
(\ref{eq: chiral transf for phi's}),
and
(\ref{eq: trsf law vertex operators under chiral})
is the following. If the expectation value of the left-hand side
of Eq.\ (\ref{eq: trsf law vertex operators under chiral})
is to be evaluated with the partition function 
(\ref{eq: def HWK model after integration over disorder})
at given 
$g^{\vphantom{*}}_M$ 
and 
$g^{\vphantom{*}}_A$,
then the expectation value of the right-hand side
of Eq.\ (\ref{eq: trsf law vertex operators under chiral})
is to be evaluated at the same
$g^{\vphantom{*}}_M$  
but different
$g^{\vphantom{*}}_A=0$.
Let us denote by 
$
\langle
(\cdots)
\rangle^{\vphantom{*}}_{\mathcal{L},g^{\vphantom{*}}_M,g^{\vphantom{*}}_A}
$
the expectation value with the partition function
(\ref{eq: def HWK model after integration over disorder}).
We have just proved that
\begin{eqnarray}
&&
\left
\langle
\vphantom{e^{A^A}}
\mathcal{F}
[
\phi^{\vphantom{*}}_a,
\phi^{\prime}_a,
\chi^{* }_a,
\chi^{\vphantom{*}}_a
]
\right
\rangle^{\vphantom{*}}_{\mathcal{L},g^{\vphantom{*}}_M,g^{\vphantom{*}}_A}
=
\nonumber\\
&&
{}{
\left
\langle
e^{
{i}(g^{\vphantom{*}}_A/2\pi)
{{{}}{\zeta}^{\vphantom{2}}_{\vphantom{+}}}\Sigma^{\vphantom{*}}_{-}
\left(
\sum_{\iota }
\alpha^{\vphantom{*}}_{\iota }
+
\sum_{\iota'}
\alpha^{\prime      }_{\iota'}
\right)
  }
\mathcal{F}
[
\phi^{\vphantom{*}}_a,
\phi^{\prime}_a,
\chi^{* }_a,
\chi^{\vphantom{*}}_a
]
\right
\rangle^{\vphantom{*}}_{\mathcal{L},g^{\vphantom{*}}_M,0}\!\!.
                 }
\nonumber\\
&&
\label{eq: expectation value vertex operators under chiral}
\end{eqnarray}
Let us denote by 
$
\mathcal{G}
[
\phi^{\vphantom{*}}_a,
\phi^{\prime}_a,
\chi^{* }_a,\chi^{\vphantom{*}}_a
]
$
the right-hand side of 
Eq.\ (\ref{eq: trsf law vertex operators under chiral}).
We also write
$
\langle
(\cdots)
\rangle^{\vphantom{*}}_{\mathcal{L}^{\vphantom{*}}_1,g^{\vphantom{*}}_M,0}
$
for the expectation value with 
$\mathcal{L}$ in the partition function
(\ref{eq: reorganised def bosonic int field theory N>1})
replaced by $\mathcal{L}^{\vphantom{*}}_1$ with $g^{\vphantom{*}}_A=0$.
We then define the perturbative expansion of the expectation value
$
\left
\langle
\mathcal{G}
[
\phi^{\vphantom{*}}_a,
\phi^{\prime}_a,
\chi^{* }_a,\chi^{\vphantom{*}}_a
]
\right
\rangle^{\vphantom{*}}_{\mathcal{L},g^{\vphantom{*}}_M,0}
$
around the ``unperturbed theory''
$\mathcal{L}^{\vphantom{*}}_1$ with $g^{\vphantom{*}}_A=0$
by
\begin{eqnarray}
&&
\left\langle
\vphantom{e^{A^A}}
\mathcal{G}
[
\phi^{\vphantom{*}}_a,
\phi^{\prime}_a,
\chi^{* }_a,
\chi^{\vphantom{*}}_a
]
\right\rangle_{\mathcal{L},g^{\vphantom{*}}_M,0}=
\sum_{n=0}^{\infty}
\frac{(-1)^n}{n!}\,
\mathcal{G}^{(n)}_{\mathcal{L}^{\vphantom{*}}_1,g^{\vphantom{*}}_M,0},
\nonumber\\
&&
\mathcal{G}^{(n)}_{\mathcal{L}^{\vphantom{*}}_1,g^{\vphantom{*}}_M,0}=
\int
\frac{d^2{\bm{r}^{\vphantom{*}}_1}}{2\pi}
\cdots
\int
\frac{d^2{\bm{r}^{\vphantom{*}}_n}}{2\pi}
\left\langle\vphantom{A^{A^{A}}}
\mathcal{G}
[
\phi^{\vphantom{*}}_a,
\phi^{\prime}_a,
\chi^{* }_a,
\chi^{\vphantom{*}}_a
]
\right.
\nonumber\\
&&
\hphantom{
\mathcal{G}^{(n)}_{g^{\vphantom{*}}_M,0}=
\int
d^2{\bm{r}^{\vphantom{*}}_1}
\cdots
\int
}
\times
\left.\vphantom{A^{A^{A}}}
\mathcal{L}^{\vphantom{*}}_2(\bm{r}_1)
\cdots
\mathcal{L}^{\vphantom{*}}_2(\bm{r}_n)
\right\rangle^{\vphantom{*}}_{\mathcal{L}^{\vphantom{*}}_1,g^{\vphantom{*}}_M,0}.
\nonumber\\
\label{eq: perturbation theory in L2 around L1}
\end{eqnarray}
{}{

\noindent
To go from 
Eq.\ (\ref{eq: expectation value vertex operators under chiral})
to
Eq.\ (\ref{eq: perturbation theory in L2 around L1})
                }
we made use of the fact that the perturbation
$\mathcal{L}^{\vphantom{*}}_2$ does not couple to the center of mass
$\Sigma^{\vphantom{*}}_+$ and is thus invariant under the chiral transformation
(\ref{eq: chiral transf for phi's}).
Nearly conformal invariance is a property of the theory
defined by Lagrangian $\mathcal{L}^{\vphantom{*}}_1$
with $g^{\vphantom{*}}_A=0$
as $\mathcal{L}^{\vphantom{*}}_1$ then reduces to 
two independent copies of the $N=1$ limiting theory
for which nearly conformal invariance was demonstrated in 
Ref.\ \onlinecite{Guruswamy00}.
Before we characterize in more detail
what is meant by nearly conformal invariance, 
it suffices here to say that 
any perturbative contribution to the right-hand side
of
Eq.\ (\ref{eq: expectation value vertex operators under chiral})
is to be evaluated with the following rules.
(i)  All fields obey Wick's theorem.
(ii) Any two-point function between $\Sigma^{\vphantom{*}}_-$
and the relative coordinates entering $\mathcal{L}^{\vphantom{*}}_2$
vanishes. If so, perturbation theory in powers of $\mathcal{L}_2$
does not renormalize the dependence on 
$g^{\vphantom{*}}_A$ of the expectation value of the right-hand side of 
Eq.\ (\ref{eq: perturbation theory in L2 around L1})
obtained to zeroth order in powers of $\mathcal{L}^{\vphantom{*}}_2$.
We have thus proved that
the exact dependence on $g^{\vphantom{*}}_A$
of any functional of the form (\ref{eq: def prod vertex})
is already obtained 
to zeroth order in perturbation theory in powers of 
$\mathcal{L}^{\vphantom{*}}_2$ 
around the unperturbed Lagrangian 
$\mathcal{L}^{\vphantom{*}}_1$
with $g^{\vphantom{*}}_A=0$
once the chiral transformation
(\ref{eq: trsf law vertex operators under chiral})
has been performed.

Under nearly conformal invariance of
the theory defined by the Lagrangian
$\mathcal{L}^{\vphantom{*}}_1$
with $g^{\ }_A=0$,
we understand the following properties
demonstrated in Ref.\ \onlinecite{Guruswamy00}.
\begin{widetext}

(i)
Nonvanishing two-point functions between the fields
$\phi^{\vphantom{*}}_a$,
$\phi^{\prime}_a$
[see Eq.\ (\ref{eq: def fields phi and phi'})],
their duals
\begin{eqnarray}
\widetilde\phi^{\vphantom{*}}_{a}:=
\varphi^{\vphantom{*}}_{a-}-\varphi^{\vphantom{*}}_{a+},
\quad
\widetilde\phi^{\prime}_{a}:=
\varphi^{\prime}_{a-}-\varphi^{\prime}_{a+},
\nonumber\\
&&
\label{eq: def fields dual to phi and phi'}
\end{eqnarray}
and the auxiliary fermions 
$\chi^{*}_a$ 
and 
$\chi^{\vphantom{*}}_a$
are\cite{errorguru}
\begin{subequations}
{}{
\begin{eqnarray}
\langle
\phi^{\vphantom{*}}_{a}(z,\bar z)\phi^{\vphantom{*}}_{b}(0,0)
\rangle_{\mathcal{L},g^{\vphantom{*}}_M,g^{\vphantom{*}}_A}
&=&
{{{}}{\zeta}^{\vphantom{2}}_{\vphantom{+}}}
\left[
-\delta^{\vphantom{*}}_{ab}
-\left(g^{\vphantom{*}}_A{{{}}{\zeta}^{\vphantom{2}}_{\vphantom{+}}}/\pi\right)
\right]
 \ln  \left(\frac{z\bar z}{\mathfrak{a}^2}\right)
-[g^{\vphantom{*}}_M{{{}}{\zeta}^{\vphantom{2}}_{\vphantom{+}}}/(2\pi)]^2
 {{{}}{\zeta}^{2}_{\vphantom{-}}}
 \ln^2\left(\frac{z\bar z}{\mathfrak{a}^2}\right)
+\cdots,
\nonumber\\
\langle
\phi^{\prime}_{a}(z,\bar z)\phi^{\prime}_{b}(0,0)
\rangle_{\mathcal{L},g^{\vphantom{*}}_M,g^{\vphantom{*}}_A}
&=&
{{{}}{\zeta}^{\vphantom{2}}_{\vphantom{+}}}
\left[
+\delta^{\vphantom{*}}_{ab}
-\left(g^{\vphantom{*}}_A{{{}}{\zeta}^{\vphantom{2}}_{\vphantom{+}}}/\pi\right)
\right]
 \ln  \left(\frac{z\bar z}{\mathfrak{a}^2}\right)
-[g^{\vphantom{*}}_M{{{}}{\zeta}^{\vphantom{2}}_{\vphantom{+}}}/(2\pi)]^2
{{{}}{\zeta}^{2}_{\vphantom{-}}}
 \ln^2\left(\frac{z\bar z}{\mathfrak{a}^2}\right)
+\cdots,
\nonumber\\
\langle
\phi^{\vphantom{*}}_{a}(z,\bar z)\phi^{\prime}_{b}(0,0)
\rangle_{\mathcal{L},g^{\vphantom{*}}_M,g^{\vphantom{*}}_A}
&=&
\hphantom{\zeta\big[+\delta_{ab}}
-
\left(g^{\vphantom{*}}_A{{{}}{\zeta}^{2}_{\vphantom{+}}}/\pi\right)
 \ln  \left(\frac{z\bar z}{\mathfrak{a}^2}\right)
-[g^{\vphantom{*}}_M{{{}}{\zeta}
^{\vphantom{2}}_{\vphantom{+}}}/(2\pi)]^2
{{{}}{\zeta}^{2}_{\vphantom{-}}}
 \ln^2\left(\frac{z\bar z}{\mathfrak{a}^2}\right)
+\cdots,
\label{eq: final 2-point for phi phi' N>1}
\\&&\nonumber\\
\langle
\widetilde\phi^{\hphantom{\prime}}_{a}(z,\bar z)
\widetilde\phi^{\hphantom{\prime}}_{b}(0,0)
\rangle_{\mathcal{L},g^{\vphantom{*}}_M,g^{\vphantom{*}}_A}
&=&
{{{}}{\zeta}^{\vphantom{2}}_{\vphantom{+}}}
\left[
-\delta^{\vphantom{*}}_{ab}
+\left(g^{\vphantom{*}}_A{{{}}{\zeta}^{\vphantom{2}}_{\vphantom{+}}}/\pi\right)
\right]
 \ln  \left(\frac{z\bar z}{\mathfrak{a}^2}\right)
+[g^{\vphantom{*}}_M{{{}}{\zeta}^{\vphantom{2}}_{\vphantom{+}}}/(2\pi)]^2
{{{}}{\zeta}^{2}_{\vphantom{-}}}
 \ln^2\left(\frac{z\bar z}{\mathfrak{a}^2}\right)
+\cdots,
\nonumber\\
\langle
\widetilde\phi^{         {\prime}}_{a}(z,\bar z)
\widetilde\phi^{         {\prime}}_{b}(0,0)
\rangle_{\mathcal{L},g^{\vphantom{*}}_M,g^{\vphantom{*}}_A}
&=&
{{{}}{\zeta}^{\vphantom{2}}_{\vphantom{+}}}
\left[
+\delta^{\vphantom{*}}_{ab}
+\left(g^{\vphantom{*}}_A{{{}}{\zeta}^{\vphantom{2}}_{\vphantom{+}}}/\pi\right)
\right]
 \ln  \left(\frac{z\bar z}{\mathfrak{a}^2}\right)
+[g^{\vphantom{*}}_M{{{}}{\zeta}^{\vphantom{2}}_{\vphantom{+}}}/(2\pi)]^2
{{{}}{\zeta}^{2}_{\vphantom{-}}}
 \ln^2\left(\frac{z\bar z}{\mathfrak{a}^2}\right)
+\cdots,
\nonumber\\
\langle
\widetilde\phi^{\hphantom{\prime}}_a(z,\bar z)
\widetilde\phi^{         {\prime}}_b(0,0)
\rangle_{\mathcal{L},g^{\vphantom{*}}_M,g^{\vphantom{*}}_A}
&=&
\hphantom{\zeta\big[+\delta_{ab}}
+
\left(g^{\vphantom{*}}_A{{{}}{\zeta}^{2}_{\vphantom{+}}}/\pi\right)
 \ln  \left(\frac{z\bar z}{\mathfrak{a}^2}\right)
+[g^{\vphantom{*}}_M{{{}}{\zeta}
^{\vphantom{2}}_{\vphantom{+}}}/(2\pi)]^2
{{{}}{\zeta}^{2}_{\vphantom{-}}}
 \ln^2\left(\frac{z\bar z}{\mathfrak{a}^2}\right)
+\cdots,
\label{eq: final 2-point for dual phi phi' N>1}
\\&&\nonumber\\
\langle
\chi^{* }_a(z,\bar z)
\chi^{\vphantom{*}}_b(0,0)
\rangle_{\mathcal{L}^{\vphantom{*}}_1,g^{\vphantom{*}}_M,0}&=&
-
\langle
\chi^{\vphantom{*}}_a(z,\bar z)
\chi^{* }_b(0,0)
\rangle_{\mathcal{L}^{\vphantom{*}}_1,g^{\vphantom{*}}_M,0}=
-\delta^{\vphantom{*}}_{ab}\,{{{}}{\zeta}^{\vphantom{2}}_{\vphantom{-}}} 
 \ln\left(\frac{z\bar z}{\mathfrak{a}^2}
\right)
\label{eq: Ansatz chi* chi 2 pt fct N>1}
\end{eqnarray}
                 }
\label{eq: nonvanishing 2point fct if nearly CFT}
\end{subequations}

\noindent ($a=1,2$),
respectively. There are nonvanishing correlation functions
between $\phi$'s and their dual $\widetilde\phi$'s but we will not
need them. 
Here, the ellipses means that short-distance contributions such as $\delta$
functions have been omitted.
Observe that 
Eqs.\ (\ref{eq: final 2-point for phi phi' N>1})
and   (\ref{eq: final 2-point for dual phi phi' N>1})
apply not only to $\mathcal{L}^{\vphantom{*}}_1$
with $g^{\ }_A=0$
but also to $\mathcal{L}^{\vphantom{*}}$
with $g^{\ }_A\geq0$. 
However, we stress that
Eqs.\ (\ref{eq: final 2-point for phi phi' N>1})
and   (\ref{eq: final 2-point for dual phi phi' N>1})
were first computed with $g^{\ }_A=0$. The full 
$g^{\ }_A$ 
dependence was then uncovered through the chiral transformation
(\ref{eq: chiral transf for phi's}).
Another important point to keep in mind is that the squared
logarithmic term enters with a negative sign in the two-point
functions for the fields $\phi^{\ }_a$ and $\phi^{\prime}_a$,
while it enters with a \textit{positive} sign
for the dual fields $\widetilde\phi^{\ }_a$ and
$\widetilde\phi^{\prime}_a$.

(ii)
Many-point correlation functions among
$\phi^{\vphantom{*}}_a$,
$\phi^{\prime}_a$,
$\widetilde\phi^{\vphantom{*}}_a$,
$\widetilde\phi^{\prime}_a$,
$\chi^{*}_a$,
$\chi^{\vphantom{*}}_a$
can be reduced to the two-point correlation functions
(\ref{eq: nonvanishing 2point fct if nearly CFT})
with the help of Wick's theorem.
Applicability of Wick's theorem is specific to
$\mathcal{L}^{\vphantom{*}}_1$
with $g^{\ }_A=0$.
\end{widetext}

The $\beta$ functions in
Eqs.\ (\ref{eq: beta functions in GLL})
for 
$g{\vphantom{*}}_A$ 
and 
$g{\vphantom{*}}_M$
follow from 
Eqs.\ (\ref{eq: final 2-point for phi phi' N>1})
or    (\ref{eq: final 2-point for dual phi phi' N>1})
by demanding that these two-point functions be invariant under rescaling
(\ref{eq: rescaling lattice spacing}) 
of the lattice spacing $\mathfrak{a}$.
It is the presence of the squared logarithm terms in
Eqs.\ (\ref{eq: nonvanishing 2point fct if nearly CFT})
that spoils the conformal symmetry of the MDH model, i.e.,
the theory with
$g{\vphantom{*}}_A\geq0$ 
and 
$g{\vphantom{*}}_M=0$.
The damages caused to conformal invariance are minimal, however,
since it remains possible to use Wick's theorem perturbatively
and since the dependence on $g{\vphantom{*}}_A$ in correlation
functions is exact to lowest order in perturbation theory.

\begin{table}[t]
\caption{
\label{table: bosonization rules: diag in flavor Cooper}
Abelian bosonization rules for the eight Cooper terms entering the energy
perturbation (\ref{eq: smeared energy perturbation}).
The fields $\phi_a$, 
$\varphi_{a-}$, and $\varphi_{a+}$
are related by
$\phi_a(z,\bar z)= 
 \varphi_{a-}(z)+\varphi_{a+}(\bar z)$.
The fields $\phi^{\prime}_a$,
$\varphi^{\prime}_{a-}$, and $\varphi^{\prime}_{a+}$
are related by
$\phi^{\prime}_a(z,\bar z)= 
 \varphi^{\prime}_{a-}(z)+\varphi^{\prime}_{a+}(\bar z)$.
The fields $\widetilde\phi_a$, 
$\varphi_{a-}$, and $\varphi_{a+}$
are related by
$\widetilde\phi_a(z,\bar z)= 
 \varphi_{a-}(z)-\varphi_{a+}(\bar z)$.
The fields $\widetilde\phi^{\prime}_a$,
$\varphi^{\prime}_{a-}$, and $\varphi^{\prime}_{a+}$
are related by
$\widetilde\phi^{\prime}_a(z,\bar z)= 
 \varphi^{\prime}_{a-}(z)-\varphi^{\prime}_{a+}(\bar z)$.
The lattice spacing $\mathfrak{a}$ is used as a short-distance cutoff.}
\begin{ruledtabular}
\begin{tabular}{ll}
\hbox{SUSY}
&
\hbox{Bosonized Cooper terms}
\\\hline\\
$
\psi ^{* }_{1-}
\psi ^{* }_{2+}
$
&
$
\mathfrak{a}^{-1}\,
\exp\left\{
-\frac{{i}}{2}
\left[
(
\widetilde\phi^{\ }_{1}
+
\widetilde\phi^{\ }_{2}
)
+
(
          \phi^{\ }_{1}
-
          \phi^{\ }_{2}
)
\right]
    \right\}
$
\\
$
\psi ^{* }_{1+}
\psi ^{* }_{2-}
$
&
$
\mathfrak{a}^{-1}\,
\exp\left\{
-\frac{{i}}{2}
\left[
(
\widetilde\phi^{\ }_{1}
+
\widetilde\phi^{\ }_{2}
)
-
(
          \phi^{\ }_{1}
-
          \phi^{\ }_{2}
)
\right]
    \right\}
$
\\
$
\beta^{* }_{1-}
\beta^{* }_{2+}
$
&
$
\mathfrak{a}\,
\exp\left\{
-\frac{{i}}{2}
\left[
(
\widetilde\phi^{\prime}_{1}
+
\widetilde\phi^{\prime}_{2}
)
+
(
          \phi^{\prime}_{1}
-
          \phi^{\prime}_{2}
)
\right]
    \right\}
({i}    \partial\chi^{\ }_{1})
({i}\bar\partial\chi^{\ }_{2})
$
\\
$
\beta^{* }_{1+}
\beta^{* }_{2-}
$
&
$
\mathfrak{a}\,
\exp\left\{
-\frac{{i}}{2}
\left[
(
\widetilde\phi^{\prime}_{1}
+
\widetilde\phi^{\prime}_{2}
)
-
(
          \phi^{\prime}_{1}
-
          \phi^{\prime}_{2}
)
\right]
    \right\}

({i}\bar\partial\chi^{\ }_{1})
({i}    \partial\chi^{\ }_{2})
$
\\
$
\psi ^{\ }_{2+}
\psi ^{\ }_{1-}
$
&
$
\mathfrak{a}^{-1}\,
\exp\left\{
+\frac{{i}}{2}
\left[
(
\widetilde\phi^{\ }_{1}
+
\widetilde\phi^{\ }_{2}
)
+
(
          \phi^{\ }_{1}
-
          \phi^{\ }_{2}
)
\right]
    \right\}
$
\\
$
\psi ^{\ }_{2-}
\psi ^{\ }_{1+}
$
&
$
\mathfrak{a}^{-1}\,
\exp\left\{
+\frac{{i}}{2}
\left[
(
\widetilde\phi^{\ }_{1}
+
\widetilde\phi^{\ }_{2}
)
-
(
          \phi^{\ }_{1}
-
          \phi^{\ }_{2}
)
\right]
    \right\}
$
\\
$
\beta^{\ }_{2+}
\beta^{\ }_{1-}
$
&
$
\mathfrak{a}\,
\exp\left\{
+\frac{{i}}{2}
\left[
(
\widetilde\phi^{\prime}_{1}
+
\widetilde\phi^{\prime}_{2}
)
+
(
          \phi^{\prime}_{1}
-
          \phi^{\prime}_{2}
)
\right]
    \right\}
({i}\bar\partial\chi^{* }_{2})
({i}    \partial\chi^{* }_{1})
$
\\
$
\beta^{\ }_{2-}
\beta^{\ }_{1+}
$
&
$
\mathfrak{a}\,
\exp\left\{
+\frac{{i}}{2}
\left[
(
\widetilde\phi^{\prime}_{1}
+
\widetilde\phi^{\prime}_{2}
)
-
(
          \phi^{\prime}_{1}
-
          \phi^{\prime}_{2}
)
\right]
    \right\}
({i}    \partial\chi^{* }_{2})
({i}\bar\partial\chi^{* }_{1})
$
\end{tabular}
\end{ruledtabular}
\end{table}

\subsection{Operator content at near criticality}

We have seen in 
Sec.\ \ref{subsec: Operator content at criticality and RG equations}
that the energy perturbation 
(\ref{eq: smeared energy perturbation})
induces local composite operators given by 
Eq.\ (\ref{eq: most general local comp opera})
with the anomalous scaling dimensions
(\ref{eq: scaling dim for most general local comp opera})
in the MDH model. We are going to show that the dependence
on $g^{\ }_M$ of the
anomalous scaling dimensions
of these operators can be calculated perturbatively
in the HWK model
whereas the dependence on $g^{\ }_A$ is, up to the positive
multiplicative factor ${{{}}{\zeta}^{2}_{\vphantom{+}}}$, exact
in Eq.\ (\ref{eq: scaling dim for most general local comp opera}).
Hence, it is imperative to account for the most
relevant of these operators in a consistent way
when writing the Callan-Symanzik equation for the {}{DOS}.

To begin with, we must bosonize the eight Cooper terms entering in the
energy perturbation (\ref{eq: smeared energy perturbation}).
This is done with the help of 
Table \ref{table: bosonization rules: diag in flavor Cooper}
from which we infer that only the four linear combinations
\begin{eqnarray}
\Theta^{\ }_{\pm}:=
\frac{1}{2}
\left[
(
\widetilde\phi^{\ }_{1}
+
\widetilde\phi^{\ }_{2}
)
\pm
(
          \phi^{\ }_{1}
-
          \phi^{\ }_{2}
)
\right]
\end{eqnarray}
and
\begin{eqnarray}
\Theta^{\prime}_{\pm}:=
\frac{1}{2}
\left[
(
\widetilde\phi^{\prime}_{1}
+
\widetilde\phi^{\prime}_{2}
)
\pm
(
          \phi^{\prime}_{1}
-
          \phi^{\prime}_{2}
)
\right]
\end{eqnarray}
enter in the vertex operators representing the Cooper terms.
Observe that the energy perturbation depends on the dual
fields after Abelian bosonization.

Permutation symmetry of the flavor indices at the band center
simplifies considerably 
the evaluation of two-point functions between $\Theta$'s.
In particular, two-point functions between scalar fields and their
dual cancel each other.
Alternatively, this is so because the conformal spin of the energy
perturbation vanishes; i.e., there are as many left as right
movers in any of the monomials making up the energy perturbing operator.

With the help of 
Eqs.\ (\ref{eq: final 2-point for phi phi' N>1})
and   (\ref{eq: final 2-point for dual phi phi' N>1})
we can evaluate exactly all two-point functions listed in the first column of
Table \ref{table: 2-point functions for RG equations fugacities: final}.
For the purpose of computing scaling dimensions, these
exact two-point functions are only the 
zeroth-order contribution to the expectation value of products
of vertex operators in perturbation theory in powers of
$\mathcal{L}^{\ }_2$. They are given in the second column of Table
\ref{table: 2-point functions for RG equations fugacities: final}.
Observe that the dependence on the squared logarithm and hence on
$g^{\ }_A$ (through the chiral transformation)
arises solely from the dual sector of the theory
as the contributions from the nondual sector always cancel.
This is consistent with the fact that the random vector potential
only couples to the dual field. In the present formalism,
this follows from the fact that the 
center of mass drops out in the difference of any of the coordinates
$\phi^{\ }_a$ or $\phi^{\prime}_a$.
The dependence on $g^{\ }_A$ extracted from
Table \ref{table: 2-point functions for RG equations fugacities: final}
is exact for the scaling dimensions of vertex operators in that
perturbation theory only changes the
linear dependence on the logarithm through corrections that depend
solely on $g^{\ }_M$.
Observe that vertex operators corresponding to products of $\beta$'s
follow from two-point functions with prime fields. Hence, the
contribution to the term linear in the logarithm that does not
depend on $g^{\ }_A$ has the ``wrong sign.'' In the absence of a random mass,
this wrong sign is corrected by accounting for the contribution to the
correlation functions from the $\chi$ sector 
[see Eq.\ (\ref{eq: scaling dim for most general local comp opera})
and Appendix \ref{app: sec Bosonization of the ghost system}].
In the presence of a random mass,
the correction to this ``wrong sign'' from the $\chi$ sector is only partial
[see Eq.\ (\ref{eq: scaling dim for most general local comp opera if MDU})].

\begin{table*}[t]
\caption{
\label{table: 2-point functions for RG equations fugacities: final}
Evaluation of all two-point functions governing
scaling dimensions appearing in OPE's of energy 
perturbation with itself.
The ellipses refer to $\delta$ function  contributions
to the two-point functions. Although these two-point functions are
exact, they yield an approximation to the calculation
of scaling dimensions. Nevertheless, the dependence on
$g^{\ }_A$ of scaling dimensions is exact. 
The dependence of linear logarithmic term on 
$g^{\ }_M$ that does not
depend on $g^{\ }_A$ is modified by residual interactions
in $\mathcal{L}^{\ }_2$ when computing scaling dimensions.}
\begin{ruledtabular}
{}{
\begin{tabular}{lc}
$
\left\langle
\Theta^{\ }_{+}(z,\bar z)
\Theta^{\ }_{+}(0,0)
\right\rangle_{\mathcal{L},g^{\ }_M,g^{\ }_A}
$
&
\hbox{
$
{{{}}{\zeta}^{\vphantom{2}}_{\vphantom{+}}}
\left(
-1
+\frac{g^{\ }_A{{{}}{\zeta}^{\vphantom{2}}_{\vphantom{+}}}}{\pi}
\right)
 \ln  \left(\frac{z\bar z}{\mathfrak{a}^2}\right)
+\left(
\frac{g^{\ }_M{{{}}{\zeta}^{         {2}}_{\vphantom{+}}}}{2\pi}\right)^2
 \ln^2\left(\frac{z\bar z}{\mathfrak{a}^2}
\right)
+\cdots
$
}
\\
$
\left\langle
\Theta^{\ }_{-}(z,\bar z)
\Theta^{\ }_{-}(0,0)
\right\rangle_{\mathcal{L},g^{\ }_M,g^{\ }_A}
$
&
as above
\\
$
\left\langle
\Theta^{\ }_{+}(z,\bar z)
\Theta^{\ }_{-}(0,0)
\right\rangle_{\mathcal{L},g^{\ }_M,g^{\ }_A}
$
&
\hbox{
$
\hphantom{\big[-1+}
\frac{g^{\ }_A{{{}}{\zeta}^{         {2}}_{\vphantom{+}}}}{\pi}
 \ln  \left(\frac{z\bar z}{\mathfrak{a}^2}\right)
+\left(
\frac{g^{\ }_M{{{}}{\zeta}^{         {2}}_{\vphantom{+}}}}{2\pi}\right)^2
 \ln^2\left(\frac{z\bar z}{\mathfrak{a}^2}
\right)
+\cdots
$
}
\\
$
\left\langle
\Theta^{\ }_{-}(z,\bar z)
\Theta^{\ }_{+}(0,0)
\right\rangle_{\mathcal{L},g^{\ }_M,g^{\ }_A}
$
&
as above
\\&\\
$
\left\langle
\Theta^{\prime}_{+}(z,\bar z)
\Theta^{\prime}_{+}(0,0)
\right\rangle_{\mathcal{L},g^{\ }_M,g^{\ }_A}
$
&
\hbox{
$
{{{}}{\zeta}^{\vphantom{2}}_{\vphantom{+}}}
\left(
+1
+\frac{g^{\ }_A{{{}}{\zeta}^{\vphantom{2}}_{\vphantom{+}}}}{\pi}
\right)
 \ln  \left(\frac{z\bar z}{\mathfrak{a}^2}\right)
+\left(
\frac{g^{\ }_M{{{}}{\zeta}^{         {2}}_{\vphantom{+}}}}{2\pi}\right)^2
 \ln^2\left(\frac{z\bar z}{\mathfrak{a}^2}
\right)
+\cdots
$
}
\\
$
\left\langle
\Theta^{\prime}_{-}(z,\bar z)
\Theta^{\prime}_{-}(0,0)
\right\rangle_{\mathcal{L},g^{\ }_M,g^{\ }_A}
$
&
as above
\\
$
\left\langle
\Theta^{\prime}_{+}(z,\bar z)
\Theta^{\prime}_{-}(0,0)
\right\rangle_{\mathcal{L},g^{\ }_M,g^{\ }_A}
$
&
\hbox{
$
\hphantom{\big[+1+}
\frac{g^{\ }_A{{{}}{\zeta}^{         {2}}_{\vphantom{+}}}}{\pi}
 \ln  \left(\frac{z\bar z}{\mathfrak{a}^2}\right)
+\left(
\frac{g^{\ }_M{{{}}{\zeta}^{         {2}}_{\vphantom{+}}}}{2\pi}\right)^2
 \ln^2\left(\frac{z\bar z}{\mathfrak{a}^2}
\right)
+\cdots
$
}
\\
$
\left\langle
\Theta^{\prime}_{-}(z,\bar z)
\Theta^{\prime}_{+}(0,0)
\right\rangle_{\mathcal{L},g^{\ }_M,g^{\ }_A}
$
&
as above
\\&\\
$
\left\langle
\Theta^{\prime}_{+}(z,\bar z)
\Theta^{\     }_{+}(0,0)
\right\rangle_{\mathcal{L},g^{\ }_M,g^{\ }_A}
$
&
\hbox{
$
\hphantom{\big[+1+}
\frac{g^{\ }_A{{{}}{\zeta}^{         {2}}_{\vphantom{+}}}}{\pi}
 \ln  \left(\frac{z\bar z}{\mathfrak{a}^2}\right)
+\left(
\frac{g^{\ }_M{{{}}{\zeta}^{         {2}}_{\vphantom{+}}}}{2\pi}\right)^2
 \ln^2\left(\frac{z\bar z}{\mathfrak{a}^2}
\right)
+\cdots
$
}
\\
$
\left\langle
\Theta^{\prime}_{-}(z,\bar z)
\Theta^{\     }_{-}(0,0)
\right\rangle_{\mathcal{L},g^{\ }_M,g^{\ }_A}
$
&
as above
\\
$
\left\langle
\Theta^{\prime}_{+}(z,\bar z)
\Theta^{\     }_{-}(0,0)
\right\rangle_{\mathcal{L},g^{\ }_M,g^{\ }_A}
$
&
as above
\\
$
\left\langle
\Theta^{\prime}_{-}(z,\bar z)
\Theta^{\     }_{+}(0,0)
\right\rangle_{\mathcal{L},g^{\ }_M,g^{\ }_A}
$
&
as above
\end{tabular}
                 }
\end{ruledtabular}
\end{table*}

\begin{widetext}
The central result of this section is the evaluation of the 
anomalous scaling dimensions 
$x^{\vphantom{*}}_{\boldsymbol{m},\boldsymbol{n}}$
for the local composite operators
defined by 
Eq.\ (\ref{eq: most general local comp opera}).
To zeroth order in the perturbation theory
defined in Sec. \ref{subsec: Nearly conformal field theory}, 
they are given by
{}{
\begin{eqnarray}
x^{\vphantom{*}}_{\boldsymbol{m},\boldsymbol{n}}&=&
\left[
(m^{\vphantom{*}}_{12+} - m^{\vphantom{*}}_{12-})^2
+
(m^{\vphantom{*}}_{21+} - m^{\vphantom{*}}_{21-})^2
-
(n^{\vphantom{*}}_{12+}-n^{\vphantom{*}}_{12-})^2
-
(n^{\vphantom{*}}_{21+}-n^{\vphantom{*}}_{21-})^2
\right]
{{{}}{\zeta}^{\vphantom{2}}_{\vphantom{+}}}\left[1+\mathcal{O}(g^{\ }_M)\right]
\nonumber\\
&&-
\left[
\left(
m^{\vphantom{*}}_{12+}
-
m^{\vphantom{*}}_{12-}
\right)
+
\left(
m^{\vphantom{*}}_{21+}
-
m^{\vphantom{*}}_{21-}
\right)
+
\left(
n^{\vphantom{*}}_{12+}
-
n^{\vphantom{*}}_{12-}
\right)
+
\left(
n^{\vphantom{*}}_{21+}
-
n^{\vphantom{*}}_{21-}
\right)
\right]^2
\frac{g^{\vphantom{*}}_A}{\pi}{{{}}{\zeta}^{2}_{\vphantom{+}}}
\nonumber\\
&&
+
(n^{\vphantom{*}}_{12+}-n^{\vphantom{*}}_{12-})^2
+
(n^{\vphantom{*}}_{21+}-n^{\vphantom{*}}_{21-})^2
+
|n^{\vphantom{*}}_{12+}-n^{\vphantom{*}}_{12-}|
+
|n^{\vphantom{*}}_{21+}-n^{\vphantom{*}}_{21-}|
+\mathcal{O}(g^{\ }_M)
\label{eq: scaling dim for most general local comp opera if MDU}
\end{eqnarray}
               }
\end{widetext}
\noindent
as the expectation value of any product of vertex operators
can be replaced by the exponential of the properly weighted sum
of all 12 nonvanishing two-point functions listed in
Table 
\ref{table: 2-point functions for RG equations fugacities: final}.
Anomalous scaling dimensions are then extracted from
the coefficient to the term linear in 
$\ln(\bar zz/\mathfrak{a^2})$.
One should not forget the contribution to the
anomalous scaling dimension
coming from the $\chi$'s 
which is responsible for the 
{}{
third line
in
Eq.\ (\ref{eq: scaling dim for most general local comp opera if MDU}).
                }
What makes
Eq.\ (\ref{eq: scaling dim for most general local comp opera if MDU})
remarkable is that even if 
$\mathcal{O}^{\ }_{\bm{m},\bm{n}}$
is not relevant at $g^{\ }_M=0$, 
switching on a mass perturbation renders
$\mathcal{O}^{\ }_{\bm{m},\bm{n}}$
relevant at sufficiently large length scales since
$g^{\ }_A$ is marginally relevant 
{{}}{(in the sense that it grows logarithmically with length)}
whereas $g^{\ }_M$ is exactly marginal.
Moreover, higher-order corrections from perturbation theory do not invalidate
this argument. The strategy to deal with this difficulty was described in
Sec.\ \ref{sec: DOS for the MDH model}.
We are going to generalize the RG analysis
of 
Sec.\ \ref{sec: DOS for the MDH model}
to account for the relevance of $g^{\ }_A$.
As before we select the operators (\ref{eq: selection most relevant operators})
since they are the most relevant operators 
for given $N$ in Eq.\ (\ref{eq: def N in terms m's and n's})
although it must be noted that their scaling dimensions are not
anymore degenerate for finite $g^{\vphantom{*}}_M$.
The lifting of the degeneracy 
(\ref{eq: most important scaling dimension within perturbation theory})
is negligible, however, since it is of order 
$g^{\vphantom{*}}_M/g^{\vphantom{*}}_A$
whereby $g^{\vphantom{*}}_M$ does not flow but 
$g^{\vphantom{*}}_A$ flows to infinity in the infrared limit.

\subsection{Callan-Symanzik equation for {}{DOS}}

The Callan-Symanzik equation obeyed by the {}{DOS} in the 
HWK model is given by 
\begin{subequations}
\begin{eqnarray}
&&
0\!=\!
\left(
\sum_{N=1}^{\infty}
\beta^{\vphantom{*}}_{Y^{\vphantom{*}}_{N}}
\frac{\partial}{\partial Y^{\vphantom{*}}_{N}}
\!+\!
\beta^{\vphantom{*}}_{g^{\vphantom{*}}_{M}}
\frac{\partial}{\partial g^{\vphantom{*}}_{M}}
\!+\!
\beta^{\vphantom{*}}_{g^{\vphantom{*}}_{A}}
\frac{\partial}{\partial g^{\vphantom{*}}_{A}}
\!-\!
x^{\vphantom{*}}_1
\right)
{}{\nu(\varepsilon)},
\nonumber\\
&&
\end{eqnarray}
where the $\beta$ functions for the fugacities are
\begin{eqnarray}
&&
\beta^{\vphantom{*}}_{Y^{\vphantom{*}}_N}=
(2-x^{\vphantom{*}}_N) Y^{\vphantom{*}}_N
+
\pi
\sum_{N^{\prime}=1}^{N-1}
\left(\begin{array}{c}
N^{\vphantom{*}}
\\
N^{\prime}
\end{array}\right)
Y^{\vphantom{*}}_{N^{\prime}}
Y^{\vphantom{*}}_{N^{\vphantom{*}}-N^{\prime}},
\nonumber\\
&&
x^{\vphantom{*}}_N=
{{{}}{\zeta}^{\vphantom{2}}_{\vphantom{+}}}[1+\mathcal{O}(g^{\vphantom{2}}_{M})]N
-
\left[
\frac{g^{\vphantom{*}}_A}{\pi}{{{}}{\zeta}^{2}_{\vphantom{+}}}
+
\mathcal{O}(g^{\vphantom{2}}_{M})
\right]N^2,
\nonumber\\
&&
Y^{\vphantom{*}}_{N}(l=0)=
\frac{
\varepsilon-i\eta
     }
     {
\mathfrak{a}^{x^{\vphantom{*}}_{1}-2}
     }
\delta^{\vphantom{*}}_{1,N},
\label{eq: beta fcts for fug in HWK model}
\end{eqnarray}
the $\beta$ function for the variance of the random mass vanishes,
and the $\beta$ function for the variance of the random vector potential is
\begin{eqnarray}
&&
{}{
\beta^{\vphantom{*}}_{g^{\vphantom{*}}_A}=
\frac{
(g^{\vphantom{*}}_M{{{}}{\zeta}^{\vphantom{2}}_{\vphantom{-}}})^2
     }
     {
2\pi^2
     }.
                }
\label{eq: beta functions gA}
\end{eqnarray}
\label{eq: final CZ eq for HWK model}
\end{subequations}

\noindent
[Remember that
$\beta^{\vphantom{*}}_{g^{\vphantom{*}}_M}$
and
$\beta^{\vphantom{*}}_{g^{\vphantom{*}}_A}$
follow from demanding that the two-point functions 
Eqs.\ (\ref{eq: final 2-point for phi phi' N>1})
or    (\ref{eq: final 2-point for dual phi phi' N>1})
be invariant under rescaling
(\ref{eq: rescaling lattice spacing}) 
of {}{the lattice spacing $\mathfrak{a}$}.]
The function ${{{}}{\zeta}^{\vphantom{2}}_{\vphantom{-}}}$
is given in Eq.\ (\ref{eq: beta functions in GLL}).
We made the same approximations to reach
Eq.\ (\ref{eq: final CZ eq for HWK model})
as we did to obtain
Eq.\ (\ref{eq: final CZ eq for MDH model});
i.e., we ignored renormalization effects induced on
$g^{\vphantom{*}}_M$
and
$g^{\vphantom{*}}_A$
by the fugacities and we ignored annihilation processes when
deriving the RG equations obeyed by the fugacities.
Furthermore, we work to lowest order in $g^{\vphantom{*}}_M$
as $g^{\vphantom{*}}_M$ does not flow whereas 
$g^{\vphantom{*}}_A$ flows to infinity in the infrared limit.
The infinite set of RG equations 
(\ref{eq: beta fcts for fug in HWK model})
obeyed by the fugacities can be recast
as a second-order nonlinear partial differential equation of the KPP type
\begin{subequations}
\begin{eqnarray}
&&
\partial^{\vphantom{n}}_l \widetilde G=
\left(
{{{}}{\zeta}^{\vphantom{2}}_{\vphantom{+}}}
\partial^{\vphantom{n}}_y
\!+\!
\frac{g^{\vphantom{n}}_A{{{}}{\zeta}^{2}_{\vphantom{+}}}}{\pi}
\partial^{          2}_{y}
\right)\widetilde G
+2\widetilde G(\widetilde G-1),
\nonumber\\&&\\
&&
\widetilde G(y,l)=
1
+
\frac{\pi}{2}
\sum_{n=1}^{\infty}
\frac{(-1)^n e^{-ny}}{n!}Y^{\vphantom{n}}_n(l),
\nonumber\\&&\\
&&
\widetilde G(y,0)=
\exp
\left(
-
\frac{\pi}{2}
\frac{
\varepsilon-i\eta
     }
     {
\mathfrak{a}^{x^{\vphantom{*}}_{1}-2}
     }
e^{-y}
\right).
\nonumber\\&&
\label{eq: Initial condition for widetilde G: HWK}
\end{eqnarray}
\label{eq: would be KPP for widetilde G: HWK}
\end{subequations}

\noindent
The notable difference with Eq.\ (\ref{eq: RG for widetilde G})
is the scale dependence of the diffusion term 
$g^{\vphantom{*}}_A(l){{{}}{\zeta}^{2}_{\vphantom{+}}}/\pi$.

We tentatively define the dynamical exponent $z^{\vphantom{*}}_A$ 
for the HWK model by
\begin{subequations}
\begin{eqnarray}
z^{\vphantom{*}}_{A}&:=&
\left\{
\begin{array}{ll}
2
-
{{{}}{\zeta}^{\vphantom{2}}_{\vphantom{+}}}
+
\frac{g^{\vphantom{*}}_{A}}{\pi}{{{}}{\zeta}^{2}_{\vphantom{+}}}
+
\mathcal{O}(g^{\vphantom{*}}_M),
&
\hbox{ for }
g^{\vphantom{*}}_{A}{{{}}{\zeta}^{2}_{\vphantom{+}}}<2\pi,
\\
&\\
4\sqrt{\frac{g^{\vphantom{*}}_{A}}{2\pi}}{{{}}{\zeta}^{\vphantom{2}}_{\vphantom{+}}}
-
{{{}}{\zeta}^{\vphantom{2}}_{\vphantom{+}}}
-
\mathcal{O}(g^{\vphantom{*}}_M),
&
\hbox{ for }
g^{\vphantom{*}}_{A}{{{}}{\zeta}^{2}_{\vphantom{+}}}\geq 2\pi.
\end{array}
\right.
\nonumber\\
&&
\label{eq: dynamical exponent of HWK}
\end{eqnarray}
The dynamical exponent $z^{\vphantom{*}}_A$ in 
Eq.\ (\ref{eq: dynamical exponent of HWK})
would be the velocity of the wave front of the Callan-Symanzik
equation
(\ref{eq: would be KPP for widetilde G: HWK})
if the diffusion constant was scale independent.
However, the dependence of the diffusion constant 
$\propto g_{A}(l)$ 
on the rescaling parameter $l$ is linear
whereas that of
$\widetilde{G}(y,l)$ 
is exponential.
We can thus make an adiabatic approximation by which 
the Callan-Symanzik equation for the 
{}{DOS}
(\ref{eq: Callan-Symanzik MDH typical DOS})
becomes
\begin{eqnarray}
&&
0=
\left[
z^{\vphantom{*}}_A
\varepsilon
\frac{\partial}{\partial\varepsilon}
+
\beta^{\vphantom{*}}_{g^{\vphantom{*}}_A}
\frac{\partial}{\partial g^{\vphantom{*}}_A}
-
\left(
2-z^{\vphantom{*}}_A
\right)
\right]
{}{\nu(\varepsilon)}.
\nonumber\\
&&
\label{eq: Callan-Symanzik HWK typical DOS}
\end{eqnarray}
 \label{eq: final CZ for HWK}
\end{subequations}

\noindent
The Callan-Symanzik equation 
(\ref{eq: final CZ for HWK})
is the central result of our paper.
It differs from the Callan-Symanzik equation
of Guruswamy \textit{et al.}\ in Ref.\ \onlinecite{Guruswamy00}
through the freezing transition of the dynamical exponent
in Eq.\ (\ref{eq: dynamical exponent of HWK}).

\subsection{{}{DOS}}

The formal solution to Eq.\ (\ref{eq: final CZ for HWK})
is
\begin{eqnarray}
&&
{}{\nu(l)}=
{}{\nu(0)}\,
\exp
\left[
+\int_0^l 
dl 
\left(
2-z^{\vphantom{*}}_A
\right)
\right],
\nonumber\\
&&
dl= 
\frac{2\pi^2}{(g^{\vphantom{*}}_M{{{}}{\zeta}^{\vphantom{2}}_{\vphantom{-}}})^2}
d g^{\vphantom{*}}_A,
\nonumber\\
&&
z^{\vphantom{*}}_Adl= 
\frac{d\varepsilon}{\varepsilon}.
\end{eqnarray}
With the notation 
$\varepsilon^{\vphantom{*}}_{\mathrm{ren}}\equiv\varepsilon(l)$,
$\varepsilon                              \equiv\varepsilon(0)$,
we find
\begin{subequations}
\begin{eqnarray}
\frac{
{}{\nu(\varepsilon)}
     }
     {
{}{\nu(\varepsilon^{\vphantom{*}}_{\mathrm{ren}})}
     }=
\left(
\frac{
\varepsilon^{\vphantom{*}}_{\mathrm{ren}}
     }
     {
\varepsilon
     }
\right)
\times
e^{
-
\frac{4\pi^2}{(g^{\vphantom{*}}_M{{{}}{\zeta}^{\vphantom{2}}_{\vphantom{-}}})^2}
\left[
g^{\vphantom{*}}_A(\varepsilon^{\vphantom{*}}_{\mathrm{ren}})
-
g^{\vphantom{*}}_A(\varepsilon)
\right]
  },
\nonumber\\
&&
\label{eq: formal solution for DOS in HWK}
\end{eqnarray}
where 
$
g^{\vphantom{*}}_A(\varepsilon^{\vphantom{*}}_{\mathrm{ren}})
-
g^{\vphantom{*}}_A(\varepsilon)
$
is the solution to
\begin{eqnarray}
\ln
\left(
\frac{
\varepsilon^{\vphantom{*}}_{\mathrm{ren}}
     }
     {
\varepsilon
     }
\right)
=
\frac{
2\pi^2
     }
     {
{}{
(g^{\vphantom{*}}_M{{{}}{\zeta}^{\vphantom{2}}_{\vphantom{-}}})^2
                }
     }
\int_{g^{\vphantom{*}}_A(\varepsilon)}
    ^{g^{\vphantom{*}}_A(\varepsilon^{\vphantom{*}}_{\mathrm{ren}})}
d g^{\vphantom{*}}_A\, z^{\vphantom{*}}_A.
\nonumber\\
&&
\label{eq: gA integral over za}
\end{eqnarray}
\end{subequations}

\noindent
The renormalized energy scale
$\varepsilon^{\vphantom{*}}_{\mathrm{ren}}\equiv\varepsilon(l)$
is the energy scale at which we cannot neglect
renormalization effects from the energy perturbation on the
variances of the random potentials anymore. 
Beyond the energy scale 
$\varepsilon^{\vphantom{*}}_{\mathrm{ren}}$,
the RG analysis encoded by 
the Callan-Symanzik equation
(\ref{eq: final CZ for HWK})
breaks down. We take
$\varepsilon^{\vphantom{*}}_{\mathrm{ren}}$ 
as given, but we allow $l$ to be tuned, or, equivalently, 
we take the bare energy
$\varepsilon\equiv\varepsilon(0)$
to be tunable. How large the rescaling scale $l$ is thus
dictates how close the bare energy scale
$\varepsilon$
is to the band center. 
The infrared limit $l\to\infty$
gives the 
{}{DOS}
arbitrarily close to the band center.

Equation (\ref{eq: formal solution for DOS in HWK})
implies that the
{}{DOS}
is characterized by two regimes
each of which can itself be divided into two subregimes.
When
\begin{eqnarray}
&&
\frac{g^{}_A(l)-g^{}_A(0)}{g^{}_A(0)}\ll1,
\label{eq: condition to recover MDH TDOS}
\end{eqnarray}
the scale dependence of $g^{\vphantom{*}}_A$
can be neglected in the Callan-Symanzik equation
(\ref{eq: Callan-Symanzik HWK typical DOS}). 
We then recover the 
{}{DOS}
for the MDH model (\ref{eq: final typical DOS for MDH})
with $z{\vphantom{*}}_A$ given by
Eq.\ (\ref{eq: dynamical exponent of HWK}),
whereby the freezing transition can be observed as a function of 
the bare energy $\varepsilon$ depending on whether
$g^{\vphantom{*}}_A(\varepsilon){{{}}{\zeta}^{2}_{\vphantom{+}}}>2\pi$
or
$g^{\vphantom{*}}_A(\varepsilon){{{}}{\zeta}^{2}_{\vphantom{+}}}<2\pi$.
When
\begin{eqnarray}
\frac{g^{}_A(l)-g^{}_A(0)}{g^{}_A(0)}\gtrsim1,
\end{eqnarray}
the scale dependence of $g^{\vphantom{*}}_A$
in the Callan-Symanzik equation 
(\ref{eq: Callan-Symanzik HWK typical DOS})
is dominant. In the infrared limit $l\to\infty$,
i.e., arbitrarily close to the band center,
the integral on the right-hand side of 
Eq.\ (\ref{eq: gA integral over za})
is always dominated by the contribution
\begin{eqnarray}
&&
\int_{2\pi/{{{}}{\zeta}^{2}_{\vphantom{+}}}}
    ^{g^{\vphantom{*}}_A(\varepsilon^{\vphantom{*}}_{\mathrm{ren}})}
d g^{\vphantom{*}}_A\, z^{\vphantom{*}}_A
\approx
\\
&&\hphantom{AAAA}
\left[
\frac{8}{3}
\sqrt{
\frac{
g^{\vphantom{*}}_{A}(\varepsilon^{\vphantom{*}}_{\mathrm{ren}})
     }
     {
2\pi}
     }
{{{}}{\zeta}^{\vphantom{2}}_{\vphantom{+}}}
-
{{{}}{\zeta}^{\vphantom{2}}_{\vphantom{+}}}
-
\mathcal{O}(g^{\vphantom{*}}_M)
\right]
g^{\vphantom{*}}_{A}(\varepsilon^{\vphantom{*}}_{\mathrm{ren}}).
\nonumber
\end{eqnarray}
After solving
Eq.\ (\ref{eq: gA integral over za})
for the leading dependence of 
$g^{\vphantom{*}}_{A}$ 
on
$\varepsilon^{\vphantom{*}}_{\mathrm{ren}}$,
this gives the 
{}{DOS}
\begin{eqnarray}
\frac{
{}{\nu(\varepsilon)}
     }
     {
{}{\nu(\varepsilon^{\vphantom{*}}_{\mathrm{ren}})}
     }\sim
\left(
\frac{
\varepsilon^{\vphantom{*}}_{\mathrm{ren}}
     }
     {
\varepsilon
     }
\right)
\exp
\left[
-c\left|
\ln
\left(
\frac{\varepsilon^{\vphantom{*}}_{\mathrm{ren}}}{\varepsilon}
\right)
\right|^{2/3}
\right],
\nonumber\\
&&
\label{eq: final asymptotic solution for DOS in HWK}
\end{eqnarray}
where $c$ is some nonuniversal positive constant.
Observation of the Gade singularity
\begin{eqnarray}
\frac{
{}{\nu(\varepsilon)}
     }
     {
{}{\nu(\varepsilon^{\vphantom{*}}_{\mathrm{ren}})}
     }\sim
\left(
\frac{
\varepsilon^{\vphantom{*}}_{\mathrm{ren}}
     }
     {
\varepsilon
     }
\right)
\exp
\left[
-c\left|
\ln
\left(
\frac{\varepsilon^{\vphantom{*}}_{\mathrm{ren}}}{\varepsilon}
\right)
\right|^{1/2}
\right],
\nonumber\\
&&
\label{eq: Gade crossover solution for DOS in HWK}
\end{eqnarray}
demands fine-tuning of bare and renormalized 
$g^{\vphantom{*}}_A$'s so as to avoid the freezing transition
in Eq.\ (\ref{eq: dynamical exponent of HWK}).
At best, the Gade singular DOS can be observed 
as a crossover in a finite window of energy.

\section{Conclusions}
\label{sec: Conclusions}

Two recent theoretical works have addressed the issue of the
distribution of the DOS 
in the HWK model.
Guruswamy \textit{et al.}\ in Ref.\ \onlinecite{Guruswamy00}
considered a narrow distribution of disorder
at the microscopic level for which the use of field-theoretical 
methods at longer length scales is justified.
They computed the {}{DOS within a SUSY} field theory 
and found that it displays the Gade singularity 
$\varepsilon^{-1}\exp\left(-c|\ln \varepsilon|^{1/2}\right)$
in the neighborhood of  the band center.
{{}}{Motrunich \textit{et al.}\ in Ref.\ \onlinecite{Motrunich02},
{{}}{bypassing a purely field-theoretical approach},
came to the conclusion that the 
{}{DOS}
is given by
$\varepsilon^{-1}\exp\left(-c|\ln \varepsilon|^{2/3}\right)$
instead of the Gade singularity.
{}{
Our field-theoretical analysis agrees with the prediction
of Motrunich \textit{et al}. 
               }

We find that the RG analysis performed on
the {}{DOS} 
by Guruswamy \textit{et al.}\ is not consistent as it
overlooked the existence of infinitely many operators with
negative anomalous scaling dimensions. The physical interpretation of these 
operators is that the LDOS
is a very broadly distributed random
variable close to the band center. 
The situation here shares many similarities with the scaling with system size
of the distribution of the conductance in 2D mesoscopic samples
which is known to broaden with increasing system size
(i.e., by moving away from the diffusive regime).\cite{Altshuler86} 
A parallel with quasi-one-dimensional wires is particularly instructive.
In quasi-one-dimensions, it is known that the infinite set of
scaling equations for the moments of the conductance can be traded for
a Fokker-Planck equation obeyed by the underlying distribution 
of the conductance.\cite{Dorokhov82}
The advantage of the Fokker-Planck equation over the scaling equations
for the moments is that the Fokker-Planck equation can be solved 
in the localized regime (i.e., when the length of the sample diverges).
It is then possible to show that the typical conductance is self-averaging
(at least in the standard universality classes of Anderson localization) 
in the localized regime.\cite{Beenakker97} 
Similarly, all moments of the LDOS are here strongly coupled
through infinitely many RG equations encoding the dependence on energy
arbitrarily close to the band center. 
The Gade singularity results
from an inconsistent truncation of this infinite set of coupled equations 
to a finite subset.
{}{
As with the quasi-one-dimensional conductance, 
it is possible to trade the infinite set of scaling equations for 
the {}{DOS} in favor
of a second-order partial differential equation obeyed by the
underlying probability distribution for the normalization of zero modes
from which one can extract the dependence on energy of the
DOS. As with the quasi-one-dimensional typical conductance
in the standard universality classes of Anderson localization,
the DOS is well behaved statistically;
i.e., it is self-averaging in the infrared limit.
                }

A classification of random Hamiltonians with short-range-correlated 
disorder in terms of intrinsic symmetries
in conjunction with scaling concepts has emerged,
since its inception in the late 1970's,
as perhaps one of the deepest insight into the physics of
Anderson localization.
The HWK model is invariant under time reversal and spin rotation. 
Those are the standard symmetries in Anderson localization. 
The HWK model also preserves a bipartite
lattice symmetry, making it a member of the chiral orthogonal
universality class. The HWK model is nevertheless special since
the disorder is parametrized (in a way consistent with rescaling
of the lattice spacing) by only two independent variances
as opposed to four for a generic member of the 
chiral orthogonal universality class.
Since the works of Gade and Wegner, it is believed that
the {}{DOS} in the chiral orthogonal universality class
displays the Gade singularity. If universality is indeed the
only ingredient that determines the scaling exponent
$\upsilon$ and $\kappa$
for the singular density
\begin{eqnarray}
|\varepsilon|^{-\upsilon}
\exp\left(-c|\ln\varepsilon|^{\kappa}\right)
\end{eqnarray}
in the chiral orthogonal universality class,
we must conclude that the RG analysis made by Gade
in Ref.\ \onlinecite{Gade93} 
suffers from the same deficiencies as the RG analysis
of Guruswamy \textit{et al}. 

This possibility is perhaps not so surprising if we take
the existence of infinitely many operators with 
negative anomalous scaling dimensions as the ``smoking gun'' for
broadly distributed random variables. 
There have been indications since the early 1980's
that replicated 
NLSM might support infinitely many operators with negative
anomalous scaling dimensions.\cite{Wegner80}
A perhaps stronger clue comes from the
interplay between multifractality of zero modes
and the broad distribution of the LDOS in the HWK model.
As we have shown explicitly in this paper, essentially
the same spectrum of
negative scaling dimensions controls the multifractal spectrum
of zero modes in the MDH model
and the Callan-Symanzik equation for the {}{DOS}
near the band center in the HWK model.
Fal'ko and Efetov have searched for and found 
in the diffusive regime of 2D disordered metals 
the signature of multifractal states.\cite{Falko95} 
There are states, dubbed prelocalized states, that are
characterized by very large amplitude fluctuations whose statistics
are well described (in a finite window of length scales)
by a log-normal distribution, i.e.,
by the multifractal spectrum of zero modes for a Dirac spinor subjected 
to a random vector potential.\cite{Kogan96,Falko95} 
It could very well be that the gl($1|1$) current algebra at the heart
of the freezing transition in the multifractal spectrum of zero modes
in the MDH model and for the 
{}{DOS}
of the HWK model, although
broken by all the additional degrees of freedom
encoded by NLSM in the standard or chiral universality classes,
is sufficiently robust to reveal itself in those latter models
in connection with prelocalized states and broadly distributed
LDOS, respectively. To put it differently, the critical statistics of
wave functions might still be log normal even though the wave functions
have a much more complicated structure than
a single exponential of a field with logarithmic spatial correlations.
To add credit to this speculation, we know of one example where
critical modes have a much more complex structure than that of a
single exponential of a two-dimensional scalar field 
(a Gaussian random surface in the terminology of 
Motrunich \textit{et al}.)\ 
although their multifractal spectrum is still of the 
Gaussian random surface type: $N$ flavors of Dirac fermions
subjected to a random SU($N$) gauge potential.\cite{Mudry96,Caux98}
An interesting direction that might be worth exploring is
the construction of variational methods 
that would generalize to the NLSM
the one devised by Horovitz and Le Doussal in Ref.\ \onlinecite{Horovitz02}
to capture the nonperturbative physics of freezing
in the 2D random phase $XY$ model. 

Another intriguing question is whether multifractality of critical
states at a mobility edge or at the critical energy 
in the quantum Hall plateau transition necessarily reveals itself
in a broad distribution of the LDOS. If the answer is positive
as is suggested by numerical simulations in Refs.\
\onlinecite{Dohmen96}
and
\onlinecite{Huckestein97},
the LDOS might be more interesting than commonly thought. Indeed,
it has been proposed that the freezing transition in the multifractal
spectrum of zero modes in the HWK model 
is generic to multifractal wave functions at a mobility edge
and, in particular, also applies to critical
states in the plateau transition.\cite{Mirlin00}
The corollary would then be that a counterpart to this freezing
transition characterizes the statistical distribution of the LDOS---
say, that typical and average LDOS disagree. 
So far, relatively little attention has been paid to the 
distribution of the LDOS in the plateau transition since
the calculation by Wegner of the {}{DOS}
(Ref.\ \onlinecite{Wegner83})
in spite of great theoretical efforts to identify the critical theory
governing the plateau transition 
(see Refs.\ \onlinecite{Zirnbauer99} and \onlinecite{Bhaseen00} 
for the latest endeavors).

\textit{Note added in proof.}
Within the SUSY representation of the Gade model (a NLSM) 
introduced by Guruswamy et \textit{al.} 
it can be seen that it is the so-called Gade term which is,
ultimately, responsible for the freezing transition of the DOS. 
The Gade term in the Lagrangian of the NLSM
results from the fact that the superdeterminant 
of the supermatrix $T$ entering the NLSM is not constrained to 
some fixed numerical value as is the case in all other universality classes.
The superdeterminant of $T$ is thus a dynamical variable
of its own whose fluctuations are controlled by the Gade term
$\left(\partial _\mu\ln\mathrm{Sdet} T\right)^2$.
A similar role is played by the Gade term 
in the bosonic and fermionic
replicated NLSM representing the chiral universality classes.

\section*{Acknowledgments}

{{}}{
We would like to thank A.\ W.\ W.\ Ludwig for pointing out to us
how the ambiguity in the bosonization rules of the gl($1|1$) super current
algebra is lifted by imposing boundary conditions on the bosonized fields.
                }
This work was supported in part
by a Grant-in-Aid for Scientific Research on Priority Areas (A) from
the Ministry of Education, Culture, Sports, Science and Technology   
(Grant No. 12046238) (A.F.)
and JSPS (S.R.).
C.M.\ thanks the Yukawa Institute for Theoretical Physics in Kyoto
for its hospitality during the initial stage of this work.

\appendix

\section{The process of complete annihilation}
\label{app: sec The process of complete annihilation}

In this appendix, we give the full OPE's obeyed
by the composite operators that control the {}{DOS}
in the MDH model of Sec.\  \ref{sec: DOS for the MDH model}.
Although this paper deals with two flavors of Dirac fermions 
subjected to an imaginary random vector potential and a complex-valued
random mass, 
we also consider the case of one flavor subjected to a real-valued
random vector potential.\cite{footnote purely real A} 
Indeed, as we have argued, 
the problem of the {}{DOS} 
in the two-flavor problem reduces to a large
extent to the one-flavor problem. Moreover, the one-flavor problem is 
interesting in its own right as it is related to 
the plateau transition in the integer quantum Hall effect\cite{Ludwig94}
and to the 2D random phase 
$XY$ model.\cite{Mudry96,Mudry99,Guruswamy00,Carpentier00}

\subsection{Case of one flavor}

The problem of a single flavor of Dirac fermions coupled to
a {real}-valued random vector potential 
can be shown to reduce to the action\cite{Mudry96}
\begin{eqnarray}
S_{*}
&=&\frac{1}{2g^{\vphantom{*}}_{A}}\int d^{2}\bm{r}
(\partial_{\mu}\Phi_{1})^{2}
\nonumber\\
&&
+
\int \frac{d^{2}\bm{r}}{\pi}
\left(
\psi^{*}_{-}\bar{\partial}\psi^{\vphantom{*}}_{-}
+
\psi^{*}_{+}\partial\psi^{\vphantom{*}}_{+}
\right)
\nonumber\\
&&
+
\int \frac{d^{2}\bm{r}}{\pi}
\left(
\beta^{*}_{-}\bar{\partial}\beta^{\vphantom{*}}_{-}
+
\beta^{*}_{+}\partial\beta^{\vphantom{*}}_{+}
\right).
\label{eq: def action one flavor}
\end{eqnarray}
Here, $\Phi_{1}$ is the transversal component of 
the random vector potential,
with variance $g^{\vphantom{*}}_{A}$. For simplicity,
the longitudinal part $\Phi_{2}$ and the Jacobian 
that arise when trading the vector potential by
its transversal and longitudinal components
(conformal sector with net central charge $-1$) 
have been dropped from the action
as we shall only seek the scaling dimensions of gauge invariant
composite operators. 
Action (\ref{eq: def action one flavor})
is critical for arbitrary 
value of $g^{\vphantom{*}}_{A}$ as shown in Ref.\ \onlinecite{Ludwig94}
and defines a CFT with gl($1|1$) symmetry
as shown in Ref.\ \onlinecite{Mudry96}.

It was shown in Ref.\ \onlinecite{Mudry96} that
there exists an infinite set of operators
with negative anomalous scaling dimensions.
These operators are composite operator of the form
\begin{eqnarray}
\mathcal{O}_{\boldsymbol{m},\boldsymbol{n}}&=&
\left( {\mathcal F}_{+} \right)^{m_{+}}
\left( {\mathcal B}_{+} \right)^{n_{+}}
\left( {\mathcal F}_{-} \right)^{m_{-}}
\left( {\mathcal B}_{-} \right)^{n_{-}}
\nonumber\\
&&
\times
\exp\left[2\left(m_{+}+n_{+}-m_{-}-n_{-}\right)\Phi_{1} \right],
\nonumber\\
\end{eqnarray}
where ${\cal F}_{\pm}$ and ${\cal B}_{\pm}$ are defined as
{}{
\begin{eqnarray}
{\cal F}_{+}:=\psi^{*}_{-}\psi^{\vphantom{*}}_{+},
&&\qquad
{\cal F}_{-}:=\psi^{*}_{+}\psi^{\vphantom{*}}_{-},
\nonumber\\
{\cal B}_{+}:=\beta^{*}_{-}\beta^{\vphantom{*}}_{+},
&&\qquad
{\cal B}_{-}:=\beta^{*}_{+}\beta^{\vphantom{*}}_{-}.
\end{eqnarray}
                }
Their scaling dimensions 
are given by 
\begin{eqnarray}
x_{\boldsymbol{m},\boldsymbol{n}}&=&
\left(m_{+}-m_{-}\right)^{2}+|n_{+}-n_{-}|
\nonumber\\
&&
-\frac{g^{\vphantom{*}}_{A}}{\pi}
\left[
\left(m_{+}-m_{-}\right)
+
{}{
\left(n_{+}-n_{-}\right)
                }
\right]^{2},
\end{eqnarray}
respectively. We thus see that for any value of the disorder strength
$g^{\vphantom{*}}_{A}$, there are infinitely many operators
with negative scaling dimensions as $\left(n_{+}-n_{-}\right)^2$
dominates over $\left|n_{+}-n_{-}\right|$
for large $\left|n_{+}-n_{-}\right|$.
Of all the composite operators
$\left\{ 
{\mathcal O}_{\boldsymbol{m},\boldsymbol{n}}
\right\}$,
we focus on the subset spanned by
\begin{eqnarray}
\left(
\mathcal{B}^{\vphantom{*}}_{\pm}
\right)^{N}
e^{\pm2N\Phi^{\vphantom{*}}_{2}},
\quad
\left(
\mathcal{F}^{\vphantom{*}}_{\pm}
\right)
\left(
\mathcal{B}^{\vphantom{*}}_{\pm}
\right)^{N-1}
e^{\pm2N\Phi^{\vphantom{*}}_{2}},
\end{eqnarray}
where $N$ is a positive integer.
For any given $N$,
these two operators are the most relevant and share 
the same scaling dimension, forming a two-dimensional representation
of gl($1|1$). We then construct gl($1|1$) singlets\cite{Mudry96}
\begin{eqnarray}
{\mathcal A}_{+;N}:=
\frac{e^{+2N\Phi_{1}}}{N!}
\left(
{\mathcal B}_{+}^{N}
+N{\mathcal F}_{+}{\mathcal B}_{+}^{N-1}
\right),
\nonumber\\
{\mathcal A}_{-;N}:=
\frac{e^{-2N\Phi_{1}}}{N!}
\left(
{\mathcal B}_{-}^{N}
+N{\mathcal F}_{-}{\mathcal B}_{-}^{N-1}
\right),
\label{eq: def Apm;N}
\end{eqnarray}
whose scaling dimensions read
\begin{eqnarray}
x_{\pm;N}=x_{N}=N-\frac{g^{\vphantom{*}}_{A}}{\pi}N^{2}.
\end{eqnarray}
Operators (\ref{eq: def Apm;N})
are always induced in a RG analysis of the DOS if one is after
localization properties or in a RG analysis of the relevance of 
charge 2 vortices if one is after the fate of 
the Kosterlitz-Thouless transition in the 2D random phase $XY$ model. 
Indeed, they are induced through
repeated OPE's of the operator
\begin{eqnarray}
{\mathcal O}
=e^{+2\Phi_{1}}
\left(
{\mathcal B}_{+}+{\mathcal F}_{+}
\right)
+e^{-2\Phi_{1}}
\left(
{\mathcal B}_{-}+{\mathcal F}_{-}
\right)
\end{eqnarray}
with itself. Hence, any RG analysis of the DOS or
of the stability of the spin-wave phase in the 2D random phase $XY$ model
is controlled by the OPE's within the set
$\left\{\mathcal{A}_{s;N}\right\}$.
These OPE's can be classified into three different processes
which we call
``fusion,''
``partial annihilation,'' 
and 
``complete annihilation''
and are given by
\begin{widetext}
\begin{subequations}
{}{
\begin{eqnarray}
{\mathcal A}_{s;N}(z,\bar{z})\,
{\mathcal A}_{s;N^{\prime}}(w,\bar{w})
&=&
\hphantom{(-)^{N_{-}}}
|z-w|^{x_{N+N^{\prime}}
       -x_{N^{\vphantom{\prime}}}-x_{N^{\prime}}}
\left(\begin{array}{cc} N+N^{\prime}\\ N \end{array}\right)
{\mathcal A}_{s;N+N^{\prime}}(w,\bar{w})
+\cdots,
\qquad
\nonumber\\
&&
\hphantom{AAAAAAAAAAAAAAAAAAAAAAAAAAAA}
0<N^{\vphantom{*}},N^{\prime}\in\mathbb{N},
\ s=\pm,
\label{eq: OPE for N=1; fusion}
\\
{\mathcal A}_{+;N_{+}}(z,\bar{z})\,
{\mathcal A}_{-;N_{-}}(w,\bar{w})
&=&
(-)^{N_{-}}
|z-w|^{x^{\vphantom{\prime}}_{N_{+}-N_{-}}
     -x^{\vphantom{\prime}}_{N_{+}}-x^{\vphantom{\prime}}_{N_{-}}}
\left(\begin{array}{cc}N_{+}-1 \\ N_{-} \end{array}\right)
{\mathcal A}_{+;N_{+}-N_{-}}(w,\bar{w})
+\cdots,
\qquad
\nonumber\\
&&
\hphantom{AAAAAAAAAAAAAAAAAAAAAAAAAAAA}
0<N^{\vphantom{*}}_{-}<N^{\vphantom{*}}_{+}\in\mathbb{N},
\label{eq: OPE for N=1; annihilation}
\\
{\cal A}_{+;N_{+}}(z,\bar{z})
{\cal A}_{-;N_{-}}(w,\bar{w})
&=&
(-)^{N\hphantom{s}}
|z-w|^{2-2x_{N}^{\vphantom{\prime}}}
\Big\{
+(N-1){\cal O}_{A}(w,\bar{w})
+N{\cal O}_{M}(w,\bar{w})
\nonumber\\
&&
\hphantom{(-)^{N}}
-
2N\partial \Phi_{1}(w)
\left[
-\bar{J}^{\prime}(\bar{w})+\bar{J}(\bar{w})
\vphantom{\bar{J}}
\right]
+
2N\bar{\partial}\Phi_{1}(\bar{w})
\left[
-J^{\prime}(w)
+J(w)
\vphantom{\bar{J}}
\right]
\Big\}
+\cdots,
\nonumber\\
&&
\hphantom{AAAAAAAAAAAAAAAAAAAAAAAAAAAA}
0<N^{\vphantom{*}}_{-}=N^{\vphantom{*}}_{+}=N^{\vphantom{*}}\in\mathbb{N},
\label{eq: OPE for N=1; complete annihilation}
\end{eqnarray}
                 }
\label{eq: OPE for N=1; A x A}
\end{subequations}
\end{widetext}
respectively. These OPE's are the central result of this section.
We have used the shorthand notation
\begin{eqnarray}
{\cal O}_{A}&:=&
-\left(
 \psi^{*}_{-}\psi^{\vphantom{*}}_{-}
+\beta^{*}_{-}\beta^{\vphantom{*}}_{-}
\right)
\left(
 \psi^{*}_{+}\psi^{\vphantom{*}}_{+}
+\beta^{*}_{+}\beta^{\vphantom{*}}_{+}
\right)
\nonumber\\
&=&
-(J^{\prime}-J)(\bar{J}^{\prime}-\bar{J})
\label{eq: the random vector perturbation;one-flavor}
\end{eqnarray}
for the random vector potential perturbation
and
\begin{eqnarray}
{\cal O}_{M}&:=&
-\left(
 \psi^{*}_{+}\psi^{\vphantom{*}}_{-}
+\beta^{*}_{+}\beta^{\vphantom{*}}_{-}
\right)
\left(
 \psi^{*}_{-}\psi^{\vphantom{*}}_{+}
+\beta^{*}_{-}\beta^{\vphantom{*}}_{+}
\right)
\nonumber\\
&=&
+
\left(
 J\bar{J}
-J^{\prime}\bar{J}^{\prime}
-G_{-}\bar{G}_{+}
+G_{+}\bar{G}_{-}
\right)
\nonumber\\
&=&
-
\left(
 {\cal F}_{+}{\cal F}_{-}
+{\cal B}_{+}{\cal B}_{-}
+{\cal F}_{+}{\cal B}_{-}
+{\cal F}_{-}{\cal B}_{+}
\right)
\nonumber\\
&&
\label{eq: the random mass perturbation;one-flavor}
\end{eqnarray}
for the random mass perturbation.
The gl($1|1$) currents are defined by
\begin{eqnarray}
J:=\psi^{\vphantom{*}}_{-}\psi^{*}_{-},&&
J^{\prime}:=\beta^{\vphantom{*}}_{-}\beta^{*}_{-},
\nonumber\\
G_{+}:=\beta^{*}_{-}\psi^{\vphantom{*}}_{-},&&
G_{-}:=\beta^{\vphantom{*}}_{-}\psi^{*}_{-},
\end{eqnarray}
with similar definitions in the antiholomorphic sectors for 
$\bar{J},\bar{J}^{\prime},\bar{G}_{+},\bar{G}_{-}$.
Operators $\mathcal{O}_A$ and $\mathcal{O}_M$ 
are induced by disorder averaging over a 
{}{Gaussian-distributed}
real-valued random vector potential and  random mass, respectively.
We see that the process of complete annihilation induces a renormalization
of $g_A$ through the presence of the terms
$
\partial \Phi_{1}\left(-J^{\prime}+J\right)
$,
$
\bar{\partial}\Phi_{1}\left(-\bar{J}^{\prime}+\bar{J}\right)
$,
and ${\mathcal O}_{A}$
on the right-hand side of Eq.\ (\ref{eq: OPE for N=1; complete annihilation}).
We can close the OPE's (\ref{eq: OPE for N=1; A x A})
by inclusion of the OPE's ($s=\pm$)
\begin{subequations}
\begin{eqnarray}
{\mathcal O}_{M}(z,\bar{z})
{\mathcal A}_{s;N}(w,\bar{w})
&=&\frac{N(N-1)}{|z-w|^{2}}{\mathcal A}_{s;N}(w,\bar{w})+\cdots,
\nonumber\\
\\
{\mathcal O}_{M}(z,\bar{z})
{\mathcal O}_{M}(w,\bar{w})
&=&\frac{-2}{|z-w|^{2}}{\mathcal O}_{A}(w,\bar{w})+\cdots.
\nonumber\\
\label{N=1_ope_2}
\end{eqnarray}
\label{eq: OPE for N=1; Om x A, Om x Om}
\end{subequations}

\par\noindent
It is consistent to neglect the process of complete annihilation
(\ref{eq: OPE for N=1; complete annihilation})
if one neglects OPE's (\ref{eq: OPE for N=1; Om x A, Om x Om}).
This is the approximation that we make to calculate the DOS
arbitrarily close to the band center.

\subsection{Case of two flavors}

We now return to the OPE's of the MDH model
in Eq.\ (\ref{eq: closed subOPE's})
which we supplement by specifying the process of complete annihilation 
\begin{widetext}
\begin{subequations}
{}{
\begin{eqnarray}
{\mathcal A}_{s;N}(z,\bar{z})\,
{\mathcal A}_{s;N^{\prime}}(w,\bar{w})
&=&
|z-w|^{x_{N+N^{\prime}}
      -x_{N}^{{\vphantom{\prime}}}-x_{N^{\prime}}}
\left(\begin{array}{cc}N+N^{\prime} \\ N \end{array}\right)
{\mathcal A}_{s;N+N^{\prime}}(w,\bar{w})
+\cdots,
\qquad
\nonumber\\
&&
\hphantom{AAAAAAAAAAAAAAAAAAAAAAAAAAAAA}
0<N,N^{\prime}\in\mathbb{N},\
s=\pm,
\\
{\mathcal A}_{+;N_{+}}(z,\bar{z})\,
{\mathcal A}_{-;N_{-}}(w,\bar{w})
&=&
|z-w|^{x^{\vphantom{\prime}}_{N_{+}-N_{-}}
      -x^{\vphantom{\prime}}_{N_{+}}-x^{\vphantom{\prime}}_{N_{-}}}
\left(\begin{array}{cc}N_{+}-1 \\ N_{-}\end{array}\right)
{\mathcal A}_{+;N_{+}-N_{-}}(w,\bar{w})
+\cdots,
\qquad
\nonumber\\
&&
\hphantom{AAAAAAAAAAAAAAAAAAAAAAAAAAAAA}
0<N_{-}<N_{+}\in\mathbb{N},
\\
{\mathcal A}_{+;N_{+}}(z,\bar{z})\,
{\mathcal A}_{-;N_{-}}(w,\bar{w})
&=& 
|z-w|^{2-2x^{\vphantom{\prime}}_{N}}
\Big\{
-(N-1){\mathcal O}_{A}(w,\bar{w})
+N{\mathcal O}_{0}(w,\bar{w})
\nonumber\\
&&
+2N
\partial\Phi_{2}(w)
\sum_{a=1}^{2} \left[
-\bar{J}_{aa}^{\prime}(\bar{w})
+\bar{J}_{aa}(\bar{w})
\right]
+2N
\bar{\partial}\Phi_{2}(\bar{w})
\sum_{a=1}^{2} \left[
-J_{aa}^{\prime}(w)
+J_{aa}(w)
\right]
\Big\}
+\cdots,
\nonumber\\
&&
\hphantom{AAAAAAAAAAAAAAAAAAAAAAAAAAAAA}
0< N_{+}= N_{-}=N\in\mathbb{N}.
\end{eqnarray}
                 }
\end{subequations}
\end{widetext}
Here,
\begin{eqnarray}
{\cal O}_{A}
&=&
-\sum_{a,b=1}^{2}
\left(
\psi^{*}_{a-}\psi^{\vphantom{*}}_{a-}+
\beta^{*}_{a-}\beta^{\vphantom{*}}_{a-}
\right)
\left(
\psi^{*}_{b+}\psi^{\vphantom{*}}_{b+}+
\beta^{*}_{b+}\beta^{\vphantom{*}}_{b+}
\right)
\nonumber\\
&=&\sum_{a,b=1}^{2}
\left[
-J^{          \prime }_{aa}\bar{J}^{          \prime }_{bb}
+J^{          \prime }_{aa}\bar{J}^{\vphantom{\prime}}_{bb}
+J^{\vphantom{\prime}}_{aa}\bar{J}^{          \prime }_{bb}
-J^{\vphantom{\prime}}_{aa}\bar{J}^{\vphantom{\prime}}_{bb}
\right]
\hphantom{AA}
\nonumber\\
\end{eqnarray}
is the random vector potential perturbation
for the two-flavor case, and
${\cal O}_{0}$ is an operator generated by 
taking the OPE 
$
{\mathcal A}_{+;1}
\times
{\mathcal A}_{-;1}
$,
whose explicit form is given by
\begin{eqnarray}
&&
{\cal O}_{0}
=
\nonumber\\
&&
+J_{11}^{\prime}\bar{J}_{22}^{\prime}
+J_{12}^{\prime}\bar{J}_{21}^{\prime}
+J_{21}^{\prime}\bar{J}_{12}^{\prime}
+J_{22}^{\prime}\bar{J}_{11}^{\prime}
\nonumber\\
&&
{}{
-G_{11+}\bar{G}_{22+}
                } 
-G_{12+}\bar{G}_{21+} 
+G_{21+}\bar{G}_{12+}
+G_{22+}\bar{G}_{11+}
\nonumber\\
&&
-G_{11-}\bar{G}_{22-}
-G_{12-}\bar{G}_{21-}
+G_{21-}\bar{G}_{12-}
+G_{22-}\bar{G}_{11-}
\nonumber\\
&&
-J_{11}\bar{J}_{22} 
+J_{12}\bar{J}_{21}
+J_{21}\bar{J}_{12} 
-J_{22}\bar{J}_{11}.
\end{eqnarray}
Observe that the multiplicative factor $(-)^{N}$
present in Eq.\ (\ref{eq: OPE for N=1; A x A})
does not appear here. This is so because we find operators with
negative scaling dimensions in the gauge-variant sector, i.e., through 
``Cooper-like'' terms, when the random vector potential is imaginary.
As was the case for a real-valued random vector potential,
a renormalization of $g_A$ is encoded by the complete fusion process
through the contributions
$
\partial \Phi_{2}\sum_{a}
\left(
-\bar{J}_{aa}^{\prime}
+\bar{J}_{aa}
\right),
$
$
\bar{\partial} \Phi_{2}\sum_{a}
\left(
-J_{aa}^{\prime}
+J_{aa}
\right)
$,
and ${\mathcal O}_{A}$.
For the MDH model, we neglect this effect in the same way as we
neglect the renormalization of
$g^{\vphantom{*}}_A$ induced by 
the source term needed to compute the DOS.
This approximation should be good if one is after the DOS asymptotically
close to the band center but fails if one is interested in the DOS at
large energy scales. For the HWK model, we neglect complete annihilation
relative to the renormalization of
$g^{\vphantom{*}}_A$ induced by $g^{\vphantom{*}}_M$.

\section{Bosonization of the ghost system}
\label{app: sec Bosonization of the ghost system}

\subsection{Bosonization}

{{}}{The purpose of this section is to 
review the steps of Ref.\ \onlinecite{Guruswamy00}
by which} the {}{SUSY} action
\begin{eqnarray}
&&
S_{0}=\int \frac{d^{2}\bm{r}}{\pi}
\left(
\psi^{*}_{-}\bar{\partial }\psi^{\vphantom{*}}_{-}
+
\psi^{*}_{+} \partial \psi^{\vphantom{*}}_{+}
\right)
\nonumber\\
&&
\hphantom{S_{0}=}
+
\int \frac{d^{2}\bm{r}}{\pi}
\left(
\beta^{*}_{-}\bar{\partial }\beta_{-}
+
\beta^{*}_{+} \partial \beta^{\vphantom{*}}_{+}
\right),
\label{eq: psi-beta}
\end{eqnarray}
where $\psi^{*},\psi$ are fermionic spinors
and $\beta^{*},\beta$ are bosonic spinors,
can be ``bosonized,'' i.e., represented by a pair of independent 
scalar fields
$\phi$ and $\phi^{\prime}$ and a pair of independent fermionic fields 
$\chi^{*}$ and $\chi$.
The fermionic sector in the SUSY action (\ref{eq: psi-beta})
describes a {}{CFT} with central charge $c=+1$.
The bosonic sector  in the SUSY action (\ref{eq: psi-beta})
describes a CFT with central charge $c=-1$.
Taken together, both sectors describe a CFT with vanishing central charge.
The {}{OPE} for the spinors are
\begin{eqnarray}
\psi^{\vphantom{*}}_{-}(z)\psi^{*}_{-}(0)
&=&
\hphantom{+}\psi^{*}_{-}(z)\psi^{\vphantom{*}}_{-}(0)
=\frac{1}{z}+\cdots,
\nonumber\\
\psi^{\vphantom{*}}_{+}(\bar{z})\psi^{*}_{+}(0)
&=&
\hphantom{+}
\psi^{*}_{+}(\bar{z})\psi^{\vphantom{*}}_{+}(0)
=
\frac{1}{\bar{z}}+\cdots,
\nonumber\\
\beta^{\vphantom{*}}_{-}(z)\beta^{*}_{-}(0)
&=&-
\beta^{*}_{-}(z)\beta^{\vphantom{*}}_{-}(0)\,
=\frac{1}{z}+\cdots,
\nonumber\\
\beta^{\vphantom{*}}_{+}(\bar{z})\beta^{*}_{+}(0)
&=&-
\beta^{*}_{+}(\bar{z})\beta^{\vphantom{*}}_{+}(0)\,
=
\frac{1}{\bar{z}}+\cdots.
\label{eq: OPEs for psi-beta}
\end{eqnarray}
After defining the dimension-1 currents
\begin{eqnarray}
&&
J:=
\psi ^{\ }_-\psi ^{* }_-,
\qquad
J^{\prime}:=
\beta^{\ }_-\beta^{* }_-,
\nonumber\\
&&
\bar
J:=
\psi ^{\ }_+\psi ^{* }_+,
\qquad
\bar
J^{\prime}:=
\beta^{\ }_+\beta^{* }_+,
\nonumber\\
&&
G^{\ }_+:=
\beta^{* }_-\psi^{\ }_-,
\qquad
G^{\ }_-:=
\beta^{\ }_-\psi^{* }_-,
\nonumber\\
&&
\bar G^{\ }_+:=
\beta^{* }_+\psi^{\ }_+,
\qquad
\bar G^{\ }_-:=
\beta^{\ }_+\psi^{* }_+,
\end{eqnarray}
we see that they generate
the gl($1|1$) super current algebra
\begin{eqnarray}
J(z)J(w)&=& \frac{1}{(z-w)^{2}}+\cdots,
\nonumber\\
J^{\prime}(z)J^{\prime}(w)&=&
\frac{-1}{(z-w)^{2}}
+\cdots,
\nonumber\\
J(z)G_{\pm}(w)&=&
\pm\frac{1}{z-w}G_{\pm}(w)+\cdots,
\nonumber\\
J^{\prime}(z)G_{\pm}(w)&=&
\pm\frac{1}{z-w}G_{\pm}(w)+\cdots,
\nonumber\\
G_{\mp}(z)G_{\pm}(w)&=&
 \frac{\pm 1}{(z-w)^{2}}
\nonumber\\
&&
+
\frac{-1}{z-w}
[J(w)-J^{\prime}(w)]
+\cdots.
\nonumber\\
&&
\label{eq: gl(l|1) OPEs}
\end{eqnarray}

The idea behind bosonization is to represent the
SUSY action (\ref{eq: psi-beta}) in terms of a new set of fields
so as to preserve the
OPE's (\ref{eq: OPEs for psi-beta}) and (\ref{eq: gl(l|1) OPEs}).
For example, it has been known since the early 1970's that
the currents for the fermionic $c=1$ sector
can be mimicked by a scalar field $\phi(z,\bar{z})$
through the identifications
\begin{eqnarray}
J\equiv
+({}{i}    \partial\phi),
\qquad
\bar
J\equiv
-({}{i}\bar\partial\phi),
\label{eq: bosoniation for J}
\end{eqnarray}
which preserve the $J-J$ OPE
in Eq. (\ref{eq: gl(l|1) OPEs})
provided $\phi(z,\bar{z})$ obeys the OPE
\begin{eqnarray}
\phi(z,\bar{z}) \phi(0)
=-\ln z\bar{z}+\cdots.
\end{eqnarray}
In turn,
fermionic spinors can be expressed by 
the Mandelstam formulas
\begin{eqnarray}
&&
\psi^{\vphantom{*}}_{-}\equiv 
{{}}{u}^{-1}\hphantom{()^*}
\mathfrak{a}^{-1/2}e^{+i\varphi_{-}},
\nonumber\\
&&
\psi^{*}_{-}\equiv 
{{}}{u}^{+1}\hphantom{()^*}
\mathfrak{a}^{-1/2}e^{-i\varphi_{-}},
\nonumber\\
&&
\psi^{\vphantom{*}}_{+}\equiv 
({{}}{u}^*)^{-1}
\mathfrak{a}^{-1/2}e^{-i\varphi_{+}},
\nonumber\\
&&
\psi^{*}_{+}\equiv 
({{}}{u}^*)^{+1}
\mathfrak{a}^{-1/2}e^{+i\varphi_{+}},
\label{eq: Mandelstam for psi}
\end{eqnarray}
whereby $\phi(z,\bar{z})$ is decomposed into a holomorphic 
and antiholomorphic part according to
$
\phi(z,\bar{z})=
\varphi_{-}(z)+\varphi_{+}(\bar{z})
$.
Here, an arbitrary complex number ${{}}{u}$ has been introduced,
as it is not fixed by the OPE's (\ref{eq: OPEs for psi-beta}).

Although the $\beta^{*}-\beta$ subsystem is already bosonic,
it can be advantageous to trade these spinors for another scalar
field
$
\phi^{\prime}(z,\bar{z})=
\varphi^{\prime}_{-}(z)+\varphi^{\prime}_{+}(\bar{z}).
$\cite{Friedan86}
As in the fermionic case,
identification
\begin{eqnarray}
J^{\prime}\equiv
+({}{i}    \partial\phi),
\qquad
\bar
J^{\prime}\equiv
-({}{i}\bar\partial\phi),
\label{eq: bosoniation for Jprime}
\end{eqnarray}
is compatible with the $J^{\prime}-J^{\prime}$ OPE
in Eq. (\ref{eq: gl(l|1) OPEs})
provided $\phi^{\prime}$ satisfies 
\begin{eqnarray}
\phi^{\prime}(z,\bar{z}) \phi^{\prime}(0)&=&
+\ln z\bar{z}+\cdots.
\end{eqnarray}
Observe that this OPE 
differs from the one in the $\phi$ sector by a sign.
Since the $\beta^*-\beta$ sector can be thought of
as a CFT with central charge $c=-1$ whereas it is natural to assign
the central charge $c=+1$ to the $\phi^{\prime}$ sector, 
a conformal sector with central charge $c=-2$ must be supplemented 
to the $\phi^{\prime}$ sector.
The need for an additional conformal sector can also be seen when
tentatively writing down Mandelstam formulas
for the bosonic spinors $\bar{\beta}_{\pm},\beta_{\pm}$
of the form (\ref{eq: Mandelstam for psi})
with the substitution of 
$\varphi^{\vphantom{\prime}}_{\pm}$ 
by 
$\varphi^{\prime}_{\pm}$
as the exponentials $\exp(\pm i\varphi^{\prime}_{\pm})$ 
obey Fermi statistics.
The correct bosonic statistics in the $\beta^*-\beta$ sector
is implemented after the
introduction of two pairs of independent fermions
$\eta_{\pm}$
and
$\xi_{\pm}$ in the holomorphic and antiholomorphic sectors,
respectively, through the bosonization rules 
\begin{eqnarray}
&&
\beta^{\vphantom{*}}_{-}\equiv 
{{}}{v}^{+1}\hphantom{()^*}
\mathfrak{a}^{+1/2}e^{+i\varphi^{\prime}_{-}}\eta^{\vphantom{*}}_{-},
\nonumber\\
&&
\beta^{*}_{-}\equiv 
{{}}{v}^{-1}\hphantom{()^*}
\mathfrak{a}^{+1/2}\,
e^{-i\varphi^{\prime}_{-}}\partial\xi^{\vphantom{*}}_{-},
\nonumber\\
&&
\beta^{\vphantom{*}}_{+}\equiv 
({{}}{v}^*)^{+1}
\mathfrak{a}^{+1/2}\,
e^{-i\varphi^{\prime}_{+}}\eta^{\vphantom{*}}_{+},
\nonumber\\
&&
\beta^{*}_{+}\equiv 
({{}}{v}^*)^{-1}
\mathfrak{a}^{+1/2}\,
e^{+i\varphi^{\prime}_{+}}\bar{\partial}\xi^{\vphantom{*}}_{+}.
\label{eq: Mandelstam for beta}
\end{eqnarray}
As before,
the arbitrary complex number ${{}}{v}$ has been introduced,
as it is not fixed by the OPE's (\ref{eq: OPEs for psi-beta}),
whereas
$\eta_{-}$,
$\xi_{-}$,
$\eta_{+}$,
and
$\xi_{+}$ 
are anticommuting fields with the conformal weights 
$(1,0)$,
$(0,0)$,
$(0,1)$,
and
$(0,0)$,
respectively. 
Their OPE's are thus given by
\begin{eqnarray}
\xi_{-}(z)\eta_{-}(0)&=&
\eta_{-}(z)\xi_{-}(0)=\frac{1}{z}+\cdots,
\nonumber\\
\xi_{+}(\bar{z})\eta_{+}(0)&=&
\eta_{+}(\bar{z})\xi_{+}(0)=\frac{1}{\bar{z}}+\cdots.
\label{ope_phi_eta_xi}
\end{eqnarray}
Finally, the action (\ref{eq: psi-beta}) has the bosonized representation
\begin{eqnarray}
&&
S_{0}=\int \frac{d^{2}\bm{r}}{2\pi}
\left(
\partial \phi \bar{\partial}\phi
-
\partial \phi^{\prime} \bar{\partial}\phi^{\prime}
\right)
\nonumber\\
&&
\hphantom{S_{0}=}
+
\int \frac{d^{2}\bm{r}}{\pi}
\left(
\eta_{-}\bar{\partial}\xi_{-}
+
\eta_{+}\partial \xi_{+}
\right).
\label{eq: def S_{0}}
\end{eqnarray}

Bosonization of the $\beta^{*}-\beta$ sector is not unique.\cite{Gaberdiel96} 
Instead of the $\eta-\xi$ sector we could have chosen to introduce
(symplectic) fermions through the action
\begin{eqnarray}
-\int \frac{d^{2}\bm{r}}{2\pi}
\left(
\partial \chi^{*}
\bar{\partial}\chi^{}
+
\bar{\partial} \chi^{*}
\partial \chi^{}
\right),
\label{eq: action chi's}
\end{eqnarray}
the OPE's
\begin{eqnarray}
\chi^{*}(z,\bar{z})
\chi^{}(0)
=
-\chi^{}(z,\bar{z})
\chi^{*}(0)
=-\ln z\bar{z}+\cdots
\label{eq: OPE chi's}
\end{eqnarray}
and the identifications
\begin{eqnarray}
\eta_{-}(z)\equiv i\partial \chi^{*}(z),&&
\partial\xi_{-}(z)\equiv i\partial \chi^{}(z),
\nonumber\\
\eta_{+}(\bar{z})\equiv i\bar{\partial} \chi^{*}(\bar{z}),&&
\bar{\partial} \xi_{+}(\bar{z}) \equiv i\bar{\partial} \chi^{}(\bar{z}).
\end{eqnarray}
[The minus sign in Eq.\ (\ref{eq: action chi's})
 is necessary to get the minus sign on the right-hand side of the OPE 
 (\ref{eq: OPE chi's}) with the convention we are using for
 Grassmann integration.]
With the Mandelstam formulas,
fermionic currents are bosonized as
\begin{eqnarray}
G_{-}
&\equiv&
\hphantom{(}{{}}{(uv)} \hphantom{^{-1}}
\hphantom{)^{*}}
\left[
 e^{+{}{i}(\varphi_{-}^{\prime}-\varphi_{-}^{\     })}
 ({}{i}    \partial\chi^{* })
 \right],
\nonumber\\
\bar{G}_{-}
&\equiv&
\hphantom{(}{{}}{(uv)}^{*}\hphantom{{}^{-1}}
\hphantom{)}
 \left[
 e^{-{}{i}(\varphi_{+}^{\prime}-\varphi_{+}^{\     })}
 ({}{i}\bar\partial\chi^{* })
 \right],
\nonumber\\
G_{+}
&\equiv&
\hphantom{(} {{}}{(uv)}^{-1}
\hphantom{)^{*}}
 \left[
 e^{-{}{i}(\varphi_{-}^{\prime}-\varphi_{-}^{\     })}
 ({}{i}    \partial\chi^{\ })
 \right],
\nonumber\\
\bar{G}_{+}
&\equiv&
[{{}}{(uv)}^{-1}]^{*}\,\,
 \left[
 e^{+{}{i}(\varphi_{+}^{\prime}-\varphi_{+}^{\     })}
 ({}{i}\bar\partial\chi^{\ })
 \right].
\end{eqnarray}
This bosonization rule,
together with 
Eqs.\ (\ref{eq: bosoniation for J}) 
and  (\ref{eq: bosoniation for Jprime}),
is compatible 
with the gl($1|1$) super current algebra OPE's
(\ref{eq: gl(l|1) OPEs})
for any value of ${{}}{u},{{}}{v}\in \mathbb{C}$.

The ambiguity in the choice of the phase
of ${{}}{(uv)}$ 
when bosonizing fermionic currents
is, however, lifted in the presence of 
gl($1|1$) current-current interactions 
by imposing boundary conditions on the 
bosonized fields that maintain SUSY.
To see this, consider the action
[recall
Eq.\ (\ref{eq: reorganised def bosonic int field theory N>1 b})] 
\begin{eqnarray}
S&=&S_{0}
-\frac{g_{A}}{2\pi}\int \frac{d^{2}\bm{r}}{\pi}
\mathcal{O}_{A}(\bm{r})
-\frac{g_{M}}{2\pi}\int \frac{d^{2}\bm{r}}{\pi}
\mathcal{O}_{M}(\bm{r}),
\nonumber\\
&&
\end{eqnarray}
which describes action (\ref{eq: def S_{0}})
perturbed by
the random vector potential
$\mathcal{O}_{A}$ and 
the random mass 
$\mathcal{O}_{M}$
defined in 
Eqs.\ (\ref{eq: the random vector perturbation;one-flavor}) 
and   (\ref{eq: the random mass perturbation;one-flavor}),
respectively.
Although $S$ is not conformally invariant
it is still GL($1|1$) symmetric. Consequently,
the correlator
$
\left<
\mathcal{O}_{M}(z,\bar{z})
\mathcal{O}_{M}(w,\bar{w})
\right>
$,
say, vanishes for any $g_{A}$ and $g_{M}$.
Calculation of this correlation function
can be done nonperturbatively 
as a function of $g_{A}$ and $g_{M}$
along the lines of Ref.\ \onlinecite{Guruswamy00}.
The condition that this correlator vanishes
uniquely specifies the boundary condition obeyed
by the scalar fields and, in turn,
the phase of ${{}}{(uv)}$.
In this paper we choose 
the boundary condition for which
the bosonized fields vanish at
infinity in the CFT defined by
\begin{subequations}
\begin{eqnarray}
&&
S_{0}=\int \frac{d^{2}\bm{r}}{2\pi}
\left(
\partial \phi \bar{\partial}\phi
-
\partial \phi^{\prime} \bar{\partial}\phi^{\prime}
\right)
\nonumber\\
&&
\hphantom{S_{0}=}
-\int \frac{d^{2}\bm{r}}{2\pi}
\left(
\partial \chi^{*}
\bar{\partial}\chi^{}
+
\bar{\partial} \chi^{*}
\partial \chi^{}
\right).
\label{eq: final bosonized action}
\end{eqnarray}
This amounts to the choice 
${{}}{uv}/|{{}}{uv}|=-i$
for which the Mandelstam formulas become 
(with $|{{}}{uv}|=1$, ${{}}{u=v}$, say)
\begin{eqnarray}
&&
\psi^{\vphantom{*}}_{-}\equiv 
\mathfrak{a}^{-1/2}e^{-i\pi/4+i\varphi_{-}},
\nonumber\\
&&
\psi^{*}_{-}\equiv 
\mathfrak{a}^{-1/2}e^{+i\pi/4-i\varphi_{-}},
\nonumber\\
&&
\psi^{\vphantom{*}}_{+}\equiv 
\mathfrak{a}^{-1/2}e^{+i\pi/4-i\varphi_{+}},
\nonumber\\
&&
\psi^{*}_{+}\equiv 
\mathfrak{a}^{-1/2}e^{-i\pi/4+i\varphi_{+}},
\label{eq: Mandelstam for psi final}
\end{eqnarray}
and
\begin{eqnarray}
&&
\beta^{\vphantom{*}}_{-}\equiv 
\mathfrak{a}^{+1/2}\,
e^{+i\pi/4+i\varphi^{\prime}_{-}}
(i\partial\chi^*),
\nonumber\\
&&
\beta^{*}_{-}\equiv 
\mathfrak{a}^{+1/2}\,
e^{-i\pi/4-i\varphi^{\prime}_{-}}
(i\partial\chi),
\nonumber\\
&&
\beta^{\vphantom{*}}_{+}\equiv 
\mathfrak{a}^{+1/2}\,
e^{-i\pi/4-i\varphi^{\prime}_{+}}
(i\bar\partial\chi^*),
\nonumber\\
&&
\beta^{*}_{+}\equiv 
\mathfrak{a}^{+1/2}\,
e^{+i\pi/4+i\varphi^{\prime}_{+}}
(i\bar\partial\chi).
\label{eq: Mandelstam for beta final}
\end{eqnarray}
\label{eq: final bosonization rules}
\end{subequations}

\subsection{Scaling dimensions of 
composite operators made of bosonic spinors}
The $\chi$ sector plays an essential role as
it {}{ensures} that the $\beta$ sector obeys a bosonic algebra.
It also guarantees that the central charge of action
(\ref{eq: final bosonized action}) vanishes
and {}{gives} 
composite operators from the $\beta$ sector their correct
scaling dimensions. To illustrate this last point we
calculate the scaling dimension of the composite operator
$
(\beta^{*}_{+}\beta^{\vphantom{*}}_{-})^{n}
$.
Taking repeated Wick contractions,
one verifies that the scaling dimension of 
$(\beta^{*}_{+}\beta^{\vphantom{*}}_{-})^{n}$
is $n$. On the other hand, bosonization rules 
(\ref{eq: final bosonization rules}) give
$
(\beta^{*}_{+}\beta^{\vphantom{*}}_{-})^{n}
=
\mathfrak{a}^{+n}
i^n
e^{+in\phi^{\prime}}
\left(
i\partial \chi^{*}
\vphantom{\bar{\partial}}
\right)^{n}
\left(
i\bar{\partial} \chi
\right)^{n}
$.
Since $e^{+in\phi^{\prime}}$ contributes 
the negative scaling dimension $-n^2$
in view of the OPE obeyed in the $\phi^{\prime}$ sector,
$
\left(
i\partial \chi^{*}
\vphantom{\bar{\partial}}
\right)^{n}
\left(
i\bar{\partial} \chi
\right)^{n}
$
must contribute the anomalous scaling dimension $n^{2}+n$. 
That this is indeed so follows from noting that 
$
\left(
i\partial \chi^{*}
\vphantom{\bar{\partial}}
\right)^{n}
$
is generated by repeated OPE of 
$
\left(
i\partial \chi^{*}
\vphantom{\bar{\partial}}
\right)
$ 
with itself
{}{,
\begin{eqnarray}
(i \partial \chi^{*})^{n}(z)
&=&
\frac{1}{(n-1)!}\cdots\frac{1}{1!}\frac{1}{0!}
:\partial^{n-1}(i\partial \chi^{*})
\nonumber\\
&&
\times
\partial^{n-2}(i\partial \chi^{*})
 \cdots
 \partial^{1}(i\partial\chi^{*})
 \partial^{0}(i\partial\chi^{*}):
(z).
\nonumber\\
\end{eqnarray}
                 }
[The notation $:(\cdots):$ refers to normal ordering.]
Since the conformal weight of 
$
\left(
i\partial \chi^{*}
\vphantom{\bar{\partial}}
\right)
$ 
is $(1,0)$,
the conformal weight of $(i\partial \chi^{*})^{n}$
reads
\begin{eqnarray}
\left(n+(n-1)+\cdots +1,0\right)
=
\left(\frac{n(n+1)}{2},0\right).
\end{eqnarray}
Accounting for the contribution from the antiholomorphic part
$
\left(
i\bar{\partial} \chi
\right)^{n}
$
yields the anomalous scaling dimension $n(n+1)$ for
$
\left(
i\partial \chi^{*}
\right)^{n}
\left(
i\bar{\partial} \chi
\vphantom{\bar{\partial}}
\right)^{n}.
$


\begin{thebibliography}{ourpaper}

\bibitem{Edwards71}
J.\ T.\ Edwards and D.\ J.\ Thouless,
J.\ Phys.\ C \textbf{4}, 453 (1971).

\bibitem{Wegner81}
F.\ Wegner,
Z.\ Phys.\ B: Condens.\ Matter \textbf{44}, 9 (1981).

\bibitem{Pryor92}
C.\ Pryor and A.\ Zee, 
Phys.\ Rev.\ B \textbf{46}, 3116 (1992).

\bibitem{Gruzberg01}
I.\ A.\ Gruzberg, P.\ J.\ Hirschfeld, and A.\ V.\ Shytov,
Phys.\ Rev.\ Lett.\ \textbf{87}, 239703 (2001).

\bibitem{Dyson53}  
F.\ J.\ Dyson, 
Phys.\ Rev.\ \textbf{92}, 1331 (1953).

\bibitem{Theodorou76}
G.\ Theodorou and M.\ H.\ Cohen,
Phys.\ Rev.\ B \textbf{13}, 4597 (1976).

\bibitem{Eggarter78}
T.\ P.\ Eggarter and R.\ Riedinger,
Phys.\ Rev.\ B \textbf{18}, 569 (1978).

\bibitem{Gade93} 
R.\ Gade, Nucl.\ Phys.\ B \textbf{398}, 499 (1993); 
R.\ Gade and F.\ Wegner, \textit{ibid.} \textbf{360}, 213 (1991);
F.\ J.\ Wegner, Phys.\ Rev.\ B \textbf{19}, 783 (1979).

\bibitem{Ludwig94} 
A.\ W.\ W.\ Ludwig, M.\ P.\ A.\ Fisher, R.\ Shankar, and G.\ Grinstein,
Phys.\ Rev.\ B \textbf{50}, 7526 (1994).

\bibitem{Nersesyan94}
A.\ A.\ Nersesyan, A.\ M.\ Tsvelik, and F.\ Wenger,
Phys.\ Rev.\ Lett.\ \textbf{72}, 2628 (1994);
Nucl.\ Phys.\ B \textbf{438}, 561 (1995).

\bibitem{Senthil00}
T.\ Senthil and M.\ P.\ A.\ Fisher,
Phys.\ Rev.\ B \textbf{61}, 9690 (2000).

\bibitem{Bocquet00}
M.\ Bocquet, D.\ Serban, and M.\ R.\ Zirnbauer,
Nucl.\ Phys.\  B \textbf{578}, 628 (2000).

\bibitem{Brouwer00}
P.\ W.\ Brouwer, A.\ Furusaki, I.\ A.\ Gruzberg, and C.\ Mudry,
Phys.\ Rev.\ Lett.\ \textbf{85}, 1064 (2000).

\bibitem{Altland97}
A.\ Altland and M.\ Zirnbauer,
Phys.\ Rev.\ B \textbf{55}, 1142 (1997).

\bibitem{RMT}
D.\ A.\ Ivanov,
J.\ Math.\ Phys.\ \textbf{43}, 126 (2002),
and references therein.

\bibitem{Bundschuh98} 
R.\ Bundschuh, C.\ Cassanello, D.\ Serban, and M.\ R.\ Zirnbauer,
Nucl.\ Phys.\ B \textbf{532}, 689 (1998);
Phys.\ Rev.\ B \textbf{59}, 4382 (1999).

\bibitem{Altland01} 
A.\ Altland and R.\ Merkt, 
Nucl.\ Phys.\ B \textbf{607}, 511 (2001).

\bibitem{Motrunich01}
O.\ Motrunich, K.\ Damle, and D.\ A.\ Huse,
Phys.\ Rev.\ B \textbf{63}, 224204 (2001).

\bibitem{Brouwer98}
P.\ W.\ Brouwer, C.\ Mudry, B.\ D.\ Simons, and A.\ Altland, 
Phys.\ Rev.\ Lett.\ \textbf{81}, 862 (1998).

\bibitem{Titov01}
P.\ W.\ Brouwer, C.\ Mudry, and A.\ Furusaki, Phys.\ Rev.\ Lett.\
\textbf{84}, 2913 (2000);
M.\ Titov, P.\ W.\ Brouwer, A.\ Furusaki, and C.\ Mudry, Phys.\ Rev.\
B \textbf{63}, 235318 (2001).

\bibitem{Shelton98}
D.\ G.\ Shelton and A.\ M.\ Tsvelik,
Phys.\ Rev.\ B \textbf{57}, 14242 (1998).

\bibitem{Hastings01}
M.\ B.\ Hastings and S.\ L.\ Sondhi,
Phys.\ Rev.\ B \textbf{64}, 094204 (2001).

\bibitem{Comtet95}
A.\ Comtet, J.\ Desbois, and C.\ Monthus, 
Ann.\ Phys.\ (N.Y.)\ \textbf{239}, 312 (1995).

\bibitem{Balents97}
L.\ Balents and  M.\ P.\ A.\ Fisher,
Phys.\ Rev.\ B \textbf{56}, 12970 (1997).

\bibitem{Mathur97}
H.\ Mathur,
Phys.\ Rev.\ B \textbf{56}, 15794 (1997).

\bibitem{Fukui99}   
T.\ Fukui,
Nucl.\ Phys.\  B \textbf{562}, 477 (1999).

\bibitem{Guruswamy00}
S.\ Guruswamy, A.\ LeClair, and A.\ W.\ W.\ Ludwig,
Nucl.\ Phys.\  B \textbf{583}, 475 (2000).

\bibitem{Altland99}  
A.\ Altland and B.\ D.\ Simons,
Nucl.\ Phys.\  B \textbf{562}, 445 (1999).

\bibitem{Fabrizio00}
M.\ Fabrizio and C.\ Castellani,
Nucl.\ Phys.\  B \textbf{583}, 542 (2000).

\bibitem{Motrunich02}
O.\ Motrunich, K.\ Damle, and D.\ A.\ Huse,
Phys.\ Rev. B \textbf{65}, 064206 (2002).

\bibitem{Dasgupta80}
C.\ Dasgupta and S.\ K.\ Ma,
Phys.\ Rev.\ B \textbf{22}, 1305 (1980).

\bibitem{Hatsugai97}
Y.\ Hatsugai, X.-G.\ Wen, and M.\ Kohmoto,
Phys.\ Rev.\ B\ \textbf{56}, 1061 (1997).

\bibitem{Morita97}
Y.\ Morita and Y.\ Hatsugai,
Phys.\ Rev.\ Lett.\ \textbf{79}, 3728 (1997);
Phys.\ Rev.\ B\ \textbf{58}, 6680 (1998).

\bibitem{Ryu02}
S.\ Ryu and Y.\ Hatsugai,
Phys.\ Rev.\ B\ \textbf{65}, 033301 (2002).

\bibitem{Carpentier00}
D.\ Carpentier and P.\ Le Doussal,
Nucl.\ Phys.\ B \textbf{588}, 565 (2000).

\bibitem{Chamon-Mudry96}
C.\ de C.\ Chamon, C.\ Mudry, and X.-G.\ Wen,
Phys.\ Rev.\ B \textbf{53}, R7638 (1996).

\bibitem{Mudry96}
C. Mudry, C.\ Chamon, and X.-G. Wen, 
Nucl.\ Phys.\ B \textbf{466}, 383 (1996).

\bibitem{Chamon96}
C.\ C.\ Chamon, C.\ Mudry, and X.-G. Wen, 
Phys.\ Rev.\ Lett.\ \textbf{77}, 4194 (1996).

\bibitem{Kogan96}
I.\ I.\ Kogan, C.\ Mudry, and A.\ M.\ Tsvelik,
Phys.\ Rev.\ Lett.\ \textbf{77}, 707 (1996).

\bibitem{Castillo97}
H.\ E.\ Castillo, C.\ de C.\ Chamon, 
E.\ Fradkin, P.\ M.\ Goldbart, and C.\ Mudry,
Phys.\ Rev.\ B \textbf{56}, 10668 (1997).

\bibitem{Mudry99}
C.\ Mudry and X.-G.\ Wen,
Nucl.\ Phys.\ B \textbf{549}, 613 (1999).

\bibitem{Horovitz02}
B.\ Horovitz and P.\ Le Doussal,
Phys.\ Rev.\ B \textbf{65}, 125323 (2002).

\bibitem{Bernard95}
D.\ Bernard,
Nucl.\ Phys.\ B \textbf{441}, 471 (1995);
D.\ Bernard, 
in 
\textit{Low-Dimensional Applications of Quantum Field Theory}, 
edited by L.\ Baulieu, V.\ Kazakov, M.\ Picco, P.\ Windey, 
Vol.\ 362 of
\textit{NATO Advanced Study Series B: Physics}
(Plenum Press, New York, 1997);
D.\ Bernard, hep-th/9509137 (unpublished).

\bibitem{Carpentier01}
D.\ Carpentier and P.\ Le Doussal,
Phys.\ Rev.\ E \textbf{63}, 026110 (2001).

\bibitem{Derrida88} 
B.\ Derrida and H.\ Spohn, 
J.\ Stat.\ Phys.\ \textbf{51}, 817 (1988).

\bibitem{Buffet93}
E.\ Buffet, A.\ Patrick, and J.\ V.\ Pul\'e, 
J.\ Phys.\ A \textbf{26}, 1823 (1993).

\bibitem{Fradkin80's}
M.\ P.\ A.\ Fisher and E.\ Fradkin,
Nucl.\ Phys.\ B \textbf{251}, 457 (1985);
E.\ Fradkin,
Phys.\ Rev.\ B \textbf{33}, 
3257 (1986); \textbf{33}, 3263 (1986).

\bibitem{footnote purely real A} 
If the ``purely imaginary'' gauge fields
$iA_0$ and $i\boldsymbol{A}$
are analytically continued to the
``purely real'' gauge fields
$A_0$ and $\boldsymbol{A}$,
the operator 
$D^{\vphantom{\dag}}_{\hbox{\tiny HWK}}$
in the upper right block of
of Eq.\ (\ref{eq: HWK Hamiltonian}) 
becomes Hermitian. If so, 
$D^{\vphantom{\dag}}_{\hbox{\tiny HWK}}$
can be interpreted as the Hamiltonian for a single
flavor of Dirac fermion subjected to random scalar, vector, and mass
potentials studied in Refs.\ \onlinecite{Ludwig94} and
\onlinecite{Bernard95}.

\bibitem{Mistake-Guruswamy00}
The Hermitian energy perturbation in Eq.\ (5.7)
of Ref.\ \onlinecite{Guruswamy00}
differs from the non-Hermitian energy perturbation in
Eq.\ (\ref{eq: non-Hermitian source term}).
Our choice has the advantage that the RG equations
obeyed by the DOS has the same  structure as
had we chosen to work with
the representation defined by 
{}{
Eq.\ (\ref{eq: fermionic Hermitian Lagrangian with sublattice asymmetry}).
                }

\bibitem{Mistake-Hatsugai97}
The dependences on $\Phi_1$ and $\Phi_2$
in the zero mode of $H^{\vphantom{*}}_{\hbox{\tiny MDH}}$
were inadvertently exchanged in
Ref.\ \onlinecite{Hatsugai97}.

\bibitem{Kolmogorov37}
A.\ Kolmogorov,
I.\ Petrovsky,
and N.\ Piscounov,
Moscow Univ.\ Math.\ Bull.\ (Engl.\ Transl.)\textbf{1}, 1 (1937).

\bibitem{Bramson83}
M. Bramson,
\textit{
Convergence of Solutions of the Kolmogorov Equation to Traveling
Waves},
Memoirs of the American Mathematical Society, No.\ 285 
(American Mathematical Society, Providence, RI, 1983).

\bibitem{errorguru}
{{}}{
Neither the bosonization rules (4.14) and (4.15)
nor the overall sign of the action (4.16)
are chosen in a way consistent with Eq.\ (4.26)
although this inconsistency has no bearing on the main results
in Ref.\ \onlinecite{Guruswamy00},
{{}}{as these were rederived
independently in section 4.2 of Ref.\ \onlinecite{Guruswamy00}
from a more general point of view}.
To fix this ambiguity, recall that bosonization rules follow from
the gl($2|2$) current algebra. This algebra
is independent of the complex-valued parameter ${{}}{u}$
in Tables
\ref{table: bosonization rules: diag in flavor}
and
\ref{table: bosonization rules: off-diag in flavor}.
In order to preserve the SUSY of 
Eq.\ (\ref{eq: reorganised def bosonic int field theory N>1}),
bosonization rules must be supplemented by
boundary conditions on the scalar fields
$\phi^{\vphantom{\prime}}_a$ 
and 
$\phi^{\prime}_a$
at infinity. The boundary conditions are uniquely fixed by the
phase of the complex parameter ${{}}{u}$. 
The choice ${{}}{u/|u|}=-i$
imposes the boundary condition that the bosonized fields vanish at
infinity. We are indebted to 
A.\ W.\ W.\ Ludwig for clarification of this point.
       }

\bibitem{Friedan86}
D.\ Friedan, E.\ Martinec, and S.\ Shenker,
Nucl.\ Phys.\ B \textbf{271}, 93 (1986).

\bibitem{Gaberdiel96}
M.\ R.\ Gaberdiel and H.\ G.\ Kausch,
Phys.\ Lett.\ B \textbf{386}, 131 (1996).
See also H.\ G.\ Kausch, hep-th/9510149 (unpublished).

\bibitem{Rozansky92}
L.\ Rozansky and H.\ Saleur,
Nucl.\ Phys.\ B \textbf{376}, 461 (1992).

\bibitem{Altshuler86} 
B.\ L.\ Altshuler, V.\ E.\ Kravtsov, and I.\ V.\ Lerner,
JETP Lett. \textbf{43}, 440 (1986);
Sov.\ Phys.\ JETP \textbf{64}, 1352 (1986);
Phys.\ Lett.\ A \textbf{134}, 488 (1989).

\bibitem{Dorokhov82}
O.\ N.\ Dorokhov, 
JETP Lett.\ \textbf{36}, 318 (1982);
P.\ A.\ Mello, P.\ Pereyra, and N.\ Kumar, 
Ann. Phys. (N.Y.) \textbf{181}, 290 (1988).

\bibitem{Beenakker97} 
For a review see
C.\ W.\ J.\ Beenakker, 
Rev.\ Mod.\ Phys.\ \textbf{69}, 731 (1997).

\bibitem{Wegner80}
F.\ Wegner, 
Z.\ Phys.\ B \textbf{36}, 209 (1980);
in
\textit{Localization and Metal Insulator Transitions},
edited by H.\ Fritzsche and D.\ Adler, 
Institute for Amorphous Studies Series (Plenum Press, New York, 1985).

\bibitem{Falko95} 
V.\ I.\ Fal'ko and K.\ B.\ Efetov, 
Europhys.\ Lett.\ \textbf{32}, 627 (1995);
Phys.\ Rev.\ B \textbf{52}, 17413 (1996).  

\bibitem{Caux98}
J.-S.\ Caux,
Phys.\ Rev.\ Lett.\ \textbf{81}, 4196 (1998).

\bibitem{Dohmen96}
A.\ Dohmen, P.\ Freche, and M.\ Janssen,
Phys.\ Rev.\ Lett.\ \textbf{76}, 4207 (1996).

\bibitem{Huckestein97}
B.\ Huckestein and R.\ Klesse,
Phys.\ Rev.\ B \textbf{55}, R7303 (1997).

\bibitem{Mirlin00}
A.\ D.\ Mirlin,
Phys.\ Rep.\ \textbf{326}, 259 (2000);
A.\ D.\ Mirlin and F.\ Evers,
Phys.\ Rev.\ B \textbf{62}, 7920 (2000);
F.\ Evers, A.\ Mildenberger, and A.\ D.\ Mirlin,
\textit{ibid}.\ \textbf{64}, 241303(R) (2001).

\bibitem{Wegner83}
F.\ Wegner,
Z.\ Phys.\ B: Condens.\ Matter \textbf{51}, 279 (1983).

\bibitem{Zirnbauer99}
M.\ R.\ Zirnbauer,
hep-th/9905054
(unpublished).

\bibitem{Bhaseen00}  
M.\ J.\ Bhaseen, 
I.\ I.\ Kogan, 
O.\ A.\ Soloviev, 
N.\ Taniguchi, 
and A.\ M.\ Tsvelik,
Nucl.\ Phys.\ B \textbf{580}, 688 (2000).

\end{thebibliography}
\end{document}